\documentclass[10pt]{article}
\usepackage{amsmath,amssymb,color,graphicx,tikz,MnSymbol,stmaryrd}
\usepackage[mathcal]{euscript}
\DeclareMathAlphabet{\mathpzc}{OT1}{pzc}{m}{it}
\usepackage{dsfont}
\usepackage{bbm}
\usepackage[latin1]{inputenc}
\usepackage[english]{babel}
\usepackage{mathrsfs}
\usepackage[T1]{fontenc}
\usepackage{andsupp}
\usepackage{amscd}
\usepackage{boxedminipage}
\usepackage{cancel}
\usepackage{empheq}
\usepackage[line,cmtip,all,matrix,dvips]{xy}
\usetikzlibrary{positioning}

\newtheorem{theorem}{Theorem}[section]

\newtheorem{proposition}[theorem]{Proposition}
\newtheorem{lemma}[theorem]{Lemma}
\newtheorem{corollary}[theorem]{Corollary}
\newtheorem{example}{Example}[section]
\newtheorem{definition}[example]{Definition}
\newtheorem{remark}[example]{Remark}

\def\beaq{\begin{eqnarray}}
\def\eeaq{\end{eqnarray}}

\newcommand{\beq}{\begin{equation}}
\newcommand{\eeq}{\end{equation}}
\newcommand{\bea}{\begin{eqnarray}}
\newcommand{\eea}{\end{eqnarray}}
\newcommand{\dd}{\mathrm{d}}

\newcommand{\Res}{\mathop{\,\rm Res\,}}

\renewcommand{\epsilon}{\varepsilon}
\renewcommand{\rho}{\varrho}
\renewcommand{\phi}{\varphi}

\newcommand{\e}[1]{\mathbf #1}

\newcommand{\Bosonnormalord}{\stackrel{\textstyle \circ}{ \circ}}

\newcommand{\id}{\mathop{\fam0 Id}\nolimits}

\newcommand{\Aut}{\mathop{\fam0 Aut}\nolimits}

\newcommand{\Id}{\mathop{\fam0 Id}\nolimits}

\newcommand{\R}{\mathop{\fam0 {\mathbb R}}\nolimits}

\newcommand{\C}{\mathop{\fam0 {\mathbb C}}\nolimits}

\newcommand{\Z}{{\mathbb Z}}

\newcommand{\ra}{\mathop{\fam0 \rightarrow}\nolimits}

\newcommand{\T}{{\mathcal T}}

\newcommand{\cT}{\tilde{\mathcal T}}

\def\l{\lambda}\def\m{\mu}
\def\L{\Lambda}

\newcommand{\Si}{\Sigma}
\newcommand{\Sib}{{\mathbf \Sigma}}

\def\Res{\operatornamewithlimits{Res}}

\def\:{:}

\def\Bosonnormalordconstruction#1{\vcenter{\hbox{\ooalign{%
\raise.8ex\hbox{$#1\circ$}\crcr\lower.8ex\hbox{$#1\circ$}}}}}

\newcommand{\om}{\omega}

\newcommand{\bt}{\begin{theorem}}
\newcommand{\et}{\end{theorem}}
\newcommand{\br}{\begin{remark}}
\newcommand{\er}{\end{remark}}

\def\Bosonnormalord{\,\lower.8ex \hbox{$\circ$} \llap{\raise.8ex\hbox{$\circ$}} \,}
\def\normalord{\,\lower.8ex \hbox{$\cdot$} \llap{\raise.8ex\hbox{$\cdot$}} \,}

\def\Aut{\mathop{\rm Aut}\nolimits}

\def\ra{\rightarrow}

\def\Res{\mathop{\rm Res}\limits}

\def\Id{\mathop{\rm Id}\nolimits}

\begin{document}

\sloppy

\pagestyle{empty}
\addtolength{\baselineskip}{0.20\baselineskip}
\begin{center}

\vspace{26pt}

{\Large \textbf{Modular functors, cohomological field theories, \\ and topological recursion}}

\vspace{26pt}

\textsl{J\o{}rgen Ellegaard Andersen}\footnote{Centre for Quantum Geometry of Moduli Spaces, Department of Mathematics, Ny Munkegade 118, 8000 Aarhus C, Denmark. \\ \texttt{andersen@qgm.au.dk}}, \textsl{Ga\"etan Borot}\footnote{Max Planck Institut f\"ur Mathematik, Vivatsgasse 7, 53111 Bonn, Germany. \\ \texttt{gborot@mpim-bonn.mpg.de}}, \textsl{Nicolas Orantin}\footnote{\'{E}cole Polytechnique F\'{e}d\'{e}rale de Lausanne, D\'epartement de Math\'ematiques, 1015 Lausanne, Switzerland. \\ \texttt{nicolas.orantin@epfl.ch}}
\end{center}

\vspace{20pt}

%

\begin{center}
\textbf{Abstract}
\end{center}

Given a topological modular functor $\mathcal{V}$ in the sense of Walker \cite{Walker}, we construct vector bundles over $\overline{\mathcal{M}}_{g,n}$, whose Chern classes define semi-simple cohomological field theories. This construction depends on a determination of the logarithm of the eigenvalues of the Dehn twist and central element actions. We show that the intersection indices of the Chern class with the $\psi$-classes in $\overline{\mathcal{M}}_{g,n}$ is computed by the topological recursion of \cite{EOFg}, for a local spectral curve that we describe. In particular, we show how the Verlinde formula for the dimensions $D_{\vec{\lambda}}(\mathbf{\Sigma}_{g,n}) = \dim \mathcal{V}_{\vec{\lambda}}(\mathbf{\Sigma}_{g,n})$ is retrieved from the topological recursion. We analyze the consequences of our result on two examples: modular functors associated to a finite group $G$ (for which $D_{\vec{\lambda}}(\mathbf{\Sigma}_{g,n})$ enumerates certain $G$-principle bundles over a genus $g$ surface with $n$ boundary conditions specified by $\vec{\lambda}$), and the modular functor obtained from Wess-Zumino-Witten conformal field theory associated to a simple, simply-connected Lie group $G$ (for which $\mathcal{V}_{\vec{\lambda}}(\mathbf{\Sigma}_{g,n})$ is the Verlinde bundle).

\vspace{26pt}
\pagestyle{plain}
\setcounter{page}{1}
\vfill 
\subsubsection*{Acknowledgments}

We are grateful to the organizers of the AIM workshop ``Quantum curves, Hitchin systems, and the Eynard-Orantin theory" in October 2014, where this work was initiated and to AIM for their hospitality during this meeting. We thank Bertrand Eynard, Nikolai Reshetikhin, Chris Schommer-Pries and Dimitri Zvonkine for discussions. JEA is supported in part by the center of excellence grant ``Center for Quantum Geometry of Moduli Spaces (QGM)" from the Danish National Research Foundation (DNRF95). The work of GB is supported by the Max-Planck Gesellschaft. He would like to thank the QGM at Aarhus University, Berkeley, Caltech and EPFL Mathematics Departments, and the Caltech Theoretical Physics Group for hospitality, while this work was conducted. NO would like to thank the QGM at Aarhus University and the Max-Planck Institute for their hospitality at different stages of this work.

\vspace{0.5cm}

\newpage

\setcounter{tocdepth}{2}
\tableofcontents

\newpage


\section{Introduction}

\subsection*{Background}

The pioneering works of Atiyah, Segal and Witten turned $2d$ conformal field theories (CFT) \cite{BPZCFT} into an effective machinery to design interesting  $3$-manifold invariants, known under the name of "quantum invariants". More thoroughly, it allowed the construction of $3d$ topological quantum field theories (TQFT), where the $2d$ CFT is thought of as living on the boundary of $3$-manifolds. The axioms of a $3d$ TQFT were worked out in details by Reshetikhin and Turaev, and they further constructed the first and most important example from the quantum group $U_q(\mathfrak{sl}_{2})$ at $q = $a root of unity \cite{TuraRe1,TuraRe2,Tu}.

There exist several variants of axiomatizations that embody the concept of a CFT. This article deals with one of these axiomatizations for the topological part of a CFT, called "modular functor". It was proposed by Segal in the context of rational conformal field theory \cite{Se} covering the holomorphic part of CFT, and developed in Walker's notes \cite{Walker} in the purely topological context. Namely, we consider functors from the category of marked surfaces -- with projective tangent vectors and labels in a finite set $\Lambda$ at punctures, and a Lagrangian subspace of the first homology --, to the category of finite dimensional complex vector spaces. In particular, such a functor determines a representations of a central extension of the mapping class groups. The main property required to be a modular functor is that the vector spaces attached to a surface enjoy a factorization property when the surface is pinched. The full definitions are given in Section~\ref{S2}. From the data of a modular functor in this sense, \cite{KRCFT,Grove} show how to obtain a $(2 + 1)$-dimensional TQFT.

The main source of modular functors are modular tensor categories (MTC) \cite{Tu,BB,AP}. At present, it seems that all known examples of modular functors come from a modular tensor category, but it is not known whether all modular functors are of this kind. Among examples of MTC, we find some categories of representations of quantum groups \cite{Tu}, and categories of representations of vertex operator algebras (VOA) \cite{FHL,HuangVOAMTC}.

The Wess-Zumino-Witten models form a well-studied class of examples of this type. The MTC here arises from representations of a VOA constructed from affine Ka\v{c}-Moody algebras $\widehat{\mathfrak{g}}$ \cite{Kacbook}. It gives rise to Hilbert spaces of a TQFT, which come as vector bundles (the so-called Verlinde bundles) over a family of complex curves with coordinates, equipped with a projectively flat connection \cite{TUY}, which coincides with Hitchin's connection from the point of view of geometric quantization \cite{Laszlo}. The choice of coordinates can actually be bypassed \cite{TUY,Tsuchimoto,LooiWZW} and the Verlinde bundles exist as bundles $\mathcal{V}_{g,n}^{{\rm WZW}}$ over the moduli space of curves, and extend nicely to the Deligne-Mumford compactification. The explicit construction of a modular functor from this perspective -- also called the CFT approach -- was described in \cite{AU1,AU2}. There is another approach, based on a category of representations of a quantum group $U_{q}(\mathfrak{g}_{\mathbb{C}})$ at certain roots of unity. It leads to the Witten-Reshetikhin-Turaev TQFT, constructed in \cite{TuraRe1,TuraRe2,BHMV2} for $\mathfrak{g}_{\mathbb{C}} = \mathfrak{sl}_{2}$, and in \cite{TuraWz} for any simple Lie algebra of type ABCD. This theory for $\mathfrak{sl}_{2}$ was also constructed using skein theory by Blanchet, Habegger, Masbaum and Vogel in \cite{BHMV1,BHMV2} and for $\mathfrak{sl}_{N}$ by Blanchet in \cite{Bl1} for. As anticipated by Witten, the CFT approach and the quantum group approach should give equivalent TQFTs. For instance, the equivalence of the modular functors was established by the first author of this paper and Ueno in \cite{AU3,AU4} for $\mathfrak{g} = \mathfrak{su}_N$.

The rank of the Verlinde bundle is already a non-trivial invariant, which is computed by the famous Verlinde formula \cite{Verlinde,MooreSeibergo,Faltings,Beau}. Marian et al. \cite{Zvonkin} lately showed that the Chern polynomial of $\mathcal{V}_{g,n}^{{\rm WZW}}$ defines a semi-simple cohomological field theory (CohFT). It can be characterized in terms of its R-matrix thanks to the classification results of Givental and Teleman \cite{Tele,Giventals}: from the R-matrix, one can build the exponential of a second-order differential operator, which acts on a product of several copies of the Witten-Kontsevich generating series of $\psi$-classes (the matrix Airy function/KdV tau function of \cite{Kontsevich}), and returns the generating series of the intersection of the Chern class at hand with an arbitrary product of $\psi$-classes. We call these invariants the "CohFT correlation functions". When the variable $t$ of the Chern polynomial is set to $0$, these correlation functions return the rank of the bundle, which can be thought of as the "$2d$ TQFT correlation functions".

\subsection*{Contribution of the article}

In the present article, we generalize the results of \cite{Zvonkin} to any modular functor -- hence not relying on the peculiarities of the Wess-Zumino-Witten models. For a given modular functor, we construct a trivial bundle $[\tilde{\mathcal{Z}}^0_{\vec{\lambda}}]_{g,n}$ over Teichm\"uller space, which, after twisting by suitable line bundles, descends to a bundle $[\mathcal{Z}_{\vec{\lambda}}]_{g,n}$ over $\mathcal{M}_{g,n}$ (Theorem~\ref{th2}). We can use Chern-Weil theory to compute the Chern class of $[\mathcal{Z}_{\vec{\lambda}}]_{g,n}$ (Proposition~\ref{propCH}) in terms of $\psi$-classes and the first Chern class of the Hodge bundle -- their coefficients are related respectively to the central charge $c$ and the log of Dehn twist eigenvalues $r_{\lambda}$ (aka conformal weights). Besides, we show that our bundle extends to the boundary of $\overline{\mathcal{M}}_{g,n}$ (Theorem~\ref{thaa3}). All together, this constitutes our first main result. Since $\overline{\mathcal{M}}_{g,n}$ is an orbifold, the Chern class of our bundle must have rational coefficients, hence a new (geometric) proof of Vafa's theorem \cite{Vafa} stating that $c$ and $r_{\lambda}$ must be rational. Our proof actually shows this for any modular functor, including non-unitary cases.

Since our bundle enjoys factorization implied by the axioms of a modular functor, we can conclude that ${\rm Ch}_{t}([\mathcal{Z}_{\vec{\lambda}}]_{g,n})$ defines a semi-simple cohomological field theory on a Frobenius algebra $\mathcal{A}$ whose underlying vector space is $\mathbb{C}[\Lambda]$ (Theorem~\ref{CohFTh}). Because the twists can depend on log-determinations for the central charge and the conformal weights, and because we have the Chern polynomial variable $t \in \mathbb{C}$ (introduced to keep track of each degree seperately), we actually produce a $1$-parameter family of CohFTs. The existence of an S-matrix that diagonalizes the product in $\mathcal{A}$ ensures the semi-simplicity of these theories, and we compute the $R$-matrix of these CohFT in terms of the $S$-matrix (Proposition~\ref{Rcomput}).

Then, from the general result of \cite{DBOSS}, we know that the correlation functions of these CohFTs is computed by the topological recursion of \cite{EOFg} for a local spectral curve. We describe explicitly this local spectral curve and the relevant initial data $(\omega_{0,1},\omega_{0,2})$, and it depends solely on the genus $1$ representation of the modular functor (Proposition~\ref{TH42}). If the variable $t$ of the Chern polynomial is set to $0$, we retrieve Verlinde's formula for the rank of our bundle as a special case of the topological recursion (Proposition~\ref{P2p2}); in general, we obtain that the $\omega_{g,n}$'s of the topological recursion for this spectral curve are expressed in terms of integrals of the Chern polynomial and $\psi$-classes (Equation~\ref{CMUGN}). This formula is our second main result. The initial data 

We illustrate our findings on two classes of Wess-Zumino-Witten models. In Section~\ref{ex:finiteMF}, we address the modular functors associated to a finite group $G$ \cite{WV,DVVV,Freed}. They are also called "orbifold holomorphic models". In the "untwisted case", their Frobenius algebra contains simultaneously the fusion rules of the representation ring of $G$ and (albeit undirectly) the decomposition of product of conjugacy classes. The dimensions of the TQFT vector spaces count certain $G$-principle bundles over the surface in question. We therefore find -- in a rather trivial way -- a topological recursion for these numbers, where the induction concerns the Euler characteristics of the base. In the "untwisted case" we obtain a degree $0$ CohFT, which only remembers the dimension of the vector spaces, but in the twisted case, the CohFT is in general non-trivial (see Lemma~\ref{fdkm}). In Section~\ref{ex:WZW}, we examine the Wess-Zumino-Witten models based on a compact Lie group $G$ at level $\ell$, for which $[\mathcal{Z}_{\vec{\lambda}}]_{g,n}$ is the Verlinde bundle studied in \cite{Zvonkin}. Remarkably, for $SU(N)_{\ell}$, we find that $\omega_{0,2}$ for the local spectral curve is expressible in terms of a suitable truncation of double Hurwitz numbers (number of branched coverings over the Riemann sphere). This poses the question of the combinatorial interpretation of the correlation function of these CohFTs, maybe in relation with number of coverings over a surface of arbitrary topology.

To summarize, from a physical point of view, we have associated to a modular functor a CohFT that should encode Gromov-Witten theory of a target space $X$. Having a modular functor means that the worldsheet (a surface of genus $g$ with $n$ boundaries) roughly speaking carries a CFT. As we comment in Section~\ref{Globals}, the local spectral curve used in Section~\ref{S3} for the topological recursion, should describe the vicinity of isolated singularities in a Landau-Ginzburg model $\tilde{X}$. As of now, the description of the geometry of $X$ and $\tilde{X}$ is unclear to us.

\begin{figure}
\begin{center}
\includegraphics[width=\textwidth]{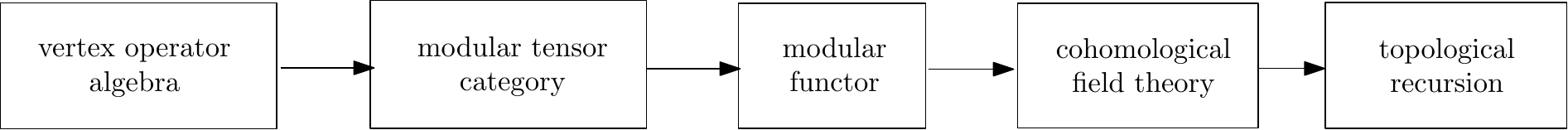}
\caption{\label{mum} $A \rightarrow B$ means that $B$ (or a set of quantities fulfilling the axioms of $B$) can be obtained from the data of $A$ ; these are not equivalences, i.e. in general $A$ cannot be fully retrieved from $B$, and neither all $B$ come from an $A$. A wealth of definitions sometimes fall under the same names. To avoid ambiguities: VOA are as in \cite{Huangannon} ; the notion of modular functor we adopt is described in Section~\ref{S23}, following the approach of Walker \cite{Walker} ; the notion of CohFT is described in Section~\ref{Gen}, simplifying the axiomatization given by Kontsevich and Manin \cite{KMCohFT} ; the topological recursion is described in Section~\ref{Locals}, following Eynard and the third author \cite{EORev}.}
\end{center}
\end{figure}

\section{Construction of vector bundles from a modular functor}

We introduce the category of marked surfaces, their automorphism groups, and review the axioms of a modular functor. The target category is that of finite dimensional vector spaces over the field of complex numbers. The motivation to introduce marked surfaces is explained e.g. in \cite{Walker}: in quantum field theory, if one works with a naive category of surfaces, the states have a phase ambiguity, so that the target category would rather be that of projective vector spaces. The marking allows the resolution of phase ambiguities and working with vector spaces. 

We then explain, in Section~\ref{fki}, how to obtain from any given modular functor, a family of vector bundles over the moduli space of curves. We compute its Chern class in terms of the basic data of the modular functor (Theorem~\ref{propCH}). A delicate but essential point for Section~\ref{S3} is to prove that the bundles extend to the Deligne-Mumford compactification of the moduli space. This is achieved in Section~\ref{fki6}, with a detour via a borderfication of the Teichm\"uller space.

\label{S2}
\subsection{The category of marked surfaces}
\label{S21}
Let us start by fixing notations. By a closed surface we mean a
smooth real $2$-dimensional, compact manifold.
For a closed oriented surface $\Si$ of genus $g$ we have the
non-degenerate skew-symmetric intersection pairing
\[(\cdot,\cdot)\,:\,\, H_1(\Si,\Z) \times H_1(\Si,\Z) \ra \Z.\]
Suppose $\Si$ is connected. In this case a Lagrangian $L\subseteq H_1(\Si,\Z)$
is by definition a subspace which is maximally isotropic with respect to the
intersection pairing. A
$\Z$-basis $(\vec \alpha, \vec \beta) = (\alpha_1,\ldots, \alpha_g,\beta_1, \ldots
\beta_g)$ for $H_1(\Si,\Z)$ is called a symplectic basis if
\[(\alpha_i,\beta_j) = \delta_{ij}, \quad (\alpha_i,\alpha_j) = (\beta_i,\beta_j) = 0,\]
for all $i,j \in \llbracket 1,g \rrbracket$. If $\Si$ is not connected, then $H_1(\Si,\Z) =
\bigoplus_i H_1(\Si_i,\Z)$, where $\Si_i$ are the connected
components of $\Si$. In this context, by definition in this paper, a Lagrangian subspace is a subspace of the form $L = \bigoplus_i L_i$, where $L_i\subset
H_1(\Si_i,\Z)$ is Lagrangian. Likewise a symplectic basis for
$H_1(\Si,\Z)$ is a $\Z$-basis of the form $((\vec \alpha^i, \vec
\beta^i))$, where $(\vec \alpha^i, \vec
\beta^i)$ is a symplectic basis for $H_1(\Si_i,\Z)$.

For any real vector space $V$, we define $\mathcal{P}V = (V-\{0\})/\R{}_+.$

\begin{definition}\label{DefPointS}
A {\em pointed surface} $(\Si,P)$ is an oriented closed
surface $\Si$ with a finite set $P\subset \Si$ of points.\end{definition}

\begin{definition}\label{DefMorPointS}
A {\em morphism of pointed surfaces} $f :(\Si_1,P_1) \ra
(\Si_2,P_2)$ is an isotopy class of orientation preserving
diffeomorphisms which maps $P_1$ to $P_2$. 
The group of automorphisms of a pointed surface $(\Si,P)$ is the mapping class group, and will be denoted
$\Gamma_{\Si,P}$. It consists of the isotopy classes of orientation preserving diffeomorphisms of $\Sigma$ which are the identity on $P$. The {\em framed mapping class group} of a pointed surface is denoted $\tilde \Gamma_{\Si,P}$ and it consists of isotopy classes of orientation preserving diffeomorphisms which are the identity on $P$ as well as on the tangent spaces at $p \in P$.
\end{definition}

We stress that for the (framed) mapping class group, the isotopies allowed in the equivalence relation must also be the identity on $P$ (and on the tangent spaces at $P$). We clearly have the following Lemma.

\begin{lemma}
There is a short exact sequence
$$ 1 \longrightarrow  \Z^{P} \longrightarrow \tilde \Gamma_{\Si,P} \longrightarrow \Gamma_{\Si,P} \ra 1,$$
where the generator of the factor of $\Z$ corresponding to $p\in P$ is given by the Dehn twist $\delta_p$ in the boundary of an embedded disk in $\Sigma - (P-\{p\})$ and centred in $p$.
\end{lemma}

\begin{definition} \label{msurface}

A {\em marked surface\/} $ {\Sib} = (\Si, P, V, L)$ is an oriented
closed surface $\Si$ with a finite subset $P \subset \Si$
of points with projective tangent vectors $V\in \prod_{p \in
P}\mathcal{P}T_{p}\Si$ and a Lagrangian subspace $L \subseteq H_1(\Si,\Z)$.
\end{definition}

\begin{definition} \label{mmorphism}

A {\em morphism\/} $\e f : {\Sib}_1 \to {\Sib}_2$ of marked
surfaces ${\Sib}_i = (\Si_i,P_i,V_i,L_i)$ is an isotopy class of
orientation preserving diffeomorphisms $f : \Si_1 \to \Si_2$ that
maps $(P_1,V_1)$ to $(P_2,V_2)$ together with an integer $s$.
Hence we write $\e f = (f,s)$.
\end{definition}

Let $\sigma$ be Wall's signature cocycle for triples of Lagrangian
subspaces of $H_1(\Si,\R)$ \cite{Wall}.

\begin{definition} \label{composition}
Let $\e f_1 = (f_1,s_1) : {\Sib}_1 \to {\Sib}_2$ and $\e f_2 =
(f_2,s_2) : {\Sib}_2 \to {\Sib}_3$ be morphisms of marked surfaces
${\Sib}_i = (\Si_i,P_i,V_i,L_i)$. Then, the {\it composition\/} of
$\e f_1$ and $\e f_2$ is $$ \e f_2 \e f_1 = (f_2 f_1, s_2 + s_1 -
\sigma((f_2f_1)_*L_1, f_{2*}L_2,L_3)). $$
\end{definition}

With the objects being marked surfaces and the morphisms and their
composition being defined as above, we have
constructed the category of marked surfaces. 

The mapping class group $\Gamma_{\Sib}$ of a marked surface
${\Sib} = (\Si,L)$ is the group of automorphisms of ${\Sib}$. $\Gamma_{\Sib}$ is a central extension of the
framed mapping class group $\tilde \Gamma_{\Si, P}$ of the pointed surface $(\Si,P)$ 
$$ 1 \longrightarrow \Z \longrightarrow \Gamma_{\Sib} \longrightarrow \tilde \Gamma_{\Si, P} \longrightarrow 1$$
defined by
the 2-cocycle $\gamma : \Gamma_{\Sib} \to \mathbb Z$, $\gamma(f_2,f_1) =
\sigma((f_2f_1)_*L,f_{2*}L,L)$. It is known that this
cocycle is equivalent to the cocycle obtained by considering $2$-framings on mapping cylinders, see \cite{At1} and \cite{A}. Briefly, the relation is as follows: A  $2$-framing is determined by the first Pontryagin number $p_1$; Hirzebruch's formula says that $p_1$ is three times the signature of the $4$-manifold and, by construction \cite{Wall}, $\sigma$ expresses the non-additivity of the signature.

Notice also that for any morphism $(f,s) : \Sib_1 \to \Sib_2$, we
can factor
\begin{eqnarray*}
(f,s) &=& \left((\id,s') : \Sib_2 \to \Sib_2\right) \circ
(f,s-s')\\ &=& (f,s-s') \circ \left((\id,s') : \Sib_1 \to
\Sib_1\right).
\end{eqnarray*}
In particular $(\id,s) : {\Sib} \to {\Sib}$ is $(\id,1)^s$.

\subsection{Operations on marked surfaces}
\label{S22} 
\begin{definition} \label{disjunion}
The operation of {\em disjoint union of marked surfaces} is $$
(\Si_1,P_1,V_1,L_1)
 \sqcup (\Si_2,P_2,V_2,L_2) = (\Si_1 \sqcup \Si_2,P_1 \sqcup P_2,V_1\sqcup V_2,L_1 \oplus
L_2). $$
Morphisms on disjoint unions are accordingly $(f_1,s_1) \sqcup
(f_2,s_2) = (f_1 \sqcup f_2,s_1 + s_2)$.
\end{definition}

We see that the disjoint union is an operation on the category of
marked surfaces.

\begin{definition}\label{or}
Let ${\Sib}$ be a marked surface. We denote by $- {\Sib}$ the
marked surface obtained from ${\Sib}$ by the {\em operation of
reversal of the orientation}. For a morphism $\e f = (f,s) :
{\Sib}_1 \to {\Sib}_2$ we let the orientation reversed morphism be
given by
 $- \e f = (f,-s) : -{\Sib}_1 \to -{\Sib}_2$.
\end{definition}

We also see that orientation reversal is an operation on the
category of marked surfaces.

Let us now consider glueing of marked
surfaces. Let $(\Si, \{p_-,p_+\}\sqcup P,\{v_-,v_+\}\sqcup V,L)$ be a marked
surface, where we have selected an ordered pair $(p_-,p_+)$ of marked points
with projective tangent vectors $((p_-,v_-),(p_+,v_+))$, at which
we will perform the glueing. Let $C : \mathcal{P}(T_{p_-}\Si) \ra \mathcal{P}(T_{p_+}\Si)$ be an orientation
reversing projective linear isomorphism such that $C(v_-) = v_+$.
Such a $C$ is called a {\em glueing map} for $\Si$. Let
$\tilde{\Si}$ be the oriented surface with boundary obtained from
$\Si$ by blowing up $p_-$ and $p_+$, i.e.
\[\tilde{\Si} = (\Si -\{p_-,p_+\})\sqcup \mathcal{P}(T_{p_-}\Si)\sqcup \mathcal{P}(T_{p_+}\Si),\]
with the natural smooth structure induced from $\Si$. Let now
$\Si_C$ be the closed oriented surface obtained from $\tilde{\Si}$
by using $C$ to glue the two boundary components of $\tilde{\Si}$ corresponding to $p_\pm$. We
call $\Si_C$ the glueing of $\Si$ at the ordered pair
$((p_-,v_-),(p_+,v_+))$ with respect to $C$.

Let now $\Si'$ be the topological space obtained from $\Si$ by
identifying $p_-$ and $p_+$. We then have natural continuous maps
$q : \Si_C \ra \Si'$ and $n : \Si \ra \Si'$. On the first homology
group $n$ induces an injection and $q$ a surjection, so we can
define a Lagrangian subspace $L_C \subseteq H_1(\Si_C,\Z)$ by $L_C =
q_*^{-1}(n_*(L))$. We note that the image of $\mathcal{P}(T_{p_-}\Si)$ (with
the orientation induced from $\tilde{\Si}$) induces naturally a line  in $H_1(\Si_C,\Z)$ and as such it is contained in $L_C$.

\begin{remark}{\em \label{remarkglue2}
If we have two glueing maps $C_i : \mathcal{P}(T_{p_-}\Si) \ra
\mathcal{P}(T_{p_+}\Si),$ $i=1,2$, we note that there is a diffeomorphism
$f\,:\,\Si \ra \Si$ inducing the identity on
$(p_-,v_-)\sqcup(p_+,v_+)\sqcup(P,V)$ which is isotopic to the
identity among such maps, and such that $(\dd f_{p_+})^{-1} C_2 \dd f_{p_-}
= C_1$. In particular $f$ induces a diffeomorphism $f : \Si_{C_1}
\ra \Si_{C_2}$ compatible with $f : \Si \ra \Si$, which maps
$L_{C_1}$ to $L_{C_2}$. Any two such diffeomorphisms of $\Si$
induce isotopic diffeomorphisms from $\Si_1$ to
$\Si_2$.}\end{remark}

\begin{definition} \label{glueing}
Let ${\Sib} = (\Si, \{p_-,p_+\}\sqcup P,\{v_-,v_+\}\sqcup V,L)$ be
a marked surface. Let $$C : \mathcal{P}(T_{p_-}\Si) \ra \mathcal{P}(T_{p_+}\Si)$$ be a
glueing map and $\Si_C$ the glueing of $\Si$ at the ordered pair
$((p_-,v_-),(p_+,v_+))$ with respect to $C$. Let $L_C \subseteq
H_1(\Si_C,\Z)$ be the Lagrangian subspace constructed above from
$L$. Then the marked surface ${\Sib}_C = (\Si_C,P,V,L_C)$ is
defined to be the {\em glueing} of ${\Sib}$ at the ordered pair
$((p_-,v_-),(p_+,v_+))$ with respect to $C$.
\end{definition}

We observe that glueing also extends to morphisms of marked
surfaces which preserves the ordered pair $((p_-,v_-),(p_+,v_+))$,
by using glueing maps which are compatible with the morphism in
question.

\subsection{The axioms for a modular functor}
\label{S23}
We now give the axioms for a modular functor.
This notion is due to G.~Segal and appeared first in \cite{Se}. We
present them here in a topological form, which is due to Walker
\cite{Walker}. We note that similar, but
different, axioms for a modular functor are given in \cite{Tu}, relying on modular tensor categories. At present, it is not known whether the definition in \cite{Tu} is equivalent to ours.

\begin{definition} \label{DefLS}

A {\em label set\/} $\L$ is a finite set equipped with an
involution $\l \mapsto \l^{\dagger}$ and a trivial element $1$ such
that $1^{\dagger} = 1$.
\end{definition}

\begin{definition} \label{lmsurface}

Let $\L$ be a label set. The category of {\em $\L$-labeled marked
surfaces\/} consists of marked surfaces with an element of $\L$
assigned to each of the marked point. An
assignment of elements of $\L$ to the marked points of ${\Sib}$ is
called a labeling of ${\Sib}$ and we denote the labeled marked
surface by $({\Sib},\vec{\l})$, where $\vec{\l}$ is the labeling. Morphisms of labeled
marked surfaces are required to preserve the labelings. 
\end{definition}

We define a labeled pointed surface similarly.

\begin{remark}{\em
The operation of disjoint union clearly extends to labeled marked
surfaces. When we extend the operation of orientation reversal to
labeled marked surfaces, we also apply the involution ${}^{\dagger}$ to all the labels. }\end{remark}

\begin{definition} \label{DefMF}
A {\em modular functor\/} based on the label set $\L$ is a functor
$\mathcal{V}$ from the category of labeled marked surfaces to the category
of finite dimensional complex vector spaces satisfying the axioms
MF1 to MF5 below.
\end{definition}

\subsubsection*{MF1} {\it Disjoint union axiom\/}. The operation of disjoint
union of labeled marked surfaces is taken to the operation of
tensor product, i.e. for any pair of labeled marked surfaces there
is an isomorphism $$ \mathcal{V}(({\Sib}_1,\vec{\l_1}) \sqcup ({\Sib}_2,\vec{\l_2}))
\cong \mathcal{V}({\Sib}_1,\vec{\l_1}) \otimes \mathcal{V}({\Sib}_2,\vec{\l_2}). $$ The
identification is associative.

\subsubsection*{MF2} {\it Glueing axiom\/}. Let ${\Sib} $ and ${\Sib}_C$ be
marked surfaces such that ${\Sib}_C$ is obtained from ${\Sib} $ by
glueing at an ordered pair of points and projective tangent
vectors with respect to a glueing map $C$. Then there is an
isomorphism $$ \mathcal{V}({\Sib}_C,\vec{\lambda}) \cong \bigoplus_{\m \in \L}
\mathcal{V}({\Sib},\m,\m^{\dagger},\vec{\l}), $$ which is associative, compatible with
glueing of morphisms, disjoint unions and it is independent of the
choice of the glueing map in the obvious way (see Remark
\ref{remarkglue2}).

\subsubsection*{MF3} {\it Empty surface axiom\/}. Let $\emptyset$ denote
the empty labeled marked surface. Then $$ \dim \mathcal{V}(\emptyset) = 1.
$$

\subsubsection*{MF4} {\it Once punctured sphere axiom\/}. Let $\Sib = (S^2,
\{p\},\{v\},0)$ be a marked sphere with one marked point. Then
$$
\dim \mathcal{V}(\Sib,\l) = \left\{ \begin{array}{ll} 1,\qquad &\l = 1\\
0,\qquad & \l \ne 1.\end{array}\right. $$

\subsubsection*{MF5} {\it Twice punctured sphere axiom\/}. Let $\Sib = (S^2,
\{p_1,p_2\},\{v_1,v_2\},\{0\})$ be a marked sphere with two marked
points. Then $$ \dim \mathcal{V}(\Sib,\l,\mu) = \left\{ \begin{array}{ll}
1, \qquad &\l = \mu^{\dagger} \\ 0,\qquad &\l \ne 
\mu^{\dagger}.\end{array}\right. $$

In addition to the above axioms one may require extra properties,
namely:

\subsubsection*{MF-D} {\it Orientation reversal axiom\/}.
The operation of orientation reversal of labeled marked surfaces
is taken to the operation of taking the dual vector space, i.e for
any labeled marked surface $({\Sib},\vec{\l})$ there is a pairing
\beq
\label{PPO}\langle \cdot,\cdot\rangle : \mathcal{V}({\Sib},\vec{\l}) \otimes \mathcal{V}(-{\Sib},\vec{
\l}^{\dagger}) \longrightarrow \C{},
\eeq
compatible with disjoint unions, glueings and
orientation reversals (in the sense that the induced isomorphisms
$ \mathcal{V}({\Sib},\vec{\l}) \cong \mathcal{V}(-{\Sib},\vec{\l}^{\dagger})^*$ and $\mathcal{V}(-{\Sib},\vec{\l}^{\dagger})
\cong \mathcal{V}({\Sib},\vec{\l})^*$ are adjoints).

\subsubsection*{MF-U} {\it Unitarity axiom\/}. Every vector
space $\mathcal{V}({\Sib},\vec{\l})$ is furnished with a hermitian inner product
$$ ( \cdot,\cdot ) : \overline{\mathcal{V}({\Sib},\vec{\l})} \otimes \mathcal{V}({\Sib},\vec{\l}) \longrightarrow {\mathbb C} $$ so that morphisms induce unitary
transformations. The hermitian structure must be compatible with
disjoint union and glueing. If we have the orientation reversal
property, then compatibility with the unitary structure means that
we have commutative diagrams $$\begin{CD} \mathcal{V}({\Sib},\vec{\l}) @>>\cong>
\mathcal{V}(-{\Sib},\vec{\l}^{\dagger})^*\\ @VV\cong V @V\cong VV\\
\overline{\mathcal{V}({\Sib},\vec{\l})^*} @>\cong>> \overline{\mathcal{V}(-{\Sib},\vec{\l}^{\dagger})},
\end{CD}$$
where the vertical identifications come from the hermitian
structure and the horizontal identifications from the pairing \eqref{PPO}.

\subsection{The Teichm\"uller space of marked surfaces}

Let us first review some basic Teichm\"{u}ller theory. Let $\Si$
be a closed oriented smooth surface and let $P$ be a finite set of
points on $\Si$. The usual Teichm\"uller space for a pointed surface $(\Sigma,P)$ consists of equivalence classes of diffeomorphisms $ \phi\,:\,\,(\Sigma,P) \rightarrow (C,Q)$, where $C$ is a Riemann surface and $Q\subset C$ is finite set of points, namely
\beq
\label{TSP} \T_{\Sigma,P} = \left\{\phi\,:\,\,(\Sigma,P) \rightarrow (C,Q)\right\}/\sim
\eeq
where we declare that two diffeomorphisms $\phi_i\,:\,(\Sigma,P) \rightarrow (C_i,Q_i)$ for $i = 1,2$ are equivalent if there exists a biholomorphic map $\Phi\,:\,(C_1,Q_1) \rightarrow (C_2,Q_2)$ such that $\Phi\circ \phi_1$ is isotopic to $\phi_2$ by diffeomorphisms preserving $P$. If $P = \emptyset$, this space is simply denoted $\mathcal{T}_{\Sigma}$. We will also consider the ``decorated Teichm\"uller space" consisting of equivalence classes of diffeomorphisms $ \phi\,:\,\,(\Sigma,P) \rightarrow (C,Q)$, where $C$ is a Riemann surface, $Q\subset C$ is finite set of points, and $W\in T_QC$ non-zero tangent vectors
$$
\cT_{\Si,P} = \left\{\phi\,:\,\,(\Sigma,P) \rightarrow (C,Q,W) \right\}/\approx
$$
where $\approx$ is now the equivalence relation where we ask that the isotopies preserve $(P, (\dd_P\varphi)^{-1}(W))$. 
We have natural projection maps $\tilde \pi_{\Sigma,P} : \tilde{\T}_{\Sigma,P}  \ra \T_{\Sigma,P}$, 
$\pi_P : \T_{\Sigma,P}  \ra T_P\Sigma$ and $\pi_\Si : \T_{\Sigma,P}  \ra \T_{\Sigma}$.

\begin{theorem}[Bers]
There is a natural structure of a finite dimensional complex analytic manifold on
the Teichm\"{u}ller spaces $\T_{\Si,P}$ and $ \cT_{\Si,P}$.
The mapping class group $\Gamma_{\Si,P}$ acts biholomorphically on $\T_{\Si,P}$, as does  $\tilde \Gamma_{\Si,P}$ on $\cT_{\Si,P}$.
\end{theorem}

\begin{proposition}
\label{Teichbun}
$\cT_{\Sigma,P}$ is a principal $\C^{P}$-bundle over $\T_{\Sigma,P}$ on which $\tilde \Gamma_{\Si,P}$ acts, covering the action of $\Gamma_{\Si,P}$  on $\T_{\Sigma,P}$.
Moreover $\T'_{\Sigma,P} = \cT_{\Sigma,P}/\langle \delta_p\mid p\in P\rangle$ is a principal $(\C^*)^P$-bundle over $\T_{\Sigma,P}$, such that the induced projection $\tilde \pi'_{\Sigma,P} : \cT_{\Sigma,P} \ra \T'_{\Sigma,P}$ is the fiberwise universal cover with respect to the projection $\pi'_{\Sigma,P} : \T'_{\Sigma,P} \ra \T_{\Sigma,P}$, compatible with the exponential map $e^{2\pi {\rm i} \cdot} : \C^{P} \ra (\C^*)^P$ on the structure groups.

\end{proposition}

\noindent\textbf{Proof.} For every $\varphi : (\Si,P) \ra (C,Q,W)$ representing a point in $\T_{\Si,P}$, we have the map
$$\pi_P|_{\tilde\pi_{\Sigma,P}^{-1}([\varphi])} : \tilde\pi^{-1}_{\Sigma,P}([\varphi]) \ra T_P\Sigma,$$
induced by assigning $(\dd_P\varphi')^{-1}(W)$ to a diffeomorphism $\varphi' : (\Si,P) \ra (C,Q,W)$ which represents a point $\tilde\pi_{\Sigma,P}^{-1}([\varphi])$. By the very definition of $\cT_{\Si,P}$ this map is independent of the representative $\varphi'$ of a point in $\tilde\pi^{-1}_{\Sigma,P}([\varphi])$. Now pick a $\varphi' : (\Si,P) \ra (C,Q,W)$ representing a point in $\tilde\pi^{-1}_{\Sigma,P}([\varphi])$ and let $V\in T_P\Si$ be given by

$$ V = (\dd_P\varphi')^{-1}(W).$$
Then consider the set $(\pi_P|_{\tilde\pi^{-1}([\varphi])})^{-1}(V)$. We claim that the group $\langle \delta_p\,\, | \,\, p\in P \rangle$ acts transitively on this set. To see this, let $\varphi'' : (\Si,P) \ra (C,Q,W)$ represent another point in $(\pi_P|_{\tilde\pi^{-1}_{\Sigma,P}([\varphi])})^{-1}(V)$.  Since the two diffeomorphisms $\varphi'$ and $\varphi''$ must represent the same element in $\T_{\Si,P}$, there exists $\Phi$ such that 
$$ (\varphi'')^{-1} \circ \Phi \circ \varphi' : (\Si,P) \ra (\Si,P)$$
is isotopic to the identity. But then there exists a diffeomorphism $\psi: (\Si,P) \ra (\Si,P)$ which represents an element in $ \langle \delta_p\,\, |\,\, p\in P \rangle$ such that 
$$ (\varphi'')^{-1} \circ \Phi \circ (\varphi' \circ \psi): (\Si,P) \ra (\Si,P)$$
is isotopic to the identity within diffeomorphisms of $(\Si,P,V)$. This means that $\varphi' \circ \psi$ represents the same point in $\cT_{\Si,P}$ as $\varphi''$ does.

It now follows that $\cT_{\Sigma,P}$ is a principal $\C^{P}$-bundle over $\T_{\Sigma,P}$ and that $\T'_{\Sigma,P} = \cT_{\Sigma,P}/\langle \delta_p\mid p\in P\rangle$ is a principal $(\C^*)^P$-bundle over $\T_{\Sigma,P}$. \hfill $\Box$

\subsection{Construction of line bundles over Teichm\"uller spaces}
\label{Cuns}
\subsubsection{Fiber products}

For each $p\in P$, we consider the representation $\rho_p : (\C^*)^P \ra \Aut(\C)$, which is obtained by projection on the factor corresponding to $p$. We denote $\tilde{\mathcal{L}}_p$ the line bundle over $\mathcal{T}_{\Sigma,P}$ associated to $\mathcal{T}'_{\Sigma,P}$ and the representation $\rho_p$. 

Take $\alpha \in \mathbb{C}$. We now show how to construct the $\alpha$'s power of $\tilde{\mathcal{L}}_p$ over $\T_{\Sigma,P}$. To this end consider the map $\tilde \rho_p : \C^P \ra \C$, which is simply the projection onto the factor corresponding to $p\in P$. Now we define
$$ \tilde{\mathcal{L}}_p^\alpha =  \cT_{\Sigma,P}\times_{\alpha \tilde \rho_p}\C,$$
where the action of $\C^P$ on $\cT_{\Sigma,P}\times \C$ is given by
$$ w(\varphi, z) = (\varphi w, \mathbf{e}(\alpha \tilde \rho_p(w))z),\qquad \mathbf{e}(t) = \exp(2{\rm i}\pi t).$$
We observe that $\tilde \Gamma_{\Si,P}$ acts on $ \tilde{\mathcal{L}}_p^\alpha$ covering the action of $\Gamma_{\Sigma,P}$ on $\T_{\Sigma,P}$, and that $\delta_p$ acts by multiplication by $e^{2\pi i \alpha}$ while $\delta_{p'}$ acts trivially for $p'\in P-\{p\}$.

\subsubsection{Determinant of the Hodge bundle}
The Hodge bundle is the vector bundle over $\T_\Si$, whose fiber at the class of $\varphi : \Si \ra C$ is $H^0(C,K_C)$. There is a natural action of $\Gamma_\Si$ on the Hodge bundle. We denote the determinant line bundle associated to this bundle by $\tilde{\mathcal{L}}_D$. It is isomorphic to the line bundle ${\mathcal V}^\dagger_{{\rm ab}}(\Sigma)$, which the first author and Ueno constructs over $\T_\Si$ using a certain abelian CFT, namely the CFT associated to the $bc$-ghost system \cite{AU1}.  We observe that for a marked surface $\Sib = (\Si, P, V, L)$, the Lagrangian $L$ induces a section $s_L$ of $\mathcal{L}_D^2$ over $\T_\Si$,
given by
$$
s_{L} = (u_1 \wedge \cdots \wedge u_{g})^{\otimes 2}
$$
where $(u_1,\ldots,u_{g})$ is normalized on an integral basis of $L$. The isomorphism between $\tilde{\mathcal{L}}_D$ and ${\mathcal V}^\dagger_{{\rm ab}}(\Sigma)$ takes this section of $\tilde{\mathcal{L}}_D^2$ to the preferred section of $({\mathcal V}^\dagger_{{\rm ab}}(\Sigma))^2$ over $\T_{\Si}$ as described in \cite{AU1}.
By \cite[Theorem 11.3]{AU2}, this section allows us to construct $\tilde{\mathcal{L}}_D^{-c/2}$ for any $c\in \mathbb{C}$, on which $\Gamma_\Sib$ acts,  such that $(\Id,1)$ acts by $e^{-{\rm i}\pi c/2}$. We continue to denote by $\tilde{\mathcal{L}}^{-c/2}_{D}$ the pullback to ${\mathcal{T}}_{\Sigma,P}$ of $\tilde{\mathcal L}_D^{-c/2}$.

\begin{remark} 
\label{remKC}We recall that the Hodge bundle over $\mathcal{T}_{\Sigma}$ has a natural hermitian structure, which is $\Gamma_\Si$ invariant. Hence it induces a hermitian structure on the holomorphic bundle $\mathcal{L}_D$ which is also $\Gamma_\Si$ invariant. Hence we get a unique unitary Chern connection in $\tilde{\mathcal{L}}_D$ compatible with the holomorphic structure on this line bundle, which by uniqueness is also $\Gamma_\Si$ invariant. By the proof of \cite[Theorem 11.3]{AU2} we see that for any $c \in \mathbb{R}$, we get an induced $\Gamma_\Sib$-invariant  unitary connection in $\tilde{{\mathcal L}}_D^{-c/2}$, whose curvature is $-c/2$ times the curvature of the Chern connection in $\tilde{\mathcal{L}}_D$.
\end{remark}

The moduli space of $(\Si,P)$ is by definition
\beq
\label{MSP} \mathcal{M}_{\Si,P} = \mathcal{T}_{\Si,P}/\Gamma_{\Si,P}
\eeq
When $\chi(\Si,P) < 0$, the stabilizers are finite, so there is a natural structure of an orbifold on $\mathcal{M}_{\Si,P}$. We see that there is a natural action of $\Gamma_{\Si,P}$ on $\tilde{\mathcal{L}}_p$, also acting with finite stabilizers and we define the orbifold line bundle 
$$ \mathcal{L}_p = \tilde{\mathcal{L}}_p/\Gamma_{\Si,P}$$
over $\mathcal{M}_{\Si,P}$.

\begin{remark} 
\label{remKC2} By picking an orbifold Hermitian structure on the holomorphic bundle $\mathcal{L}_p$ and pulling it back to $\tilde{\mathcal{L}}_p$, we see that the corresponding Chern connection in $\tilde{\mathcal{L}}_p$ is $\Gamma_\Si$ invariant. Hence, for any non-zero $r_p \in \mathbb{R}$, we get an induced unitary connection in $\tilde{\mathcal{L}}^{r_p}$, which is $\Gamma_{\Sib}$ invariant, and whose curvature is $r_p$ times that of the Chern connection of $\tilde{\mathcal{L}}_p$
\end{remark}

The moduli space $\mathcal{M}_{\Sigma,P}$ also carries the Hodge bundle, simply by pulling the Hodge bundle over $\mathcal{M}_{\Sigma}$ back to $\mathcal{M}_{\Sigma,P}$. We also denote the determinant bundle of this pull back of the Hodge bundle over $\mathcal{M}_{\Sigma,P}$ by $\mathcal{L}_{D}$ and it remains of course an orbifold line bundle over $\mathcal{M}_{\Sigma,P}$. We denote $\psi_{D} = c_1(\mathcal{L}_{D})$ and $\psi_p = c_1(\mathcal{L}_{p})$ and think of them as rational cohomology classes over $\mathcal{M}_{\Sigma,P}$.

\subsection{Modular functor and vector bundles over Teichm\"uller space}
\label{fki}

We now fix a modular functor, without assuming the unitarity and orientation reversal axioms.

\subsubsection{Scalars}

$\bullet$ The modular functor gives a morphism
$$
\mathcal{V}_{\vec{\lambda}}({\rm id},1) \in {\rm GL}(\mathcal{V}_{\vec{\lambda}}(\mathbf{\Sigma}),\mathbb{C})
$$
and the axioms imply that it acts as multiplication by a scalar $\tilde{c} \in \mathbb{C}^{*}$ independent of $\vec{\lambda}$ and the topology of $\Sib$. This is proved by factoring the $(\Id_\Sigma,1)$ along the boundary of an embedded disk $\Sigma_0$ in $\Si-P$ and writing it as $(\Id_\Sigma,1) = (\Id_{\Sigma_0},1) \cup (\Id_{\Sigma'},0) $, where $\Sigma' = \Sigma -  \Sigma_0$.

\bigskip

\noindent $\bullet$ For the sphere with 2 points, we have
$$
{\rm Aut}(\mathbb{S}^2,\{p_1,p_2\},\{v_1,v_2\},L) = {\rm Aut}(\mathbb{S}^2,\{p_1,p_2\},\{v_1,v_2\})\times \mathbb{Z},\qquad {\rm Aut}(\mathbb{S}^2,\{p_1,p_2\},\{v_1,v_2\}) = \langle \delta \rangle .
$$
Here $\delta$ is the Dehn twist along the equator. The modular functor gives a morphism:
$$
\mathcal{V}_{\lambda}(\delta) \in {\rm Aut}(\mathcal{V}_{\lambda,\lambda^{\dagger}}(\mathbb{S}^2,P,V,0)) .
$$
Since $\mathcal{V}_{\lambda,\lambda^{\dagger}}(\mathbb{S}^2,P,V,0)$ has dimension $1$, this is the multiplication by a scalar $\tilde{r}_{\lambda} \in \mathbb{C}^{*}$.

\begin{lemma} \label{Dehnlemma} The Dehn twists around the punctures have the following properties.
\begin{itemize}
\item[$(i)$] $\tilde{r}_{1} = 1$.
\item[$(ii)$] For any marked surface $(\mathbf{\Sigma},\vec{\lambda})$, the Dehn twist around a marked point $p$ with label $\lambda_p$ acts on $\mathcal{V}_{\vec{\lambda}}(\mathbf{\Sigma})$ by multiplication with $\tilde r_{\lambda_p}$.
\item[$(iii)$] For any $\lambda \in \Lambda$, $\tilde{r}_{\lambda} = \tilde{r}_{\lambda^{\dagger}}$.
\end{itemize}
\end{lemma}
\textbf{Proof.} For $(i)$ and $(ii)$, we use the factorization axiom. $(i)$ -- The Dehn twist is trivial on $\mathbb{S}^2$ and we can factor $\mathbb{S}^2$ along two curves, which are respectively the equator pushed a small amount into the northern (resp. southern) hemisphere. It then follows by considering this factorisation that $\tilde r_1 = 1$. $(ii)$ -- We factor in the boundary of an embedded disk in $\Sigma- (P-\{p\})$, centred in $p$ and then we do the Dehn twist around $p$ inside the disk. The result only depends on the label $\lambda_{p}$ at $p$, and not on the topology of the remaining surface neither on the labels at other punctures. $(iii)$ -- We can choose the points $(p_1,p_2)$ and vectors $(v_1,v_2)$ such that the (orientation preserving) map $z \ra - 1/z$ takes $(\mathbb{S}^2,  \{p_1,p_2\},\{v_1,v_2\})$ to itself, and takes the equator to the itself. Hence, it commutes with the Dehn twist, and that implies $\tilde{r}_{\lambda} = \tilde{r}_{\lambda^{\dagger}}$ for all $\lambda \in \Lambda$. \hfill $\Box$

\subsubsection{Trivial vector bundles over $\T_{\Si,P}$} \label{tvboT}

To a given $\Lambda$-marked surface $(\mathbf{\Sigma},\vec{\lambda})$, we associate the trivial vector bundle
$$
\tilde{\mathcal{Z}}_{(\mathbf{\Sigma},\vec{\lambda})}^0 := {\mathcal{T}}_{\Sigma,P} \times \mathcal{V}_{\vec{\lambda}}(\mathbf{\Sigma}) .
$$
If the modular functor is unitary, this bundle carries a $\Gamma_\Sib$-invariant unitary structure. In any case, we can equip it with the trivial flat connection, which is of course  $\Gamma_{\Sib}$-equivariant. Subsequently, the $\Gamma_\Sib$-equivariant Chern character is trivial. According to the above discussion the Dehn twist $\delta_p$ around a point $p$ acts by multiplication by $\tilde{r}_{\lambda}$. Moreover $(\Id,1)$ in $\Gamma_\Sib$ acts by multiplication by $\tilde{c}$.

\subsubsection{Vector bundle over the moduli space}

Pick up $c,r_{\lambda}\in \C$ such that\footnote{With these convention, $c$ is the Virasoro central charge, and $r_{\lambda}$ the conformal dimension.} $\tilde{c} = \mathbf{e}(c/4)$ and $\tilde{r}_{\lambda} = \mathbf{e}(r_{\lambda})$. We observe that $\delta_p$, $p\in P$ as well as $(\Id_\Si,1)$ act trivially on the vector bundle
$$
\tilde{\mathcal{Z}}_{(\mathbf{\Sigma},\vec{\lambda})} = \tilde{\mathcal{Z}}_{(\mathbf{\Sigma},\vec{\lambda})}^0 \otimes \tilde{\mathcal{L}}_{D}^{-c/2} \bigotimes_{p \in P} \tilde{\mathcal{L}}_{p}^{r_{\lambda_p}} .
$$
Therefore we get that
\begin{theorem}
\label{th2}The action of $\Gamma_\Sib$ factors to an action of $\Gamma_{\Si,P}$ and hence we can define
$$\mathcal{Z}_{(\mathbf{\Sigma},\vec{\lambda})} = \tilde{\mathcal{Z}}_{(\mathbf{\Sigma},\vec{\lambda})} /\Gamma_{\Si,P}$$ 
as an orbifold bundle over the moduli space $\mathcal{M}_{\Si,P}$.
\end{theorem}

\begin{proposition}
\label{propCH} For any modular functor $\mathcal{V}$  the total Chern class of this vector bundle is
\beq
\label{Chformula} {\rm Ch}_{t}(\mathcal{Z}_{(\mathbf{\Sigma},\vec{\lambda})}) = \dim \mathcal{V}_{\vec{\lambda}}(\mathbf{\Sigma})\,\exp\Big\{t\Big(-\frac{c}{2}\,\varLambda_{1} + \sum_{p \in P} r_{\lambda_{p}}\,\psi_{p}\Big)\Big\}
\eeq
where $\varLambda_1$ is the first Chern class of the Hodge bundle.
\end{proposition}

\noindent \textbf{Proof.}  We consider the tensor product connection of the trivial connection in $\tilde{\mathcal{Z}}^0_{(\mathbf{\Sigma},\vec{\lambda})} $ and then the $\Gamma_\Sib$-invariant unitary connections constructed in the bundles $\tilde{\mathcal{L}}_{D}^{-c/2}$ and  $\tilde{\mathcal{L}}_{p}^{r_{\lambda_p}}$ in Remarks~\ref{remKC}-\ref{remKC2}. By Chern-Weil theory we therefore have that
\bea
{\rm Ch}_{t}(\mathcal{Z}_{(\mathbf{\Sigma},\vec{\lambda})}) & = & \dim \mathcal{V}_{\vec{\lambda}}(\mathbf{\Sigma})\,\exp\Big\{t\Big(-\frac{c}{2}\,c_1(\mathcal{L}_{D}) + \sum_{p \in P} r_{\lambda_p}\,c_1(\mathcal{L}_{p})\Big)\Big\} \nonumber \\
& = & \dim \mathcal{V}_{\vec{\lambda}}(\mathbf{\Sigma})\,\exp\Big\{t\Big(-\frac{c}{2}\,\varLambda_1+ \sum_{p \in P} r_{\lambda_p} \psi_{p}\Big)\Big\} . \nonumber
\eea
\hfill $\Box$

Since Chern classes of vector bundles over the orbifold $\mathcal{M}_{\Sigma,P}$ are rational, $c$ and $r_{\lambda}$ must all be rationals. We thus get an alternative proof and generalisation of Vafa's theorem \cite{Vafa}:

\begin{corollary}
\label{cor1}For any modular functor, $\tilde{c}$ and $\tilde{r}_{\lambda}$ for any $\lambda \in \Lambda$ are roots of unity.
\end{corollary}

We recall that Vafa's original proof was written for unitary modular functors, based on an arithmetic argument following from relations in the mapping class group. Corollary~\ref{cor1} can be improved to show that

\begin{corollary}
For any $\gamma \in {\rm Aut}(\mathbf{\Sigma})$ which is the product of Dehn twists in non-intersecting curves, the element $\mathcal{V}_{\vec{\lambda}}(\gamma) \in {\rm Aut}(\mathcal{V}_{\vec{\lambda}}(\mathbf{\Sigma}))$ has finite order.
\end{corollary}

This follows immediately by factoring along two simple closed curves on either side of the simple closed curve of the Dehn twist. We recall that in general, not all elements in the representations of the mapping class group provided by $\mathcal{V}$ have finite order.

\subsection{Extension to the boundary}
\label{fki6}

\label{thm:extension}

We justify in this section the following theorem.
\begin{theorem}
\label{thaa3}
$[\mathcal{Z}_{(\mathbf{\Sigma},\vec{\lambda})}]$ extends to an orbifold bundle over the Deligne-Mumford compactification $\overline{\mathcal{M}}_{\Sigma,P}$.
\end{theorem}

In order to extend the above constructions to the Deligne-Mumford compactification of the moduli space, we introduce the augmented Teichmüller space and extend all our constructions in a mapping class group equivariant way to the augmented Teichmüller space. 

Let $(\Sigma,P)$ be a pointed surface. We introduce the set of contraction cycles ${\mathcal C}_{(\Sigma,P)}$ on $(\Sigma,P)$. It consists of isotopy classes of $1$-dimensional submanifolds $C$ of $\Sigma-P$, such that connected components of $C$ are non-contractible, nor are any two connected components of $C$ isotopic in $\Sigma-P$, nor are any of the components contractible into any of the points in $P$. We remark that $C=\emptyset$ is allowed.

The augmented Teichm\"uller space $\T^{a}_{\Sigma,P}$ for a pointed surface $(\Sigma,P)$ consists of equivalence classes of continuous maps $\phi\,:\,\,(\Sigma,C,P) \rightarrow (X,N,Q)$, where $C\in {\mathcal C}_{(\Sigma,P)}$, $X$ is a nodal Riemann surface with nodes $N$, and $Q\subset X$ is finite set of points of $X-N$, such that $\phi(C) = N$ and the restricted map $\phi : (\Sigma-C,P) \rightarrow (X-N, Q)$ is a diffeomorphism.
$$
\T^{a}_{\Sigma,P} = \left\{\phi\,:\,\,(\Sigma,C,P) \rightarrow (X,N,Q)\right\}/\sim^{a}
$$
where we declare that two continuous maps $\phi_i\,:\,(\Sigma,C,P) \rightarrow (X_i,N_i,Q_i)$ for $i = 1,2$ are $\sim^{a}$-equivalent if there exists a biholomorphic map $\Phi\,:\,(X_1,N_1,Q_1) \rightarrow (X_2,N_2, Q_2)$ such that $\Phi\circ \phi_1$ is isotopic to $\phi_2$ via continuous maps from $(\Sigma,C,P)$ to $(X_2,N_2, Q_2)$ which are diffeomorphisms from $(\Sigma-C,P)$ to $(X_2-N_2,Q_2)$. If $P = \emptyset$, this space is simply denoted $\mathcal{T}^{a}_{\Sigma}$. 

The augmented Teichm\"{u}ller space $\T^{a}_{\Sigma,P}$ has the topology uniquely determined by following property. Suppose $\pi : Z \rightarrow D$ is a holomorphic map from a complex $2$-dimensional manifold $Z$ to the unit disk $D$ in the complex plane, such that $\pi^{-1}(x)$ is a nodal Riemann surface for all $x\in D$. Suppose further that we are given a continuous map $\Phi : \Sigma \times D \rightarrow Z$, satisfying the two conditions
\begin{itemize}
\item[$\bullet$] $\Phi(\Sigma\times\{x\}) \subseteq \pi^{-1}(\{x\})$
\item[$\bullet$] if $N_x$ are the nodes of $\pi^{-1}(x)$ then $C_x=\Phi^{-1}(N_x)$ is a submanifold of $\Sigma-P$ such that $[C_x] \in {\mathcal C}_{(\Sigma,P)}$, and the restricted map $\Phi : \Sigma -C_x \rightarrow \pi^{-1}(x) - N_x$ is a diffeomorphism for all $x \in D$.
\end{itemize}
Then, the map from $D$ to $\T^{a}_{\Sigma,P}$, which sends $x$ to the restricted map $\Phi : (\Sigma, C_x,P) \rightarrow (\pi^{-1}(x), N_x, \Phi(P\times\{x\}))$ is continuous.

We observe that the mapping class group $\Gamma_{\Sigma, P}$ acts on $\T^{a}_{\Sigma,P}$, and the quotient is the Deligne-Mumford compactification $\overline{{\mathcal M}}_{g,n}/\mathfrak{S}_{n}$ of the moduli space of genus $g$ curves with $n=|P|$ marked unordered points (see section \ref{order}).

We will also consider the ``decorated augmented Teichm\"uller space" consisting of equivalence classes of continuous maps $ \phi\,:\,\,(\Sigma,C,P) \rightarrow (X,N,Q)$, where $X$ is a Riemann surface, $Q\subset X$ is a finite set of points, and $W\in T_QX$ non-zero tangent vectors.
$$
\cT^{a}_{\Si,P} = \left\{\phi\,:\,\,(\Sigma,C,P) \rightarrow (X,N,Q,W)\right\}/\approx^{a}
$$
where $\approx^{a}$ is now the equivalence relation where we ask that the isotopies preserve $(P, (\dd_P\varphi)^{-1}(W))$. 
We have the following natural projection maps among the augmented Teichm\"uller spaces.
$$
\tilde \pi^{a}_{\Sigma,P} : \tilde{\T}^{a}_{\Sigma,P}  \ra \T^{a}_{\Sigma,P},\qquad \pi^{a}_P : \T^{a}_{\Sigma,P}  \ra T_P\Sigma,\quad {\rm and}\quad \pi^{a}_\Si : \T^{a}_{\Sigma,P}  \ra \T^{a}_{\Sigma}.
$$

The proof of Proposition~\ref{Teichbun} applies word for word to extend the Proposition to the augmented setting. But then we get that all the constructions of Section \ref{fki} extend to constructions over augmented Teichm\"{u}ller spaces and hence also to the Deligne-Mumford compactification $\overline{{\mathcal M}}_{g,n}/\mathfrak{S}_{n}$ of the moduli spaces of genus $g$ curves with $n$ marked unordered points.

Concerning the extension of the bundle $\tilde{\mathcal{L}}_D$ and the section $s_L$ of $\tilde{\mathcal{L}}_D^2$ to augmented Teichmüller space, we appeal to the constructions of the first author and Ueno presented in \cite{AU1}. By the constructions of \cite[Section 5]{AU1} we see that the bundle ${\mathcal V}^\dagger_{{\rm ab}}(\Sigma)$ extends to a holomorphic bundle over augmented Teichmüller space. Moreover the preferred section of ${\mathcal V}^\dagger_{{\rm ab}}(\Sigma)$ extends to a nowhere vanishing
section of the extension of ${\mathcal V}^\dagger_{{\rm ab}}(\Sigma)$ to augmented Teichmüller space as is proved in \cite[Section 6]{AU1}. From this we conclude that the bundle $\tilde {\mathcal L}_D^{-c/2}$ extends to augmented Teichmüller space and that the action of $\Gamma_\Sib$ also extends.

\subsection{Remark on ordering of punctures}
\label{order}

In the usual definition of the moduli space $\mathcal{M}_{g,n}$, it is assumed that the marked points are ordered from $1$ to $n$. 

If $(\Sigma,P)$ is a pointed surface such that $\Sigma$ has genus $g$ and $|P| = n$, in our definition of the Teichm\"uller space $\mathcal{T}_{\Sigma,P}$ in \eqref{TSP}, the permutations of the $n$ points are divided out. We therefore have
$$
\mathcal{M}_{\Sigma,P} \simeq  \mathcal{M}_{g,n}/\mathfrak{S}_{n}
$$
and likewise for the Deligne-Mumford compactifications. Later in the text, we work only with the pull-back to $\overline{\mathcal{M}}_{g,n}$ of the bundle $\mathcal{V}_{\vec{\lambda}}$ that was so far obtained over $\overline{\mathcal{M}}_{\Sigma,P}$. The formula for the Chern classes in Theorem~\ref{thaa3} is the same for the bundle over $\mathcal{M}_{g,n}$.

\section{Cohomological field theories}
\label{S3}

\subsection{Generalities} 
\label{Gen}
\subsubsection{Frobenius algebras}
\label{defFb}
A Frobenius algebra is a finite dimensional complex vector space $\mathcal{A}$, equipped with a symmetric, non-degenerate bilinear form $b\,:\,\mathcal{A} \otimes \mathcal{A} \rightarrow \mathbb{C}$, and an associative, commutative $\mathbb{C}$-linear morphism $\times\,:\,\,\mathcal{A} \otimes \mathcal{A} \rightarrow \mathcal{A}$ such that 
$$
\forall a_1,a_2,a_3 \in \mathcal{A},\qquad b(a_1,a_2 \times a_3) = b(a_1\times a_2,a_3) .
$$
We require the existence of a unit for the product, denoted $\mathds{1}$. 

$\mathcal{A}$ is semi-simple if there exists a $\mathbb{C}$-linear basis $(\tilde{\epsilon}_1,\ldots,\tilde{\epsilon}_n)$ such that 
\beq
\label{ija1}\forall (i,j) \in \llbracket 1,n\rrbracket ,\qquad  \; \tilde{\epsilon}_i \times \tilde{\epsilon}_j = \delta_{ij}\,\tilde{\epsilon}_i .
\eeq
The unit is then $\mathds{1} = \sum_{i} \tilde{\epsilon}_i$, and \eqref{ija1} implies that the bilinear form is diagonal in this basis and reads
$$
b(\tilde{\epsilon}_i,\tilde{\epsilon}_j) = \frac{\delta_{ij}}{\Delta_i} .
$$
We say that $(\tilde{\epsilon}_i)_i$ is a canonical basis. It is sometimes more convenient to work with the orthonormal basis $\underline{\epsilon}_i = \Delta_i^{1/2}\tilde{\epsilon}_i$ which satisfies that
$$
\underline{\epsilon}_i \times \underline{\epsilon}_j = \delta_{ij}\Delta_{i}^{1/2}\,\underline{\epsilon}_i,\qquad b(\underline{\epsilon}_i,\underline{\epsilon}_j) = \delta_{ij} .
$$
Then $b$ induces a bivector $b^{\dagger} \in \mathcal{A}\otimes\mathcal{A}$, that will play an important role. In a canonical or an orthonormal basis it reads
$$
b^{\dagger} = \sum_{i} \underline{\varepsilon}_i \otimes \underline{\varepsilon}_i .
$$

\subsubsection{CohFTs}

A cohomological field theory (CohFT) is the data of a finite dimensional complex vector space $\mathcal{A}$ with a symmetric bilinear non-degenerate $b\,:\,\mathcal{A} \otimes \mathcal{A} \rightarrow \mathbb{C}$ and a sequence $\Omega_{g,n} \in H^{\bullet}(\overline{\mathcal{M}}_{g,n})\otimes \mathcal{A}^{\otimes n}$ indexed by integers $g \geq 0$ and $n \geq 0$ such that $2g - 2 + n > 0$, satisfying the axioms given below. Since $b$ gives a canonical identification of $\mathcal{A}$ with its dual $\mathcal{A}^*$, we can equivalently consider $\Omega^*_{g,n} \in H^{\bullet}(\overline{\mathcal{M}}_{g,n}) \otimes (\mathcal{A}^{*})^{\otimes n}$. The axioms are.
\begin{itemize}
\item[$\bullet$] There is a non-zero element $\mathds{1} \in \mathcal{A}$ such that the pairing is given by
$$
\forall a_1,a_2 \in \mathcal{A},\qquad b(a_1,a_2) = \int_{\overline{\mathcal{M}}_{0,3}} \Omega_{0,3}^*(a_1 \otimes a_2 \otimes \mathds{1}) .
$$
\item[$\bullet$] $\Omega_{g,n}^*$ is symmetric by simultaneous permutations of the $n$ factors in $(\mathcal{A}^{*})^{\otimes n}$ and the $n$ punctures in $\overline{\mathcal{M}}_{g,n}$.
\item[$\bullet$] Pulling back by the glueing map $\pi\,:\,\overline{\mathcal{M}}_{g,2 + n} \rightarrow \overline{\mathcal{M}}_{1 + g,n}$, we should have that
$$
\pi^*\Omega_{1 + g,n}^*(\cdot)  = \Omega_{g,2 + n}^*(b^\dagger \otimes \cdot ) .
$$

\item[$\bullet$] Pulling back by the glueing map $\pi\,:\,\overline{\mathcal{M}}_{g_1,1 + n_1} \times \overline{\mathcal{M}}_{g_2,1 + n_2} \rightarrow \overline{\mathcal{M}}_{g_1 + g_2,n_1 + n_2}$, we should have that
$$
\pi^*\Omega_{g_1 + g_2,n_1 + n_2}^*(\cdot) = \sum_{i,j} (\Omega_{g_1,1 + n_1}^* \otimes \Omega_{g_2,1 + n_2})(p_{n_1,n_2}(b^{\dagger}\otimes \cdot))
$$
where $p_{n_1}\,:\,(\mathcal{A}^*)^{\otimes 2} \otimes (\mathcal{A}^{*})^{\otimes n_1} \otimes (\mathcal{A}^{*})^{\otimes n_2} \rightarrow (\mathcal{A}^*)^{\otimes (1 + n_1)} \otimes (\mathcal{A}^*)^{\otimes(1 + n_2)}$ puts the second factor of $\mathcal{A}^*$ into the $(n_1 + 2)$-th position in the target space.
\item[$\bullet$] Pulling back by the forgetful map $\pi\,:\,\overline{\mathcal{M}}_{g,1 + n} \rightarrow \overline{\mathcal{M}}_{g,n}$, we should have that
$$
\pi^*\Omega_{g,n}^*(\cdot) = \Omega_{g,1 + n}^*(\mathds{1}\otimes \cdot) .
$$
\end{itemize}
The axioms imply that $\mathcal{A}$ is a Frobenius algebra with the product
$$
b(a_1 \times a_2,a_3) = \int_{\overline{\mathcal{M}}_{0,3}} \Omega_{0,3}^{*}(a_1\otimes a_2 \otimes a_3) .
$$

Givental \cite{Giv1} describes two basic actions on the set of CohFTs over the same Frobenius algebra. We follow the presentation of \cite{Zvonkin}.

\subsubsection{Translations}
\label{Sec:trans}
Let $T(u) \in u^2\mathcal{A}[[u]]$, and consider the forgetful maps $\pi_{m}\,:\,\overline{\mathcal{M}}_{g,m + n} \rightarrow \overline{\mathcal{M}}_{g,n}$. One can define a new CohFT by the formula
$$
(\hat{T}\Omega^*)_{g,n}(a_1 \otimes \ldots \otimes a_n) = \sum_{m \geq 0} \frac{1}{m!} \sum_{k_1,\dots, k_m \geq 2} (\pi_{m})_*\Big\{\Omega^*_{g,n + m}(T_{k_1}\otimes \cdots \otimes T_{k_m}\otimes a_1 \otimes \cdots \otimes a_n) \psi_1^{k_1} \dots \psi_m^{k_m}\Big\}
$$ 
where $T(u) = \sum_{m \geq 2} T_m\,u^m$ for $T_m \in \mathcal{A}$. 
Here $(\pi_m)_*$ is the push forward in cohomology classes, induced on smooth forms on the smooth compect orbifolds $\overline{\mathcal{M}}_{g,n}$.

\subsubsection{$R$-matrix actions}
\label{Rmatriact}
Let $R(u) \in {\rm End}(\mathcal{A})[[u]]$ such that $R(0) = {\rm id}$ and satisfying the symplectic condition
$$
R(u)R^{\dagger}(-u) = {\rm id}
$$
where $R^{\dagger}(u) \in {\rm End}(\mathcal{A})[[u]]$ is the adjoint for the pairing $b$. One then defines\footnote{Let us remark that this notation differs from the one used in the topological recursion literature. In the topological recursion setup, $B$ refers usually to the so-called Bergman kernel while our $B$ corresponds to its Laplace transform often denoted by $\check{B}$.} 
$$
B(u_1,u_2) := \frac{b^{\dagger} - R(u_1) \otimes R(u_2) \cdot b^{\dagger}}{u_1 + u_2} \in \left(\mathcal{A}\otimes\mathcal{A}\right)[[u_1,u_2]] .
$$
The symplectic condition guarantees that $B$ is a formal power series in $u_1$ and $u_2$. One can define a new CohFT by the formula
\beq
\label{ushf}(\hat{R}\Omega)_{g,n} := \sum_{\substack{\Gamma \\ {\rm stable}\,\,{\rm graph}}} \frac{1}{|{\rm Aut}\,\Gamma|}\,\,(\pi_{\Gamma})_*\bigg(\prod_{l = {\rm leaf}} R(\psi_{l}) \!\! \prod_{\substack{e = {\rm edge} \\ \{v'_e,v''_e\}}}B(\psi_{v_e'},\psi_{v_e''}) \!\!\!\prod_{\substack{v  = {\rm vertex}}} \Omega_{g(v),n(v)}\bigg)
\eeq
The sum is over stable graphs of topology $(g,n)$, namely $\Gamma$ meeting the following requirements.
\begin{itemize}
\item[$\bullet$] vertices $v$ are trivalent, carry an integer label $g(v) \geq 0$ (the genus), and their valency $n(v)$ satisfy $2g(v) - 2 + n(v) > 0$.
\item[$\bullet$] there are $n$ leaves (1-valent vertices), labeled from $1$ to $n$.
\item[$\bullet$] ${\rm rank}(H_1(\Gamma,\mathbb{Z}))+ \sum_{v = {\rm vertex}} g(v) = g$.
\end{itemize}
In \eqref{ushf}, the endomorphisms are naturally composed along the graph, and we use the pushforward by the following glueing map along the graph
$$
\pi_{\Gamma}\,:\,\prod_{v} \overline{\mathcal{M}}_{g(v),n(v)} \rightarrow \overline{\mathcal{M}}_{g,n}.
$$

\subsubsection{Classification of semi-simple CohFT}

Given a CohFT defined by correlators $\Omega_{g,n}$, its restriction to the degree 0 part $\omega_{g,n} \in H^0(\mathcal{M}_{g,n})\otimes {\cal A}^{\otimes n}$ is sometimes called (abusively) a topological quantum field theory ($2d$ TQFT). Teleman \cite{Tele}, building upon the work of Givental \cite{Giv1,Giventals}, has classified semi-simple CohFT whose underlying Frobenius algebra has dimension $k$.
\begin{theorem} \cite{Tele}
Any semi-simple CohFT can be obtained from a degree $0$ CohFT by the composition of the action of an $R$-matrix, and a translation such that
\beq
\label{TRmat} T(u) = u\big(\mathds{1} - R(u)\cdot\mathds{1}\big).
\eeq
More precisely, if $\Omega$ denote the correlators of the CohFT, and $\Omega^{{\rm deg}\,\,0}$ the correlators of its underlying $2d$ TQFT, we have $\Omega = \hat{R}\hat{T}\Omega^{{\rm deg}\,\,0}$.
\end{theorem}

This reconstruction is a powerful tool since the correlators $\Omega_{g,n}^{{\rm deg}\,\,0}$ of a degree $0$ CohFT with canonical basis $(\tilde{\epsilon}_i)_{i}$ such that $b(\tilde{\epsilon}_i,\tilde{\epsilon}_j) = \delta_{ij}/\Delta_i$ and $\tilde{\epsilon}_i \times \tilde{\epsilon}_j = \delta_{ij}\,\tilde{\epsilon}_i$ are
\beq
\Omega_{g,n}^{{\rm deg}\,\,0} =  \sum_{i = 1}^k \Delta_i^{g-1}\, [\overline{\mathcal{M}}_{g,n}] \otimes \tilde{\epsilon}_{i} \otimes \cdots \otimes \tilde{\epsilon}_{i},
\eeq
where the Poincar\'e duality is implicitly used in this formula. The knowledge of the $R$-matrix is enough for reconstructing the correlators of a semi-simple CohFT.

\subsection{Reference vector spaces attached to a modular functor}
\label{Referencespace}
We shall now describe the CohFT defined by a modular functor, starting by the definition of the reference vector spaces underlying it. We start with a general modular functor $\mathcal{V}$. We shall work with marked surfaces of reference, and choose basis in their corresponding vector spaces.

\subsubsection{Once-punctured sphere}

$\mathbf{\Sigma}_{0,1}$ is the $2$-sphere with $P = \{0\}$ and $V = \{v_0\}$ with $v_0$ pointing to the positive real axis. $\mathcal{V}_{1}(\mathbf{\Sigma}_{0,1})$ is a line, and we pick up a generator $\tilde{\zeta}[1]$.  This induces an isomorphism $\mathcal{V}_{1}(\mathbf{\Sigma}_{0,1}) \cong \mathcal{V}_{1}(\mathbf{\Sigma}_{0,1})^{\star}$, and allows us to project the isomorphism from propagation of vacua 
$$
\mathcal{V}_{\lambda}(\Sigma,P,V,L) \cong \mathcal{V}_{\lambda ,1}(\Sigma,P \sqcup \{p'\},V \sqcup \{v'\},L) \otimes \mathcal{V}_{1}(\mathbf{\Sigma}_{0,1})
$$
to an isomorphism
$$
\mathcal{V}_{\lambda}(\Sigma,P,V,L) \cong \mathcal{V}_{\lambda , 1}(\Sigma,P\sqcup\{p'\},V\sqcup \{v'\},L) .
$$
Later on, this will be used systematically.

\subsubsection{Twice-punctured sphere}

$\mathbf{\Sigma}_{0,2}$ is the $2$-sphere with $P = \{0,\infty\}$ and $V = \{v_0,v_{\infty}\}$ with $v_{\infty}$ pointing to the negative real axis. $\mathcal{V}_{\lambda,\lambda^{\dagger}}(\mathbf{\Sigma}_{0,2})$ is a line. The property of propagation of vacua provides an isomorphism
\beq
\label{firstiso}\mathcal{V}_{1, 1}(\mathbf{\Sigma}_{0,2}) \cong \mathcal{V}_{1}(\mathbf{\Sigma}_{0,1}) \otimes \mathcal{V}_{1,1}(\mathbf{\Sigma}_{0,2})
\eeq
and the property of factorization provides isomorphisms 
\beq
\label{secondiso}\mathcal{V}_{\lambda,\lambda^{\dagger}}(\mathbf{\Sigma}_{0,2}) \cong \mathcal{V}_{\lambda,\lambda^{\dagger}}(\mathbf{\Sigma}_{0,2})\otimes \mathcal{V}_{\lambda^{\dagger},\lambda}(\mathbf{\Sigma}_{0,2}) = \mathcal{V}_{\lambda,\lambda^{\dagger}}(\mathbf{\Sigma}_{0,2})^{\otimes 2} .
\eeq
They determine for each $\lambda \in \Lambda$ a unique generator $\zeta[\lambda]$ of $\mathcal{V}_{\lambda,\lambda^{\dagger}}(\mathbf{\Sigma}_{0,2})$ such that $\zeta[1] = \zeta[1] \otimes \tilde{\zeta}[1]$ using \eqref{firstiso}, and $\zeta[\lambda] = \zeta[\lambda]\otimes \zeta[\lambda]$ using \eqref{secondiso}. 

\subsubsection{Thrice-punctured sphere}

$\mathbf{\Sigma}_{0,3}$ is the $2$-sphere with $P = \{0,1,\infty\}$ and $V = \{v_0,v_1,v_{\infty}\}$ with $v_1$ pointing in real direction to $\infty$. We denote the dimension of $\mathcal{V}_{\lambda,\mu,\nu}(\mathbf{\Sigma}_{0,3})$ by
$$
N_{\lambda\mu\nu} =  \dim \,\mathcal{V}_{\lambda,\mu,\nu}(\mathbf{\Sigma}_{0,3}) .
$$
Since the homomorphism from the mapping class group of $\mathbf{\Sigma}_{0,3}$ to the permutation group of the marked points is surjective, this symbol is invariant under permutation of $(\lambda,\mu,\nu)$. Propagation of vacua gives an isomorphism
\beq
\label{iso03}\mathcal{V}_{\lambda,\lambda^{\dagger}}(\mathbf{\Sigma}_{0,2}) \cong \mathcal{V}_{\lambda,\lambda^{\dagger},1}(\mathbf{\Sigma}_{0,3}) \otimes \mathcal{V}_{1}(\mathbf{\Sigma}_{0,1}) .
\eeq
We fix a basis $\zeta_i[\lambda\mu\nu]$ of the space $\mathcal{V}_{\lambda,\mu,\nu}(\mathbf{\Sigma}_{0,3})$, indexed by $i = 1,\ldots,N_{\lambda\mu\nu}$. Without loss of generality, we can require that, under \eqref{iso03}, we have
$$
\zeta[\lambda] = \zeta_1[\lambda\lambda^{\dagger}1] \otimes \tilde{\zeta}[1] .
$$

\subsubsection{Torus}

$\mathbf{\Sigma}_{1}$ is the torus $\mathbb{C}/(\mathbb{Z} \oplus {\rm i}\mathbb{Z})$. We denote $\alpha$ -- resp. $\beta$ -- the closed, oriented, simple curve based at $o$ and following the positive real axis -- resp. the positive imaginary axis.

\subsection{The Frobenius algebra of a modular functor}
\label{FMF}
\subsubsection{As a vector space}
\label{FMFV}
If $\gamma$ is a simple oriented closed curve on $\mathbf{\Sigma}_{1}$, we obtain a marked surface $\mathbf{\Sigma}^{(\gamma)}_{1}$ 
by considering the Lagrangian spanned by the homology class of $\gamma$. In this paragraph, we shall define a structure of Frobenius algebra on the vector space of a torus. For convenience, we choose a torus with a marked point:
\beq
\label{Adef}\mathcal{A} := \mathcal{V}_{1}(\mathbf{\Sigma}_{1}^{(\alpha)}) 
\eeq
By applying a diffeomorphism that takes $(\alpha,\beta)$ to $(\beta,-\alpha)$, we also have a natural isomorphism
\beq
\label{Bdef}\mathcal{A} \cong \mathcal{V}_{1}(\mathbf{\Sigma}_{1}^{(\beta)}). 
\eeq
By propagation of vacua, then factorization along $\alpha$ and application of a suitable diffeomorphism, we obtain from \eqref{Adef} an isomorphism 
$$
\mathcal{A} \cong \bigoplus_{\lambda \in \Lambda} \mathcal{V}_{\lambda,\lambda^{\dagger}}(\mathbf{\Sigma}_{0,2}) .
$$
Our previous choices of generators in the right-hand side carries to a basis $e_{\lambda}$ of $\mathcal{A}$. Similarly, the factorization along $\beta$ from \eqref{Bdef} gives another basis $\epsilon_{\lambda}$ of $\mathcal{A}$. The change of basis is called the "S-matrix". It is the linear map $S \,:\,\mathcal{A} \rightarrow \mathcal{A}$ defined by
$$
S(e_{\lambda}) := \varepsilon_{\lambda} = \sum_{\mu \in \Lambda} S_{\lambda\mu} e_{\mu}.
$$

\subsubsection{Pairing}

We define a pairing on $\mathcal{A}$ by the following formula.
\beq
\label{pairingMF} b(e_{\lambda},e_{\mu}) := \delta_{\lambda\mu^{\dagger}}
\eeq

\subsubsection{Involutions}

We define the charge conjugation, which is the involutive linear map $C\,:\,\mathcal{A} \rightarrow \mathcal{A}$ such that $C(e_{\lambda}) = e_{\lambda^{\dagger}}$. If $O \,:\,\mathcal{A} \rightarrow \mathcal{A}$ is a linear map, we denote $O^{\top}$ its adjoint with respect to the scalar product in which $e_{\lambda}$ is an orthonormal basis, and $O^{\dagger}$ its adjoint for the bilinear product $b$. We have 
$$
C = C^{\top} = C^{\dagger} = C^{-1} .
$$
We may sometimes confuse the operator $O$ with its matrix $(O_{\lambda,\mu})_{\lambda,\mu \in \Lambda}$ in the basis $(e_{\lambda})_{\lambda \in \Lambda}$. In terms of matrices, $O^{\top}$ is the transpose, while $M^{\dagger} = CO^{\top}C$.

\subsubsection{Curve operators and S-matrix}
\label{curvop}
For any marked surface $(\Sigma, P)$ and a simple oriented closed curve $\gamma$ on $\Sigma \setminus P$ and label $\lambda \in \Lambda$, following \cite{AU3}, we introduces curve operators, that we denote $\mathcal{C}[\gamma;\lambda]$. We consider in particular the curve operators $(\mathcal{C}[\beta;\lambda])_{\lambda \in \Lambda}$ acting on $\mathcal{A}$, as multiplication by $e_{\lambda}$. In the basis related to $\alpha$ they have the following expression.
$$
\mathcal{C}[\beta;\lambda](e_{\mu}) = \sum_{\nu \in \Lambda} N_{\lambda\mu\nu^{\dagger}}\,e_{\nu}
$$
In the basis related to $\beta$, they are simultaneously diagonalized:
\beq
\label{eigne1}\mathcal{C}[\beta;\lambda](\epsilon_{\mu}) = \mathfrak{c}_{\mu}[\lambda]\,\epsilon_{\mu} .
\eeq
From these two facts, the eigenvalue can easily be computed. Indeed, we compute from the definitions that we have
$$
\mathcal{C}[\beta;\lambda][e_{1}] = \sum_{\mu \in \Lambda} N_{\lambda 1 \nu^{\dagger}}\,e_{\nu} = e_{\lambda} = \sum_{\mu \in \Lambda} (S^{-1})_{\lambda\mu}\,\epsilon_{\mu},
$$
while, if we first go to the $\epsilon$-basis, we get
$$
\mathcal{C}[\beta;\lambda][e_{1}] = \sum_{\mu \in \Lambda} (S^{-1})_{1\mu}\,\mathfrak{c}_{\mu}[\lambda]\,\epsilon_{\mu} .
$$
The comparison gives that $(S^{-1})_{1\mu}$ is non-zero and the eigenvalue reads
\beq
\label{eigne2}\mathfrak{c}_{\mu}[\lambda] = \frac{(S^{-1})_{\lambda\mu}}{(S^{-1})_{1\mu}} .
\eeq
One then deduces the following standard formula.
\beq
\label{nlm}N_{\lambda\mu^\dagger\nu} = \sum_{\tau \in \Lambda} \frac{(S^{-1})_{\lambda\tau} S_{\tau\mu} (S^{-1})_{\nu\tau}}{(S^{-1})_{1\tau}} .
\eeq

This formula does not depend on the normalization of the basis diagonalizing the curve operator action, i.e. it is invariant under rescalings $S_{\tau \mu} \to a_\tau  S_{\tau \mu}$ with $(a_\tau)_{\tau \in \Lambda} \in \left[\mathbb{C}^*\right]^{\Lambda}$. In particular, one can write the formula with respect to the orthogonal basis $(\underline{\epsilon}_\lambda) = \sum_{\mu \in \Lambda} \underline{S}_{\lambda \mu} e_\mu$ defined in \eqref{orthoba} to get
\beq
\label{nlm2s}N_{\lambda\mu^\dagger\nu} = \sum_{\tau \in \Lambda} \frac{(\underline{S}^{-1})_{\lambda\tau} \underline{S}_{\tau\mu} (\underline{S}^{-1})_{\nu\tau}}{(\underline{S}^{-1})_{1\tau}} .
\eeq

\subsubsection{Relation between $S$ and $C$}

Setting $\mu = 1$ in \eqref{nlm} yields $N_{\lambda 1 \nu} = \delta_{\lambda\nu^{\dagger}}$, which gives the following relation.
\beq
\label{nlm2} C_{\lambda\mu} := \delta_{\lambda\mu^{\dagger}} = \sum_{\tau \in \Lambda} \,\frac{S_{\tau 1}}{(S^{-1})_{1\tau}}\,(S^{-1})_{\lambda\tau}(S^{-1})_{\mu\tau} .
\eeq
In terms of a rescaled\footnote{It follows from the definition that $b$ is non-degenerate, so $S_{\lambda1}/(S^{-1})_{1\lambda} = b(\varepsilon_{\lambda},\varepsilon_{\lambda})$ computed in \eqref{canonical} below cannot be $0$, i.e. $S_{\lambda1} \neq 0$. Then, $\underline{S}_{\lambda\mu}$ depends on the arbitrary choice of a sign for the squareroot, which does not affect any of the Verlinde formula since an even number of $\underline{S}^{-1}$ factors appear.} $S$-matrix, it can be rewritten
\beq
\label{SST}\underline{S}_{\lambda\mu} := \sqrt{\frac{(S^{-1})_{1\lambda}}{S_{\lambda 1}}}\,S_{\lambda\mu},\qquad C = \underline{S}^{-1}(\underline{S}^{-1})^{\top}, 
\eeq
or equivalently 
\beq
\label{SSJ}\underline{S}^{\top}\underline{S} = C .
\eeq
Whenever possible, we prefer to avoid the occurrence of $\dagger$ indices, so we will use \eqref{SSJ} to convert it in entries of the inverse $\underline{S}$-matrix.

\subsubsection{Symmetric formula}

This relation between $S$ and $C$ allows to write down the action of the curve operator in a more symmetric form avoiding the $\dagger$ indices.
$$
N_{\lambda \mu \nu} =  \sum_{\tau \in \Lambda} \frac{(S^{-1})_{\lambda\tau} (S^{-1})_{\mu \tau}  (S^{-1})_{\nu\tau}}{\left[(S^{-1})_{1\tau}\right]^2} S_{\tau 1} .
$$
Once again, one can express this symmetric formula in the orthonormal basis to get a more natural form:
\beq\label{symnlm}
N_{\lambda \mu \nu} =  \sum_{\tau \in \Lambda} \frac{(\underline{S}^{-1})_{\lambda\tau} (\underline{S}^{-1})_{\mu \tau}  (\underline{S}^{-1})_{\nu\tau}}{(\underline{S}^{-1})_{1\tau}}  .
\eeq
The rescaling from $S$ to $\underline{S}$ also does not affect the formula for the eigenvalues of the curve operators.
\beq
\label{eignde}\mathfrak{c}_{\mu}[\lambda] = \frac{(\underline{S}^{-1})_{\lambda\mu}}{(\underline{S}^{-1})_{1\mu}} .
\eeq

\subsubsection{Extra relations}

If \textbf{MF-D} is satisfied or if the modular functor comes from a modular tensor category \cite[page 97-98]{Tu}, then $S_{\lambda\mu} = S_{\mu\lambda^{\dagger}}^{-1}$. In particular, we have $(S^{-1})_{1\lambda} = S_{\lambda 1}$, therefore $\underline{S} = S$ and the basis $\varepsilon_{\lambda}$ is already orthonormal. If \textbf{MF-U} is satisfied, the matrix $S$ is unitary. In particular, $S^{-1}_{1\lambda} = S_{\lambda 1}^*$. These properties are justified in Appendix~\ref{Sextra}. In this text, we study modular functors where neither \textbf{MF-D} \textbf{MF-U}  is assumed and in following computations, we do not use duality nor unitarity properties of the $S$-matrix.

\subsubsection{As a Frobenius algebra}

We define a product on $\mathcal{A}$ by the following formula 
\beq
\label{productMF} e_{\lambda} \times e_{\mu} := \mathcal{C}[\beta;\lambda](e_{\mu}) .
\eeq
A direct check from the previous formulas shows that $\mathcal{A}$ is now a Frobenius algebra, with unit $e_{1} = \mathds{1}$. We also find respectively from \eqref{nlm} and \eqref{nlm2} that one has
$$
\epsilon_{\lambda}\times\epsilon_{\mu} = \delta_{\lambda\mu}\,\frac{\epsilon_{\lambda}}{(S^{-1})_{1\lambda}}\qquad \hbox{and} \qquad b(\epsilon_\lambda,\epsilon_\mu) = \frac{S_{\lambda 1}}{(S^{-1})_{1\lambda}}\,\delta_{\lambda\mu} .
$$
The canonical basis $(\tilde{\epsilon}_{\lambda})_{\lambda}$ is obtained by a rescaling of the $\epsilon_{\lambda}$ as follows.
$$
\tilde{\epsilon}_{\lambda} = (S^{-1})_{1\lambda}\,\epsilon_{\lambda} .
$$
This  satisfies
\beq
\label{canonical}\tilde{\epsilon}_{\lambda} \times \tilde{\epsilon}_{\mu} = \delta_{\lambda\mu}\,\tilde{\epsilon}_{\lambda}\qquad {\rm and}\qquad b(\tilde{\epsilon}_{\lambda},\tilde{\epsilon}_{\mu}) = (S^{-1})_{1\lambda} S_{\lambda1}\,\delta_{\lambda,\mu} .
\eeq
A third interesting normalization gives the orthonormal basis defined by
\beq
\label{orthoba} \underline{\epsilon}_{\lambda} =  \frac{\tilde{\epsilon}_{\lambda}}{\sqrt{(S^{-1})_{1\lambda} S_{\lambda 1}}} = \sqrt{\frac{(S^{-1})_{1\lambda}}{S_{\lambda 1}}}\,\epsilon_{\lambda} := \sum_{\mu \in \Lambda} \underline{S}_{\lambda\mu} e_{\mu}
\eeq
which satisfies
$$
\underline{\epsilon}_{\lambda} \times \underline{\epsilon}_{\mu} = \delta_{\lambda\mu} \frac{\underline{\epsilon}_{\lambda}}{\sqrt{(S^{-1})_{1\lambda}S_{\lambda 1}}}\qquad {\rm and}\qquad b(\underline{\epsilon}_{\lambda},\underline{\epsilon}_{\mu}) = \delta_{\lambda\mu}.
$$
The unit expressed in the various basis is
$$
\mathds{1} = e_{1} = \sum_{\lambda \in \Lambda} \tilde{\epsilon}_{\lambda} = \sum_{\lambda \in \Lambda} \sqrt{(S^{-1})_{1\lambda} S_{\lambda 1}}\,\underline{\epsilon}_{\lambda} = \sum_{\lambda \in \Lambda} (S^{-1})_{1\lambda} \varepsilon_{\lambda} .
$$
The norm of the canonical basis will appear in subsequent computations and is read from \eqref{canonical} as
$$
\Delta_{\lambda} := \frac{1}{b(\tilde{\epsilon}_{\lambda},\tilde{\epsilon}_{\lambda})} = \frac{1}{(S^{-1})_{1\lambda} S_{\lambda 1}} = \frac{1}{\left[\left(\underline{S}^{-1}\right)_{1\lambda}\right]^2} .
$$

\subsection{CohFTs associated to a modular functor}

We shall build, for each log-determination of the central charge and Dehn twist eigenvalues,
\beq
\label{logdet}\tilde{r}_{\lambda} = \exp(2{\rm i}\pi r_{\lambda}),\qquad \tilde{c} = \exp({{\rm i}\pi c}/2),\qquad r_{\lambda}, c \in \mathbb{R}\quad {\rm such}\,\,{\rm that}\,\,r_{\lambda} = r_{\lambda^{\dagger}}\,,
\eeq
a $1$-parameter family of CohFTs based on the Frobenius algebra $\mathcal{A}$ described in the previous paragraph. The parameter here is denoted $t \in \mathbb{C}$. 
We rely on the result of Theorem~\ref{thm:extension}: We have defined for each $n$-tuple $\vec{\lambda}$ of labels, a complex vector bundle $[\mathcal{Z}_{\vec{\lambda}}]_{g,n} \rightarrow \overline{\mathcal{M}}_{g,n}$. Then, we simply take its total Chern class.
$$
\Omega_{g,n} = \sum_{\vec{\lambda}\in \Lambda^n} {\rm Ch}_{t}([\mathcal{Z}_{\vec{\lambda}}]_{g,n})\,e_{\lambda_1}\otimes \cdots \otimes e_{\lambda_n}.
$$
\begin{theorem}
\label{CohFTh} For any $t \in \mathbb{C}$ and choice of log-determinations \eqref{logdet}, $\Omega_{g,n}$ is a semi-simple CohFT.
\end{theorem}
\textbf{Proof.} Our bundle has been defined over the boundary of the moduli space, and the axioms of a CohFT for the total Chern class immediately follow from the factorization properties of the underlying bundles.
\hfill $\Box$

To describe explicitly this CohFT, we need to find the operator which transports it to a degree 0 CohFT. The strategy is the same as in \cite{Zvonkin} which was written in the example where $\mathcal{V}_{g,n}$ is the Verlinde bundle. The task of identifying the $R$-matrix is facilitated by the following result, whose proof relies on Teleman's classification of semi-simple CohFTs \cite{Tele}. 
\begin{lemma}\cite{Zvonkin}
In a semi-simple CohFT, the restriction of $\Omega_{g,n}$ to the smooth locus $\mathcal{M}_{g,n}$ completely determines the $R$-matrix.
\end{lemma}

For any modular functor we already have computed in Proposition~\ref{propCH} the Chern character of $[\mathcal{Z}_{\vec{\lambda}}]_{g,n}$ on the smooth locus $\mathcal{M}_{g,n}$.
$$
\Omega_{g,n}^*(e_{\lambda_1}\otimes \cdots \otimes e_{\lambda_n}) = D_{\vec{\lambda}}(\mathbf{\Sigma}_{g,n})\,\exp\Big\{t\Big(-\frac{c}{2} \varLambda_1 + \sum_{i = 1}^{n} r_{\lambda_i}\psi_i\Big)\Big\},\qquad D_{\vec{\lambda}}(\mathbf{\Sigma}_{g,n}) := {\rm dim}\,\mathcal{V}_{\vec{\lambda}}(\mathbf{\Sigma}_{g,n}) .
$$
Here, $\varLambda_1$ is the first Chern class of the Hodge bundle, not to be confused with the label set $\Lambda$. $\mathbf{\Sigma}_{g,n}$ is any marked surface of genus $g$ with $n$ points. By the Verlinde formula, $D_{\vec{\lambda}}(\mathbf{\Sigma}_{g,n})$ only depends on $g$, $n$ and $\vec{\lambda}$.
\begin{proposition}
\label{Rcomput}Assume $r_1 = 0$. We have the following diagonal $R$-matrix in the $e$-basis, which can also be written non-diagonally in the $\epsilon$-basis
\beq
\label{Rmatri}R(u) = \sum_{\lambda \in \Lambda} e^{ut(r_{\lambda} + c/24)}\,{\rm id}_{e_{\lambda}}.
\eeq
The corresponding translation is $T(u) = u\big(1 - \exp(utc/24)\big) e_{1}$.
\end{proposition}
Since $u$ here is not a quantized parameter, the log-determination of $r_{\lambda}$ and $c$ do matter in \eqref{Rmatri}.

\br
We observe that, up to the scalar $e^{utc/24}$, the $R$-matrix can be identified with the action of the flow at time $ut/4{\rm i}\pi$ generated by an infinitesimal Dehn twist around a puncture, on the space
$
\mathcal{V}_{{\rm tot}}(\mathbf{\Sigma}_{g,n}) = \bigoplus_{\vec{\lambda} \in \Lambda^{n}} \mathcal{V}_{\vec{\lambda}}(\mathbf{\Sigma}_{g,n}) .
$
This awaits an interpretation in hyperbolic geometry. 
\er

\noindent \textbf{Proof.} We first remark that
\beq
\label{Dgt0} \Omega_{g,n}^*(e_{\lambda_1}\otimes \cdots \otimes e_{\lambda_n})|_{t = 0} := D_{\vec{\lambda}}(\mathbf{\Sigma}_{g,n})\,[\overline{\mathcal{M}}_{g,n}]
\eeq
are the correlators of the degree $0$ part of the CohFT of the modular functor. Denote $\iota\,:\,\mathcal{M}_{g,n} \rightarrow \overline{\mathcal{M}}_{g,n}$. With the formula $\iota^*(\varLambda_{1}) = \iota^*(\kappa_1/12)$ \cite{MumfordR}, we obtain that
\beq
\label{omea}\iota^*\Omega_{g,n}^*(e_{\lambda_1}\otimes \cdots \otimes e_{\lambda_n}) = \iota^*{\rm Ch}_{t}([\mathcal{Z}_{\vec{\lambda}}]_{g,n}) = \exp(-t\, c\, \kappa_{1}/24)\,\exp\Big(\sum_{i = 1}^n t\, r_{\lambda_i}\psi_{i}\Big) .
\eeq
Comparing with Teleman's classification theorem, one gets
$$
\Omega_{g,n} = \hat{R}\,\hat{T} \, \Omega_{g,n}|_{t = 0}.
$$
We see by definition of the $\kappa$-classes that the first factor in \eqref{omea} can only arise by the action of the translation operator, and the second factor arises from the action of the $R$-matrix given in \eqref{Rmatri}. The expression of the translation follows from the fact that we have a CohFT and the general result \eqref{TRmat}. Since $r_1 = 0$ and $e_1$ is the unit, we find $T(u) = u\big(1 - \exp(utc/24)\big) e_{1}$. \hfill $\Box$

\subsection{Remark on log-determinations}
\label{VOADK}
When the modular functor comes from a category of representations of a vertex operator algebra $\mathfrak{A}$, $\mathcal{A}$ is the character ring of $\mathfrak{A}$. Characters are functions on the upper-half plane $\{\tau \in \mathbb{C},\,\,{\rm Im}\,\tau > 0\}$, and after multiplication by a suitable rational power of $q = e^{2{\rm i}\pi \tau}$, the characters fit in a vector-valued modular form. The $S$-matrix is implemented by the transformation $\tau \mapsto -1/\tau$, and the Dehn twist is implemented by $\tau \mapsto \tau + 1$. The modularity of characters then provides canonical log-determinations $r_{\lambda}$ and $c$, since multiplying by another power of $q$ will destroy modularity.

\section{Topological recursion}

The main result of \cite{DBOSS} is that the evaluation of the classes $\Omega_{g,n}$ of a semi-simple CohFT against $\overline{\mathcal{M}}_{g,n}$  are computed by the topological recursion of \cite{EOFg}. We quickly review this theory, in the (minimal) context of local spectral curves. We will comment on the setting of global spectral curves in Section~\ref{Globals}.

\subsection{Local spectral curve and residue formula}
\label{Locals}
A local spectral curve consists in the following set of data.
\begin{itemize}
\item[$\bullet$] $U = \bigsqcup_{i} U_i$ which is a disjoint union of formal neighborhoods of complex dimension $1$ of points $o_i \in U_i$ ;
\item[$\bullet$] a branched covering $x\,:\,U \rightarrow V$ whose ramification divisor is $O = \bigsqcup_{i} \{o_i\}$. $V$ itself is also a disjoint union of formal neighborhoods of a point in $\mathbb{P}^1$.
\end{itemize}
We assume that $x$ has only simple ramifications. Then, we can choose a coordinate on $U_i$, denoted $\zeta_i$ such that $x(\zeta_i) = \zeta_i^2/2 + x(o_i)$. We make this choice once for all in each $U_i$. $U_i$ carries a holomorphic involution $\sigma_i$, which sends the point in $U_i$ with coordinate $\zeta_i$, to the point in $U_i$ with coordinate $-\zeta_i$. In the cases, where we consider cross product $U_{i_{\ell_1}}\times \ldots \times U_{i_{\ell_m}}$, we denote points in this cross product  $(z_{\ell_1}, \ldots z_{\ell_m})$ and we shall denote by $\zeta_{i_{\ell_i}}(z_{{\ell}_i})$ their respective coordinates.

Let $\Delta = \{(z,z) \in U^2 \mid z \in U\}$ be the diagonal divisor and $K_U$ be the canonical bundle of $U$. The initial data of the topological recursion is
$$
\omega_{0,1} \in H^0(U,K_{U}),\qquad \omega_{0,2} \in H^0\big(U^2,K_{U}^{\boxtimes 2}(2\Delta)\big)^{\mathfrak{S}_{2}}
$$
with the extra condition that $\omega_{0,1}(z) - \omega_{0,1}(\sigma_i(z))$ has at most double zeros at $o_i$. In coordinates this means that we have
\bea
 \label{localcorrelation} \forall z \in U_{i} \, , & \omega_{0,1}(z) = & \sum_{d \geq 0} \omega_{0,1}\left[\begin{smallmatrix}
i  \\
d \end{smallmatrix}\right]\, \zeta_i^{d}(z)\,\dd\zeta_i(z) \\
\forall (z_1,z_2) \in U_{i_1} \times U_{i_2}\, ,  & \omega_{0,2}(z_1,z_2)  = & \frac{\delta_{i_1i_2}\,t_{i_1}\,\dd\zeta_{i_1}(z_1)\,\dd\zeta_{i_2}(z_2)}{(\zeta_{i_1}(z_1) - \zeta_{i_2}(z_2))^2} \nonumber \\
& & + \sum_{d_1,d_2 \geq 0} \omega_{0,2}\left[\begin{smallmatrix}
i_1 & i_2 \\
d_1 & d_2
\end{smallmatrix}\right]\,\zeta_{i_1}^{d_1}(z_1)\zeta_{i_2}^{d_2}(z_2)\,\dd \zeta_{i_1}(z_1)\dd \zeta_{i_2}(z_2) \nonumber
\eea
with $\omega_{0,1}\left[\begin{smallmatrix} i \\ 0 \end{smallmatrix}\right] \neq 0$ or $\omega_{0,1}\left[\begin{smallmatrix} i \\ 2 \end{smallmatrix}\right] \neq 0$. We usually assume that the coefficients of the double poles are $t_i = 1$ for all $i$.

The topological recursion provides a sequence of symmetric forms in $n$ variables,
\beq
\label{admis}\omega_{g,n} \in H^0(U^n,(K_{U}(*O))^{\boxtimes n})^{\mathfrak{S}_{n}},\qquad {\rm for}\,\,2g - 2 + n > 0 ,
\eeq
which describe the unique normalized solution of the "abstract loop equations" with initial data $(\omega_{0,1},\omega_{0,2})$ \cite{BEO,BSblob}. In \eqref{admis}, we take forms that are invariant under the symmetric group $\mathfrak{S}_{n}$ acting by permutation of the $n$ factors of $U$. The definition of $\omega_{g,n}$ proceeds by induction on $2g - 2 + n > 0$. We introduce the recursion kernel
\beq
\label{kluin}\mathcal{K}_i \in H^0\big(U \times U_i,[K_U \boxtimes K_{U_i}^{-1}(O)](\Delta \sqcup \Delta_{\sigma})\big),\qquad \mathcal{K}_i(z_0,z) = \frac{1}{2}\,\frac{\int_{\sigma_i(z)}^{z} \omega_{0,2}(\cdot,z_0)}{\omega_{0,1}(z) - \omega_{0,1}(\sigma_i(z))} .
\eeq
where $\Delta_{\sigma} = \bigsqcup_{i}\{(z,\sigma_i(z))\mid z \in U_i\}$. Denote by $I = \{z_2,\ldots,z_n\}$ a set of $(n - 1)$-variables. The topological recursion formula defining the symmetric forms is
\beq
\label{toprec} \omega_{g,n}(z_1,I) = \sum_{o_i \in O} \Res_{z \rightarrow o_i} \mathcal{K}_i(z_0,z)\bigg\{\omega_{g - 1,n + 1}(z,\sigma_i(z),I) \ \  + \sum_{\substack{h + h' = g \\ J \sqcup J' = I \\ (h,J) \neq (0,\emptyset),(g,I)}} \omega_{h,1 + |J|}(z,J)\otimes\omega_{h',1 + |J'|}(\sigma_i(z),J')\bigg\} .
\eeq
The induction reduces $\omega_{g,n}$ to an expression involving $2g - 2 + n$ residues with $\mathcal{K}_i$'s and $g + n - 1$ factors of $\omega_{0,2}$. If the induction is completely unfolded, $\omega_{g,n}$ is a sum over certain trivalent graphs containing a spanning tree, having $2g - 2 + n$ vertices and $g + n - 1$ extra edges in the complement of the tree \cite{EOFg}.

The topological recursion also define a sequence of numbers $(F_g)_{g \geq 2}$ by
$$
F_g = \frac{1}{2 - 2g} \sum_{o_i \in O} \Res_{z \rightarrow o_i} \bigg(\oint_{o_i}^{z} \omega_{0,1}\bigg)\omega_{g,1}(z)
$$

\br
Sometimes, the data of the differential form $\om_{0,1}$ is replaced by the data of a germ of  functions $y$ holomorphic at all the $o_i$ such that $\om_{0,1}(z) = y(z) dx(z)$ for any $z \in U$.
\er

\subsection{Relation to cohomological field theories}

For any local spectral curve, the $\omega_{g,n}$ can be represented in terms of intersection numbers on $\overline{\mathcal{M}}_{g,n}$ \cite{Einter}. There is a partial converse presented in \cite{DBOSS}, where it is established that the correlators of a semi-simple CohFT are computed by the topological recursion for a local spectral curve prescribed by the corresponding $R$-matrix (we stress that not all local spectral curves can arise from a CohFT). We now review this correspondence.

\subsubsection{CohFT data}

Let $(\tilde{\epsilon}_i)_i$ be a canonical basis, and $\underline{\epsilon}_i = \tilde{\epsilon}_i/\Delta_i^{1/2}$ the orthonormal basis. We write the formal series expansions
\bea
\label{Rmat} T(u) = u(\mathds{1} - R(u)\cdot\mathds{1}) & = & \sum_{i} \sum_{d \geq 2} T\left[\begin{smallmatrix} i \\ d \end{smallmatrix}\right]\,u^{d}\,\underline{\epsilon}_i \\
 \label{Bmat} B(u_1,u_2) & = & \frac{b^{\dagger} - R(u_1) \otimes R(u_2)\cdot b^{\dagger}}{u_1 + u_2} = \sum_{\substack{i_1,i_2 \\ d_1,d_2 \geq 0}} B\left[\begin{smallmatrix}
i_1 & i_2 \\
d_1 & d_2 
\end{smallmatrix}\right]\,u_1^{d_1}u_2^{d_2}\,\underline{\epsilon}_{i_1} \otimes \underline{\epsilon}_{i_2}
\eea
of the $R$-matrix and translation matrix defining uniquely a CohFT with canonical basis $(\tilde\epsilon_i)_i$. We denote by $\Omega_{g,n}$ its correlators and $\Omega_{g,n}^{{\rm deg}\,\,0}$ the restriction to their degree 0 part which only depends on the norms $\Delta_i$.

\subsubsection{Local spectral curve data}

We now define a local spectral curve, in terms of the above CohFT data. Its ramification points are indexed by a canonical basis $(\tilde{\epsilon}_i)_i$ of the underlying Frobenius manifold, and we set
\bea
\label{inicohf} \forall z \in U_i & \quad\quad\quad x(z) = & \!\!\!\! x(o_i) + \zeta_i^2(z)/2  \nonumber \\
\forall z \in U_i & \,\,\,\,\,\quad\omega_{0,1}^{{\rm odd}}(z) = &\!\!\!\! - \Delta_i^{-1/2}\,\zeta_i^2(z)\dd \zeta_i(z) + \sum_{d \geq 2} \frac{T\left[\begin{smallmatrix} i \\ d \end{smallmatrix}\right]}{(2d - 1)!!}\,\zeta_i^{2d}(z)\dd\zeta_i(z) \\
\forall (z_1,z_2) \in U_{i_1}\times U_{i_2} & \omega_{0,2}^{{\rm odd}}(z_1,z_2) = & \!\!\!\!\frac{\delta_{i_1i_2} \, \dd\zeta_{i_1}(z_1) \, \dd\zeta_{i_2}(z_2)}{(\zeta_{i_1}(z_1) - \zeta_{i_2}(z_2))^2} \nonumber \\
&& \!\!\!\!+ \sum_{d_1,d_2 \geq 0} \frac{B\left[\begin{smallmatrix}
i_1 & i_2 \\
d_1 & d_2 
\end{smallmatrix}\right]}{(2d_1 - 1)!!(2d_2 - 1)!!}\,\zeta_{i_1}^{2d_1}(z_1)\,\zeta_{i_2}^{2d_2}(z_2)\,\dd\zeta_{i_1}(z_1)\dd\zeta_{i_2}(z_2). \nonumber
\eea
By convention, $(2d - 1)!! = 1$ for $d = 0$.  Then, let $\omega_{0,1}^{\cancel{{\rm odd}}}(z)$ 
be an arbitrary holomorphic $1$-form which is invariant under the local involution $\zeta_i(z) \rightarrow -\zeta_i(z)$ in the patch $z \in U_i$, and an arbitrary bidifferential of the form
$$
\forall (z_1,z_2) \in U_{i_1}\times U_{i_2},\qquad \omega_{0,2}^{\cancel{\rm odd}}(z_1,z_2) = \sum_{\substack{d_1,d_2 \geq 0 \\ d_1\,\,{\rm or}\,\,d_2\,\,{\rm even}}} \omega_{0,2}\left[\begin{smallmatrix}
i_1 & i_2 \\
d_1 & d_2 
\end{smallmatrix}\right]\,\zeta_{i_1}^{d_1}(z_1)\zeta_{i_2}^{d_2}(z_2)\,\dd\zeta_{i_1}(z_1)\dd\zeta_{i_2}(z_2) .
$$
We then consider the following quantities as initial data for the topological recursion
\beq
\label{initiw}\omega_{0,1} = \omega_{0,1}^{{\rm odd}} + \omega_{0,1}^{\cancel{{\rm odd}}},\qquad \qquad \omega_{0,2} = \omega_{0,2}^{{\rm odd}} + \omega_{0,2}^{\cancel{{\rm odd}}}.
\eeq
To sum up, in the decomposition
$$
\omega_{0,2}(z_1,z_2) = \frac{\delta_{i_1i_2}\dd\zeta_{i_1}(z_1)\dd\zeta_{i_2}(z_2)}{(\zeta_{i_1}(z_1) - \zeta_{i_2}(z_2))^2} + \sum_{d_1,d_2 \geq 0} \omega_{0,2}\left[\begin{smallmatrix}
i_1 & i_2 \\
d_1 & d_2 
\end{smallmatrix}\right]\,\zeta_{i_1}^{d_1}(z_1)\zeta_{i_2}^{d_2}(z_2)\,\dd\zeta_{i_1}(z_1)\dd\zeta_{i_2}(z_2) ,
$$
the $R$-matrix of the CohFT only fixes the following coefficients.
$$
\omega_{0,2}\left[\begin{smallmatrix}
i_1 & i_2 \\
2d_1 & 2d_2 
\end{smallmatrix}\right] = \frac{B\left[\begin{smallmatrix}
i_1 & i_2 \\
d_1 & d_2 
\end{smallmatrix}\right]}{(2d_1 - 1)!!(2d_2 - 1)!!}
$$
All the other can be arbitrarily chosen. With this $\omega_{0,2}$ at hand, we build a family of meromorphic $1$-forms $\Xi_{d,i}(z_0)$ indexed by ramification points and an integer $d \geq 0$ defined by

\beq
\label{Xidef1}\Xi_{i,d}(z_0) := \Res_{z \rightarrow o_i} \frac{(2d + 1)!!\,\dd\zeta_{i}(z)}{\zeta_{i}(z)^{2d + 2}}\,\int_{o_i}^{z} \omega_{0,2}(\cdot,z_0) .
\eeq
The only singularity of this $1$-form is a pole at $z_0 \rightarrow o_i$, and we actually have, for $z_0 \in U_{i_0}$,
\beq
\label{Xidef2}\Xi_{i,d}(z_0) = \delta_{ii_0}\,\frac{(2d + 1)!!\,\dd\zeta_{i_0}(z_0)}{\zeta_{i_0}^{2d + 2}(z_0)} + (2d - 1)!! \sum_{d_0 \geq 0} \omega_{0,2}\left[\begin{smallmatrix}
i & i_0 \\
2d & d_0 
\end{smallmatrix}\right]\,\zeta_{i_0}^{d_0}(z_0)\,\dd\zeta_{i_0}(z_0) .
\eeq
We stress that the domain of definition of $\Xi_{i,d}$ is the whole $U$ -- as for $\omega_{0,2}$ -- and not only $U_{i}$.

\begin{theorem}\cite{DBOSS}
\label{TDB} Consider the $n$-forms $\omega_{g,n}(z_1,\ldots,z_n)$ defined by
\beq
\label{comparet}\omega_{g,n}(z_1,\ldots,z_n) = \sum_{\substack{i_1,\ldots,i_n \\ d_1,\ldots,d_n \geq 0}} \int_{\overline{\mathcal{M}}_{g,n}} \Omega_{g,n}^*(\underline{\epsilon}_{i_1}\otimes \cdots \otimes \underline{\epsilon}_{i_n})\prod_{\ell = 1}^n \psi_{\ell}^{d_{\ell}}\,\prod_{\ell = 1}^n \Xi_{i_{\ell},d_{\ell}}(z_{\ell}) .
\eeq
Then, $\omega_{g,n}$ is computed by the topological recursion \eqref{toprec} with initial data \eqref{initiw}.
\end{theorem}

Note that, in the right-hand side, the only dependence in the choice of $\omega_{0,2}^{\cancel{\rm odd}}$ lies in the meromorphic forms $\Xi_{d,i}$ defined in \eqref{Xidef1} and on which the correlators are decomposed. Further, the $\omega_{g,n}$ do not depend on the constant $x(o_i)$ and the $1$-form $\omega_{0,1}^{{\rm even}}$. Allowing them to be non-zero can simplify expressions, and does matter if one is interested in building a Landau-Ginzburg model for which \eqref{inicohf} is the local expansion near ramification divisors. This matter will be discussed in Section~\ref{sec:glob}.

We observe that
\beq
\label{intL}\int_{\mathbb{R}} \frac{e^{-\zeta^2/2u}}{2(2\pi u)^{1/2}}\,\zeta^{2d}\,\dd \zeta = (2d - 1)!!\,u^{d} .
\eeq
Therefore, the relation between the initial data \eqref{Rmat} and the $R$-matrix of the CohFT is given by a Laplace transform, as it is well-known.
\bea
\mathds{1}u + T(u) & = & \sum_{i} \bigg(\int_{\gamma_i} \frac{\exp\big[-\frac{x(z) - x(o_i)}{u}\big]}{2(2\pi u)^{1/2}}\,\omega_{0,1}(z)\bigg)\,\underline{\epsilon}_i \nonumber \\
B(u_1,u_2) & = &  \sum_{i,j} \bigg(\int_{\gamma_i \times \gamma_j} \frac{\exp\big[-\frac{x(z_1) - x(o_i)}{u_1} - \frac{x(z_2) - x(o_j)}{u_2}]}{4(2\pi u_1)^{1/2}(2\pi u_2)^{1/2}}\bigg\{\omega_{0,2}(z_1,z_2) - \frac{\delta_{ij}}{(\zeta_{i}(z_1) - \zeta_{j}(z_2))^2}\bigg\}\bigg)\,\underline{\epsilon}_i \otimes \underline{\epsilon}_j  .\nonumber
\eea
Here, $\gamma_i$ is a steepest descent contour in the spectral curve passing through the ramification point $o_i$ and going to $\infty$ in the direction $\mathrm{Re}\,x(z)/u \rightarrow -\infty$. For the case of local spectral curves we consider here, the meaning of the right-hand side must be precised. We expand the non-exponential part of the integrand as power series when $z \rightarrow o_{i}$, then integrate term by term against $e^{-x(z)/u}$ using \eqref{intL}. Each term yields a monomial in $u$, and thus the right-hand side is a well-defined formal power series in $u$.

\subsubsection{$0$-point correlation functions}

The following lemma did not appear in \cite{DBOSS}, but is an easy consequence of the content of that paper

\begin{theorem}
\label{TDB2} For $g \geq 2$, we have
$$
\int_{\overline{\mathcal{M}}_{g,0}} \Omega_{g,0} = F_g:= \frac{1}{2 - 2g} \sum_{i} \Res_{z \rightarrow o_i} \bigg(\int^{z}_{o_i} \omega_{0,1}\bigg)\omega_{g,1}(z).
$$
\end{theorem}

In order to prove it, we shall use the dilaton equation for the classes produced by the action of the Givental group.

\begin{lemma}
Given a semi-simple 2d TQFT $(\Omega_{g,n}^{{\rm deg}\,\,0})_{g,n}$, a translation $T(u) \in u^2 {\cal A}[[u]]$ and an endomorphism $R(u) \in {\rm End}(\mathcal{A})[[u]]$, the cohomology classes
$\Omega = \hat{T}\hat{R}\Omega^{{\rm deg}\,\,0}$ satisfy
\beq
\label{dilatongen}(2g-2+n) \int_{\overline{\cal M}_{g,n}} \Omega_{g,n}^*(\cdot) = \int_{\overline{\cal  M}_{g,n+1}}\Omega_{g,n+1}^*(\cdot \otimes \mathds{1})  \psi_{n+1} - \sum_k \int_{\overline{\cal  M}_{g,n+1}}\Omega_{g,n+1}^*(\cdot \otimes T_k)  \psi_{n+1}^k
\eeq
for $2g-2+n>0$.
\end{lemma}
In particular, this result holds for CohFTs, where $R$ and $T$ are related by $T(u) = u({\rm Id} - R(u))\mathds{1}$. Theorem~\ref{TDB2} is then a simple consequence of \eqref{dilatongen} using a residue representation.

\medskip

\noindent \textbf{Proof.} 
We prove this formula in three steps. First, let us remark that this statement is true for a vanishing $T(u) $ and $R(u) = {\rm Id}$ by the dilaton equation on the moduli space of curves:
\beq
\label{dilaton}\int_{\overline{\mathcal{M}}_{g,n + 1}} \Big[\prod_{i = 1}^n \psi_i^{d_i}\Big] \psi_{n + 1} = (2g - 2 + n)\int_{\overline{\mathcal{M}}_{g,n}} \prod_{i = 1}^n \psi_i^{d_i}
\eeq

Next, we check the validity of \eqref{dilatongen} for arbitrary $R(u)$ with $T(u) \equiv 0$. By definition, we have that
$$
\check{\Omega}_{g,n} := (\hat{R}\Omega)_{g,n} = \sum_{\substack{\Gamma \in G_{g,n}}} \frac{1}{|{\rm Aut}\,\Gamma|}\,\,(\pi_{\Gamma})_*\bigg(\prod_{l = {\rm leaf}} R(\psi_{l}) \!\! \prod_{\substack{e = {\rm edge} \\ \{v'_e,v''_e\}}}B(\psi_{v_e'},\psi_{v_e''}) \!\!\!\prod_{\substack{v  = {\rm vertex}}} \Omega_{g(v),n(v)}^{{\rm deg}\,\,0}\bigg)
$$
where $G_{g,n}$ denotes the stable graphs of genus $g$ with $n$ legs. We can now compute 
$$
\int_{\overline{\cal  M}_{g,n+1}} \check{\Omega}_{g,n+1}^*(\cdot \otimes \mathds{1})  \psi_{n+1} = \sum_{\substack{\Gamma \in G_{g,n+1}}} \frac{1}{|{\rm Aut}\,\Gamma|}\,\, \int_{\overline{\cal  M}_{g,n+1}} \psi_{n+1} \, (\pi_{\Gamma})_*\bigg(\prod_{l = {\rm leaf}\backslash \{n+1\}} R(\psi_{l}) \!\! \prod_{\substack{e = {\rm edge} \\ \{v'_e,v''_e\}}}B(\psi_{v_e'},\psi_{v_e''}) \!\!\!\prod_{\substack{v  = {\rm vertex}}} \Omega_{g(v),n(v)}^{{\rm deg}\,\,0 \, *}\bigg) 
$$
Denoting by $v^*$ the unique vertex of $\Gamma$ adjacent to the leaf corresponding to the $(n+1)$-th marked point, we can write it as a product of integrals over the moduli spaces associated to each vertex
\bea
\int_{\overline{\cal  M}_{g,n+1}} \check{\Omega}_{g,n+1}^*(\cdot \otimes \mathds{1})  \psi_{n+1} &=& 
\sum_{\substack{\Gamma \in G_{g,n+1}}} \frac{1}{|{\rm Aut}\,\Gamma|} \,\,\,\sum_{\substack{d(h) \geq 0\,\,\,\,h = {\rm half}-{\rm edge} \\ i(v) \in \llbracket 1,k \rrbracket\,\,\,\,v = {\rm vertex}}} \,\,\prod_{\substack{e = {\rm edge} \\ \{v'_e,v''_e\}}}  B\left[\begin{smallmatrix}
i(v'_e) & i(v''_e) \\
d(h_1(e)) & d(h_2(e)) 
\end{smallmatrix}\right] \cr
&& \quad \times \prod_{v \neq v^*} \bigg(\Delta_{i(v)}^{g(v)-1} \int_{\overline{\cal  M}_{g(v),n(v)}} \prod_{j=1}^{n(v)} \psi_j^{d_j(v)}\bigg) \cr
&& \; \; \qquad \times  \Delta_{i(v^*)}^{g(v^*)-1}\int_{\overline{\cal  M}_{g(v^*),n(v^*)}}\bigg[\prod_{j=1}^{n(v^*)-1} \psi_j^{d_j(v^*)}\bigg]\,\psi_{n(v^*)}
\cr
\eea
with the following notations. Each vertex $v$ is assigned a label $i(v) \in \llbracket 1 , k \rrbracket$ where $k$ is the dimension of ${\cal A}$. Each half-edge $h$ is assigned a nonnegative integer $d(h)$. Each edge $e$ is decomposed into a pair of half edges $(h_1(e),h_2(e))$ such that $h_1(e)$ (resp. $h_2(e)$) is adjacent to $v'_e$ (resp. $v''_e$). The incident half-edges to a vertex $v$ are labeled (in some arbitrary way) from 1 to $n(v)$ and we denote by $d_j(v)$ the label $d(h)$ where $h$ is the $j$-th half edge incident to $v$. By convention, the $d(v^*)$-th half edge is the one from $v^*$ to the $(n + 1)$-th leaf.

We can now apply the dilaton equation to the vertex $v^*$ in order to remove the $(n+1)$-th marked point leading to 
\bea
\int_{\overline{\cal  M}_{g,n+1}}\check{\Omega}_{g,n+1}^*(\cdot \otimes \mathds{1})  \psi_{n+1} &=& 
\sum_{\substack{\Gamma \in G_{g,n+1}}} \frac{1}{|{\rm Aut}\,\Gamma|}\,\,\sum_{d_1(e),d_2(e) \, , \, e = {\rm edge}  } \sum_{i(v) \, , \, v  = {\rm vertex}} \, \, \prod_{\substack{e = {\rm edge} \\ \{v'_e,v''_e\}}}  B\left[\begin{smallmatrix}
i(v'_e) & i(v''_e) \\
d(h_1(e)) & d(h_2(e)) 
\end{smallmatrix}\right] \cr
&& \quad \times \prod_{v \neq v^*} \left[\Delta_{i(v)}^{g(v)-1} \int_{\overline{\cal  M}_{g(v),n(v)}} \prod_{j=1}^n(v) \psi_j^{d_j(v)}\right] \cr
&& \; \; \qquad \times  (2 g(v^*)-2+n(v^*)-1) \, \, \Delta_{i(v^*)}^{g(v^*)-1}\int_{\overline{\cal  M}_{g(v^*),n(v^*)-1}}\, \left[\prod_{j=1}^{n(v)-1} \psi_j^{d_j(v)}\right] .
\cr
\eea
Finally, recall that a graph of genus $g$ with $n + 1$ leaves is nothing but a graph a genus $g$ with $n$ leaves and a distinguished vertex -- one retrieves the first by adding an $(n+1)$-th leaf at the distinguished vertex. So, we can convert the sum into a sum over $\Gamma \in \mathcal{G}_{g,n}$, and since $\sum_{v} (2g(v) - 2 + n(v)) = 2g-2+n$, we obtain
$$
\int_{\overline{\cal  M}_{g,n+1}}\check{\Omega}_{g,n+1}^*(\cdot \otimes \mathds{1})  \psi_{n+1} = (2g-2+n) \int_{\overline{\cal  M}_{g,n}}\check{\Omega}_{g,n}^*(\cdot) .
$$

In the last step of the argument, we study the action of the translation on the dilaton equation. Let $(\Omega_{g,n}^*)_{g,n}$ be a set of cohomology classes satisfying $\int_{\overline{\cal  M}_{g,n+1}} {\Omega}_{g,n+1}^*(\cdot \otimes \mathds{1})  \psi_{n+1} = (2g-2+n) \int_{\overline{\cal  M}_{g,n}} {\Omega}_{g,n}^*(\cdot)$, and let  $T(u) = \sum_{k\geq 2} T_k u^k \in u^2 {\cal A}[[u]]$ be a  formal series. For $g \geq 2$, we have by definition that
$$
 (T \Omega^*)_{g,1}(\mathds{1})\psi_0 =   \sum_{m \geq 0} \frac{1}{m!} \sum_{k_1,\dots k_m \geq 2} (\pi_{m})_*\Big\{{\Omega}^*_{g, 1+m}(\mathds{1} \otimes T_{k_1}\otimes \cdots \otimes T_{k_m}) \psi_1^{k_1} \dots \psi_m^{k_m}\Big\}  \psi_0 $$
where we denote by $\pi_m: \overline{\cal M}_{g,1+m} \to \overline{\cal M}_{g,1}$ the map forgetting the last $m$ marked points. This implies that
$$
\int_{\overline{\cal M}_{g,1}}  (T \Omega^*)_{g,1}(\mathds{1})\psi_0 =
\sum_{m \geq 0} \frac{1}{m!} \sum_{k_1,\dots k_m \geq 2} \int_{\overline{\cal M}_{g,1+m}} {\Omega}^*_{g, 1+m}(\mathds{1} \otimes T_{k_1}\otimes \cdots \otimes T_{k_m}) \psi_1^{k_1} \dots \psi_m^{k_m}  \psi_0.
$$
Using the dilaton equation for the classes $(\Omega_{g,n})$, one can push it down to 
$$
\int_{\overline{\cal M}_{g,1}}  (T \Omega^*)_{g,1}(\mathds{1})\psi_0 =
\sum_{m \geq 0} \frac{(2g-2+m)}{m!} \sum_{k_1,\dots k_m \geq 2} \int_{\overline{\cal M}_{g,m}} {\Omega}^*_{g, m}(T_{k_1}\otimes \cdots \otimes T_{k_m}) \psi_1^{k_1} \dots \psi_m^{k_m} .
$$
This can be decomposed into a sum of two terms which reads
$$
\int_{\overline{\cal M}_{g,1}} (T \Omega^*)_{g,1}(\mathds{1})\psi_0  =
(2g-2) \int_{\overline{\cal M}_{g,0}} (T \Omega^*)_{g,0}(\mathds{1}) +
\sum_{m \geq 1} \frac{1}{(m-1)!} \sum_{k_1,\dots k_m \geq 2} \int_{\overline{\cal M}_{g,1+m-1}} {\Omega}^*_{g, 1+m}(T_{k_1}\otimes \cdots \otimes T_{k_m}) \psi_1^{k_1} \dots \psi_m^{k_m}.
$$
On can see the second term as a contribution to $(\hat{T} \Omega)_{g,1}$ by marking the 1-st point for example and making the change of variable $m \to m-1$. This proves the lemma for $n=0$. The generalization to $n > 0$ follows by similar arguments. \hfill $\Box$

\medskip

In the next paragraphs, we apply Theorems~\ref{TDB}-\ref{TDB2} to the CohFT obtained from a modular functor.

\subsection{Topological recursion and Verlinde formula}
\label{Verlindeder}

We shall not reproduce the general proof of Theorem~\ref{TDB}, but it is easy and nevertheless instructive to derive it directly for the degree 0 part of the theory, i.e. for the $2d$ TQFT associated to a modular functor. In this way, we exhibit the Verlinde formula computing the dimension $D_{\vec{\lambda}}(\mathbf{\Sigma}_{g,n}) = {\rm dim}\,\mathcal{V}_{\vec{\lambda}}(\mathbf{\Sigma}_{g,n})$ of the $3d$ TQFT vector spaces as a special case of topological recursion.

\subsubsection{Verlinde formula}

Let us start with a brief review of the Verlinde formula. In terms of matrix elements in the $(e_{\lambda})_{\lambda}$ basis, it takes the following form.
\beq
\label{Verlind1}D_{\vec{\lambda}}(\mathbf{\Sigma}_{g,n}) = \Big[\mathcal{C}[\beta;\lambda_1]\cdots \mathcal{C}[\beta;\lambda_n]\,\mathcal{W}^{g}\big]_{11} = b(e_1,e_{\lambda_1}\times \cdots \times e_{\lambda_n} \times \mathcal{W}^{g}(e_1)) \nonumber
\eeq
In this expression, we used  the curve operators introduced in \S~\ref{curvop}, and
$$
\mathcal{W} = \sum_{\mu \in \Lambda} \mathrm{Tr}\big(\mathcal{C}[\beta;\mu^{\dagger}]\big)\,\mathcal{C}[\beta;\mu] .
$$

Equation \eqref{Verlind1} -- and its equivalent form below \eqref{Verlind2} -- have been first conjectured by Verlinde \cite{Verlinde}. It was then proved by Moore and Seiberg \cite{MooreSeiberg} for modular functors coming from modular tensor categories (though the notion had not been coined yet), but the proof actually holds for any modular functor (see also \cite{Tu}). The strategy consists (a) in degenerating $\mathbf{\Sigma}_{g,n}$ in an arbitrary way to a nodal surface whose smooth components are all thrice-punctured spheres, and then use the factorization, thus expressing $D_{\bullet}(\mathbf{\Sigma}_{g,n})$ in terms of $D_{\bullet}(\mathbf{\Sigma}_{0,3})$ ; (b) to show that the $S$-matrix diagonalizes the multiplication in the Frobenius algebra, which implies \eqref{Verlind2} for $(g,n) = (0,3)$. For modular functors without further assumptions, this is for instance proved in \cite{Se} and also in \cite{AU3}. 

We can derive another expression by working in the $\tilde{\epsilon}$-basis, taking advantage of the fact that it diagonalizes simultaneously the curve operators and exploiting $e_1 = \sum_{\tau \in \Lambda} \tilde{\epsilon}_{\tau}$. The eigenvalues of the curve operators are \eqref{eigne1}
$$
\mathcal{C}[\beta;\mu](\tilde{\epsilon}_{\tau}) = \frac{(\underline{S}^{-1})_{\mu\tau}}{(\underline{S}^{-1})_{1\tau}}\,\tilde{\epsilon}_{\tau}, 
$$
so that the trace yields
\beq
{\rm Tr}\,\mathcal{C}[\beta;\mu^{\dagger}] = \sum_{\nu \in \Lambda} \frac{(\underline{S}^{-1})_{\mu^{\dagger}\nu}}{(\underline{S}^{-1})_{1\nu}} = \sum_{\nu \in \Lambda} \frac{\underline{S}_{\nu\mu}}{(\underline{S}^{-1})_{1\nu}}.
\eeq
Then the eigenvalues of $\mathcal{W}$ read, using \eqref{SST},
$$
\mathcal{W}(\tilde{\epsilon}_{\tau}) = \bigg(\sum_{\mu,\nu \in \Lambda} \frac{\underline{S}_{\nu\mu}}{(\underline{S}^{-1})_{1\nu}}\,\frac{(\underline{S}^{-1})_{\mu\tau}}{(\underline{S}^{-1})_{1\tau}}\bigg)\,\tilde{\epsilon}_{\tau} = \frac{\tilde{\epsilon}_{\tau}}{\big[(\underline{S}^{-1})_{1\tau}\big]^2} .
$$
Using that $b(\tilde{\epsilon}_{\tau},\tilde{\epsilon}_{\tau'}) = \big[(\underline{S}^{-1})_{1\tau}\big]^2$, we obtain the equivalent -- and maybe better known form -- of \eqref{Verlind1}:
\beq
\label{Verlind2}D_{\vec{\lambda}}(\mathbf{\Sigma}_{g,n}) = \sum_{\tau \in \Lambda} \frac{(\underline{S}^{-1})_{\lambda_1\tau}\,\cdots\,\,(\underline{S}^{-1})_{\lambda_n\tau}}{\big[(\underline{S}^{-1})_{1\tau}\big]^{2g - 2 + n}} .
\eeq

\subsubsection{Topological recursion for the $2d$ TQFT}
\label{Sec432} 
We denote $\omega_{g,n}^{{\rm KdV}}$ the correlation functions returned by the topological recursion for the Airy curve defined by
\beq
\label{AriY} U = \mathbb{C},\qquad x(\zeta) = \zeta^2/2,\qquad y(z) = -z,\qquad \omega_{0,2}(\zeta_1,\zeta_2) = \frac{\dd\zeta_1\dd\zeta_2}{(\zeta_1 - \zeta_2)^2} .
\eeq
Although it is not needed in the proof of Proposition~\ref{P2p2} below, we recall the following well-known formula\footnote{See \cite{Ekappa} for a proof based on matrix model techniques. It can also be obtained directly from the Virasoro constraints \cite{Witten,Kontsevich} satisfied by the intersection numbers.}.
\beq
\label{kdvsa}\omega_{g,n}^{{\rm KdV}}(\zeta_1,\ldots,\zeta_n) = \sum_{d_1,\ldots,d_n \geq 0}\int_{\overline{\mathcal{M}}_{g,n}} \prod_{i = 1}^n \psi_{i}^{d_i}\,\frac{(2d_i + 1)!!\,\dd\zeta_i}{\zeta_i^{2d_i + 2}} .
\eeq

\begin{proposition}
\label{P2p2}Consider the local spectral curve defined by the data of $U = \bigsqcup_{\lambda \in \Lambda} U_{\lambda}$ and, for $z \in U_{\lambda}$ and $(z_1,z_2) \in U_{\lambda_1} \times U_{\lambda_2}$ in the expression of the following data local coordinates.
\beq
\label{thecurve}\left\{
\begin{array}{rcl}
x(z) & = & x(o_{\lambda}) + (\zeta_{\lambda}(z))^2/2 \cr
y(z) & = & - (\underline{S}^{-1})_{1\lambda}\,\zeta_{\lambda}(z) \cr
\omega_{0,2}(z_1,z_2) & = & \delta_{\lambda_1\lambda_2}\,\frac{\dd \zeta_{\lambda_1}(z_1)\,\dd \zeta_{\lambda_2}(z_2)}{(\zeta_{\lambda_1}(z_1) - \zeta_{\lambda_2}(z_2))^2} \cr
\end{array}
\right. .
\eeq
Then, the correlation functions returned by the topological recursion for $2g - 2 + n > 0$ are, for $z_i \in U_{\lambda_i}$,
\beq
\label{omgntf}\om_{g,n}(z_1,\dots,z_n) =  \bigg(\sum_{\vec{\mu} \in \Lambda^n} \prod_{i =1}^n \underline{S}_{\lambda_i\mu_i}\,D_{\vec{\mu}}(\mathbf{\Sigma}_{g,n})\bigg)\,\omega_{g,n}^{{\rm KdV}}(\zeta_{\lambda}(z_1),\ldots,\zeta_{\lambda}(z_n)) .
\eeq
\end{proposition}
Remark that, since $\omega_{0,2}$ is purely diagonal in \eqref{thecurve}, the left-hand side in \eqref{omgntf} vanishes unless all $z_i$ belongs to the same open set $U_{\lambda_1}$. This is consistent with \eqref{Verlind2} which shows that the bracket in the right-hand side of \eqref{omgntf} vanishes unless $\lambda_i = \lambda$ for all $i$, and explains the notation used in the KdV factor. If we want to compare with Theorem~\ref{TDB}, we use \eqref{kdvsa} and the following basis of $1$-forms.
$$
z \in U_{\mu},\,\qquad \Xi_{d,\lambda}(z) = \delta_{\lambda\mu}\,\frac{(2d + 1)!!}{[\zeta_{\lambda}(z)]^{2d + 2}} .
$$

\vspace{0.2cm}

\noindent \textbf{Proof.} In \eqref{thecurve}, we recognize several copies of the Airy curve, except for a rescaling of $y(z) \rightarrow (\underline{S}^{-1})_{1\lambda}\,y(z)$ in the patch $z \in U_{\lambda}$. By the previous remark, we can assume that for $i \in \llbracket 1,n \rrbracket$, we have all $z_i \in U_{\tau}$ for some $\tau \in \Lambda$.  Since $\omega_{g,n}$ is obtained from the initial data by a sequence of $2g - 2 + n$ residues at the ramification point $o_{\tau} \in U_{\tau}$ involving the kernel \eqref{kluin}, we have that
$$
\omega_{g,n}(z_1,\ldots,z_n) = \frac{\omega_{g,n}^{{\rm KdV}}(z_1,\ldots,z_n)}{\big[(\underline{S}^{-1})_{1\tau}\big]^{2g - 2 + n}} .
$$
To conclude, we recognize with \eqref{Verlind2} that
$$
\frac{1}{\big[(\underline{S}^{-1})_{1\tau}\big]^{2g - 2 + n}} = \sum_{\tau \in \Lambda} \prod_{i = 1}^n \underline{S}_{\tau\mu_i} D_{\vec{\mu}}(\mathbf{\Sigma}_{g,n}) .
$$
\hfill $\Box$

\vspace{0.2cm}

\noindent The expression of the topological recursion in terms of intersection numbers is naturally written in terms of local coordinates around the ramification points, which are associated to the canonical basis $(\underline{\epsilon})_{\lambda}$. To retrieve the expression in the $(e_{\lambda})_{\lambda}$-basis, one has to make a linear combination which cancels the $\underline{S}$ in \eqref{omgntf}. So, the property that $(\omega_{g,n})_{g,n}$ appearing in \eqref{comparet} satisfy the topological recursion is equivalent to the Verlinde formula \eqref{Verlind2} for $D_{\bullet}(\mathbf{\Sigma}_{g,n})$.

We can reformulate the proof by saying that, in the sum over trivalent graphs that computes the $\omega_{g,n}$ of the topological recursion, the dependence in the indices $\lambda_i \in \Lambda$ of the weight of a graph is the same for all graphs\footnote{While finishing this project, we heard that Dumitrescu and Mulase arrived independently in \cite{DumiMulase} to Proposition~\ref{P2p2}.}.

\subsubsection{Change of basis}
\label{sec:basischange}
Denote $z_{\lambda}(\zeta) \in U_{\lambda}$ the point such that $\zeta_{\lambda}(z_{\lambda}(\zeta)) = \zeta$, and define the following forms
$$
\omega_{g,n}^{[\vec{\lambda}]}(\zeta_1,\ldots,\zeta_n) := \omega_{g,n}(z_{\lambda_1}(\zeta_1),\ldots,z_{\lambda_n}(\zeta_n)) .
$$
Proposition~\ref{P2p2} can be alternatively written as
\beq
\label{prov} \sum_{\vec{\lambda} \in \Lambda^n} \prod_{i=1}^n (\underline{S}^{-1})_{\mu_i \lambda_i} \om_{g,n}^{[\vec{\lambda}]}(\zeta_1,\dots,\zeta_n) =  D_{\vec{\mu}}(\mathbf{\Sigma}_{g,n})\,\om_{g,n}^{{\rm KdV}}(\zeta_1,\dots,\zeta_n) .
\eeq
In anticipation of Section~\ref{sec:glob}, we remark that if there exists a global spectral curve, it corresponds to expressing the correlation functions in a different basis of coordinates. In the Landau-Ginzburg picture, this amounts to taking a different linear combination of vanishing cycles, while in the GW theory picture, it means taking a different basis of the cohomology of the target space pulled back to the moduli space by the evaluation map. 

\subsection{Local spectral curves for modular functors}

For the CohFT's built from a modular functor, we computed the $R$-matrix in \eqref{Rmatri}. Working in the orthonormal basis $(\underline{\epsilon}_{\lambda})_{\lambda}$ introduced in \eqref{orthoba}, we deduce an expression for the local spectral curve, but we exploit the freedom to add non-odd parts to the initial data to simplify the result. Define the function $\mathcal{B}(\zeta)$ by
\beq
\label{renD}\mathcal{B}(\zeta) = \frac{{\rm cosh}(\zeta)}{\zeta^2} - \frac{{\rm sinh}(\zeta)}{\zeta}, 
\eeq
and the renormalized Dehn twist by
\beq
h_{\lambda} = 2t (r_{\lambda} + c/24) .
\eeq
From Lemma~\ref{Dehnlemma}, we deduce that $h_{1} = ct/12$ and $h_{\lambda} = h_{\lambda^{\dagger}}$.
\begin{theorem}
\label{TH42}The topological recursion applied to the initial data $(\omega_{0,1} = y\dd x\,,\,\omega_{0,2})$ with
\bea
\label{finalw02} \forall z \in U_{\lambda} & & \left\{\begin{array}{l} x(z) = x(o_{\lambda}) + \left[\zeta_{\lambda}(z)\right]^2/2 \\ y(z) = -(\underline{S}^{-1})_{1\lambda}\,\frac{\exp[(ct/12)^{1/2}\zeta_{\lambda}(z)]}{(ct/12)^{1/2}} \end{array}\right. \\
\forall (z_1,z_2) \in U_{\lambda_1} \times U_{\lambda_2} & & \omega_{0,2}(z_1,z_2) = \sum_{\tau \in \Lambda} \underline{S}_{\lambda_2\tau}\,(\underline{S}^{-1})_{\tau\lambda_1}\,h_{\tau}\,\mathcal{B}\big[h_{\tau}^{1/2}(\zeta_{\lambda_1}(z_1) - \zeta_{\lambda_2}(z_2))\big]\,\dd\zeta_{\lambda_1}(z_1)\dd\zeta_{\lambda_2}(z_2) \nonumber
\eea
computes the intersection indices of Chern classes of the bundles $[\mathcal{Z}_{\vec{\lambda}}]_{g,n}$, in the sense of Theorem~\ref{TDB}-\ref{TDB2} when decomposed on the following basis of $1$-forms.
$$
\forall z \in U_{\mu},\qquad \Xi_{d,\lambda}(z) = \sum_{\tau \in \Lambda} \underline{S}_{\lambda\tau}(\underline{S}^{-1})_{\tau\mu}\,\frac{\Gamma[2d + 2\,;\,h_{\tau}^{1/2}\zeta_{\mu}(z)]}{2^{d}\,d!\,[\zeta_{\mu}(z)]^{2d + 2}}\,\dd\zeta_{\mu}(z)
$$
where $\Gamma[a\,;\,x] := \int_{x}^{\infty} \dd v\,v^{a - 1}e^{-v}$ is the incomplete Gamma function\footnote{Since $\Gamma[a\,;\,x] = \Gamma(a) + O(x^{a - 1})$ when $x \rightarrow 0$, we see that $\Xi_{d,\lambda}(z) = \delta_{\lambda\mu}\,(2d + 1)!!\,[\zeta_{\mu}(z)]^{-(2d + 2)}\,\dd\zeta_{\mu}(z) + O(1)$ as it should be.}.
\end{theorem}

With the notational remark of Section~\ref{sec:basischange}, we therefore obtain a generalization of \eqref{prov}, which is our second main result
\beq
\label{CMUGN} \boxed{\sum_{\vec{\lambda} \in \Lambda^n} \prod_{i = 1}^n (\underline{S}^{-1})_{\mu_i\lambda_i}\,\omega_{g,n}^{[\vec{\lambda}]}(\zeta_1,\ldots,\zeta_n) = \int_{\overline{\mathcal{M}}_{g,n}} {\rm Ch}_{t}(\mathcal{Z}_{\vec{\mu}}(\mathbf{\Sigma}_{g,n})) \prod_{i = 1}^n \Xi(\psi_i,h_{\mu_i}^{1/2}\zeta_i)}
\eeq
where $\Xi(u,\zeta)$ is a formal series in $u$ of meromorphic forms in a neighborhood of $\zeta = 0$:
$$
\Xi(u,\zeta) = \sum_{d \geq 0} \frac{\Gamma[2d + 2 ; \zeta]}{2^{d}d!\,\zeta^{2d + 2}}\,u^{d} 
$$
And, for $g \geq 2$, we have as well
$$
\boxed{F_{g} = \int_{\overline{\mathcal{M}}_{g,0}} {\rm Ch}_{t}(\mathcal{Z}(\mathbf{\Sigma}_{g,0})).}
$$

\vspace{0.2cm}

\noindent \textbf{Proof.} This is pure algebra. The inverse norm of $\underline{\epsilon}_{\lambda}$ is introduced in \eqref{orthoba}, and gives $\Delta_{\lambda}^{-1/2} = \left(\underline{S}^{-1}\right)_{1\lambda}$. Inserting $r_{1} = 0$ in \eqref{Rmatri}, we obtain  $T(u) = u(1 - e^{utc/24})\mathds{1}$. Then, for $z \in U_{\lambda}$, we compute with \eqref{inicohf} that
\bea
\omega_{0,1}^{{\rm odd}}(z) & = & (\underline{S}^{-1})_{1\lambda} \bigg(-\left[\zeta_{\lambda}(z)\right]^2 - \sum_{d \geq 1} \frac{(ct/24)^{d}\,[\zeta_{\lambda}(z)]^{2d + 2}}{d!\,(2d + 1)!!}\bigg) \, \dd \zeta_\lambda(z) \nonumber \\
& = & -(\underline{S}^{-1})_{1\lambda} \bigg(\zeta_{\lambda}(z) + \sum_{d \geq 1} \frac{(ct/12)^{d}\,\left[\zeta_{\lambda}(z)\right]^{2d + 1}}{(2d + 1)!}\bigg)\zeta_{\lambda}(z)\dd\zeta_{\lambda}(z) \nonumber \\
& = & -(\underline{S}^{-1})_{1\lambda} \,\frac{{\rm sinh}[(ct/12)^{1/2}\,\zeta_{\lambda}(z)]}{(ct/12)^{1/2}}\,\dd x(z) .
\eea
By adding a suitable even part, we can choose
$$
\omega_{0,1}(z) = -(\underline{S}^{-1})_{1\lambda}\,\frac{\exp[(ct/12)^{1/2}\zeta_{\lambda}(z)]}{(ct/12)^{1/2}}\,\dd\zeta_{\lambda}(z) .
$$

We now proceed to the $2$-point function. We have $b^{\dagger} = \sum_{\lambda \in \Lambda} e_{\lambda} \otimes e_{\lambda^{\dagger}}$. Thus, we obtain from the definition \eqref{Bmat} that
$$
B(u_1,u_2) = \sum_{\lambda_1,\lambda_2 \in \Lambda} \bigg(\sum_{\tau \in \Lambda} \frac{1 - \exp\big[(u_1h_{\tau} + u_2h_{\tau^{\dagger}})/2\big]}{u_1 + u_2}\,(\underline{S}^{-1})_{\tau\lambda_1}\,(\underline{S}^{-1})_{\tau^{\dagger}\lambda_2}\bigg)\,\underline{\epsilon}_{\lambda_1} \otimes \underline{\epsilon}_{\lambda_2} .
$$
There is some simplification because $h_{\tau} = h_{\tau^{\dagger}}$, and $(\underline{S}^{-1})_{\tau^{\dagger}\lambda_2} = \underline{S}_{\lambda_2\tau}$ according to \eqref{SSJ}.
\beq
\label{bu2bu} B(u_1,u_2) = \sum_{\lambda_1,\lambda_2 \in \Lambda} \bigg(\sum_{\tau \in \Lambda} \frac{1 - \exp\big[(u_1 + u_2)h_{\tau}/2\big]}{u_1 + u_2}\,\underline{S}_{\lambda_2\tau}\,\underline{S}^{-1}_{\tau\lambda_1}\bigg)\,\underline{\epsilon}_{\lambda_1}\otimes \underline{\epsilon}_{\lambda_2} .
\eeq
Let $z_1 \in U_{\lambda_1}$ and $z_2 \in U_{\lambda_2}$. We decompose in power series of $(u_1,u_2)$ and compute with \eqref{inicohf} for the odd part:
\bea
& & \omega_{0,2}^{{\rm odd}}(z_1,z_2) - \frac{\delta_{\lambda_1\lambda_2}}{\big(\zeta_{\lambda_1}(z_1) - \zeta_{\lambda_2}(z_2)\big)^2} \nonumber \\
& = & - \sum_{\tau \in \Lambda} \underline{S}_{\lambda_2\tau}\,(\underline{S}^{-1})_{\tau\lambda_1}\,\sum_{k \geq 0}  \frac{(h_{\tau}/2)^{k + 1}}{(k + 1)!} \bigg(\sum_{\substack{d_1,d_2 \geq 0 \\ d_1 + d_2 = k}} \frac{k!}{d_1!d_2!}\,\frac{\left[\zeta_{\lambda_{1}}(z_1)\right]^{2d_1}\,\left[\zeta_{\lambda_2}(z_2)\right]^{2d_2}}{(2d_1 - 1)!!(2d_2 - 1)!!}\bigg)\,\dd\zeta_{\lambda_{1}}(z_1)\dd\zeta_{\lambda_2}(z_2) \nonumber \\
& = & - \sum_{\tau \in \Lambda} \underline{S}_{\lambda_2\tau}\,(\underline{S}^{-1})_{\tau\lambda_1}\,\sum_{k \geq 0} \frac{h_{\tau}^{k + 1}}{4(k + 1)\cdot (2k)!}\Big\{(\zeta_{\lambda_1}(z_1) + \zeta_{\lambda_2}(z_2))^{2k} + (\zeta_{\lambda_1}(z_1) - \zeta_{\lambda_2}(z_2))^{2k}\Big\}\,\dd\zeta_{\lambda_{1}}(z_1)\dd\zeta_{\lambda_2}(z_2) \nonumber \\
& = & -\frac{1}{2} \sum_{\tau \in \Lambda} \underline{S}_{\lambda_2\tau}\,(\underline{S}^{-1})_{\tau\lambda_1}\,h_{\tau}\Big\{\tilde{\mathcal{B}}[h_{\tau}^{1/2}(\zeta_{\lambda_1}(z_1) + \zeta_{\lambda_2}(z_2))]+ \tilde{\mathcal{B}}[h_{\tau}^{1/2}(\zeta_{\lambda_1}(z_1) - \zeta_{\lambda_2}(z_2))]\Big\}\,\dd\zeta_{\lambda_{1}}(z_1)\dd\zeta_{\lambda_2}(z_2) \nonumber
\eea
It is expressed in terms of the renormalized Dehn twist introduced in \eqref{renD} and the following series.
$$
\tilde{\mathcal{B}}(\zeta) = \sum_{k \geq 0} \frac{\zeta^{2k}}{2(k + 1)\cdot (2k)!} = \frac{1 - {\rm cosh}(\zeta)}{\zeta^2} + \frac{{\rm sinh}(\zeta)}{\zeta} = \frac{1}{\zeta^2} - \mathcal{B}(\zeta) .
$$
Writing $\delta_{\lambda_1\lambda_2} = \sum_{\tau \in \Lambda} \underline{S}_{\lambda_2\tau}(\underline{S}^{-1})_{\tau\lambda_1}$, the double pole can also be incorporated in the sum in the right-hand side. By adding a suitable non-odd part, we can choose a bidifferential with a rather simple expression:
\beq
\label{bas}\omega_{0,2}(z_1,z_2) := \sum_{\tau \in \Lambda} \underline{S}_{\lambda_2\tau} (\underline{S}^{-1})_{\tau\lambda_1}\,h_{\tau}\,\mathcal{B}\big[h_{\tau}^{1/2}(\zeta_{\lambda_1}(z_1) - \zeta_{\lambda_2}(z_2))\big]\,\dd\zeta_{\lambda_{1}}(z_1)\dd\zeta_{\lambda_2}(z_2) .
\eeq
This is the result announced in \eqref{finalw02}. Eventually, we compute from \eqref{Xidef1} the basis of $1$-form induced by this choice of $\omega_{0,2}$. Taking advantage of the fact that
$$
\mathcal{B}(\zeta) = \frac{{\rm cosh}(\zeta)}{\zeta^2} - \frac{{\rm sinh}(\zeta)}{\zeta} = -\frac{f'(\zeta) + f'(-\zeta)}{2},\qquad f(\zeta) = \frac{\exp(\zeta)}{\zeta}\,,
$$
we find that, for $z_0 \in U_{\mu}$,
\bea
\frac{\Xi_{d,\lambda}(z_0)}{\dd\zeta_{\mu}(z_0)} & = & \Res_{z \rightarrow o_{\lambda}} \frac{(2d + 1)!!}{\zeta_{\lambda}^{2d + 2}(z)}\,\int_{o_{\lambda}}^{z} \frac{\omega_{0,2}(\cdot,z_0)}{\dd\zeta_{\mu}(z_0)} \nonumber \\
& = & - \sum_{\tau \in \Lambda} \underline{S}_{\mu\tau}(\underline{S}^{-1})_{\tau\lambda}\,\Res_{\zeta \rightarrow 0} \frac{\exp[h_{\tau}^{1/2}(\zeta - \zeta_{\mu}(z_0))]}{\zeta - \zeta_{\mu}(z_0)}\,\frac{(2d + 1)!!\,\dd\zeta}{\zeta^{2d + 2}} \nonumber \\
& = & \sum_{\tau \in \Lambda} \underline{S}_{\mu\tau}(\underline{S}^{-1})_{\tau\lambda}\bigg(\frac{(2d + 1)!!}{[\zeta_{\mu}(z_0)]^{2d + 2}} - \sum_{m \geq 0} \frac{(2d + 1)!!}{(2d + 1)!}\,\frac{(-1)^mh_{\tau}^{d + 1 + m/2}\,[\zeta_{\mu}(z_0)]^{m}}{(2d + 2 + m)\cdot m!}\bigg)\, \nonumber \\
& = & \sum_{\tau \in \Lambda} \underline{S}_{\mu\tau}(\underline{S}^{-1})_{\tau\lambda}\bigg(\frac{(2d + 1)!!}{[\zeta_{\mu}(z_0)]^{2d + 2}} - \,\frac{h_{\tau}^{d + 1}}{2^{d}\,d!}\,\int_{0}^{1} \dd v\,v^{2d + 1}\,\exp[-h_{\tau}^{1/2}\zeta_{\mu}(z_0)v]\bigg) \nonumber \\
& = & \sum_{\tau \in \Lambda} \underline{S}_{\mu\tau}(\underline{S}^{-1})_{\tau\lambda}\,\frac{\Gamma[2d + 2\,;\, h_{\tau}^{1/2}\zeta_{\mu}(z_0)]}{2^{d}\,d!\,[\zeta_{\mu}(z_0)]^{2d + 2}}  .\nonumber
\eea
\hfill $\Box$

\begin{remark}
For $t \neq 0$, the dependence in $t$ can easily be absorbed by defining $\tilde{\zeta}_{\lambda}(z) = t^{1/2}\zeta_{\lambda}(z)$. More precisely:
$$
\omega_{g,n}|_{t}(z_1,\ldots,z_n) = t^{3g - 3 + 3n/2}\,\Big[\omega_{g,n}(z_1,\ldots,z_n)\Big|_{t = 1}\Big]_{\zeta_{\lambda}(z) \rightarrow t^{1/2}\zeta_{\lambda}(z)}
$$
and subsequently
$$
F_{g}|_{t} = t^{3g - 3}\,F_{g}|_{t = 1}
$$
The latter is also true for $t = 0$, as it is known that $F_{g}^{{\rm KdV}} = 0$ for all $g \geq 2$.
\end{remark}

In the remaining of the text, we comment on the local spectral curves obtained for two important families of modular functors, which are both related to quantum Chern-Simons theory in $3$ dimensions, with a gauge group $G$ which is either finite (Section~\ref{ex:finiteMF}) or simply-connected compact (Section~\ref{ex:WZW}). Without entering into the details of the construction, we present the main facts necessary to apply effectively the results of previous sections. We also discuss in Section~\ref{sec:glob} the problem of constructing global spectral curves in which local expansion at the ramification gives \eqref{finalw02}.

\section{Example: modular functors associated to finite groups}

\label{ex:finiteMF}

The simplest examples of modular functors are provided by quantum Chern-Simons theory with finite gauge groups $G$ in $(2 + 1)$- dimensions. These TQFT was studied by Dijkgraaf  and Witten as a particular case of quantization of Chern-Simons theory with arbitrary compact (maybe non simply connected) gauge group, and its construction depends on a cocycle $[\alpha] \in H^3(BG,{\rm U}(1))$ \cite{WV}, where $BG$ is the classifying space for $G$. When $G$ is a finite group, the path integral producing the TQFT correlation functions over a manifold $X$ is reduced to a finite sum over isomorphism classes of certain $G$-principal bundles on $X$. The theory is therefore an attractive playground to have a grasp on TQFTs. The construction of the modular functor was presented shortly after \cite{Freed}. The central extension of the mapping class group plays no role here, i.e. $\tilde{c} = 1$ and the central charge is always $0$. This model is also  known as the "holomorphic orbifold model", and fits in the framework of VOA.

We summarize below the untwisted theory $[\alpha] = 0$. The modular functor in this case can also be obtained from the modular tensor category of representations of $D(G)$, the quantum double of the finite group $G$ \cite{BB}. We have that $\mathcal{A} = {\rm Rep}(D(G))$, which is also equipped with the structure of a Frobenius manifold. Then, the Dehn twist eigenvalues are also trivial: $\tilde{d}_{\lambda} = 1$ for all $\lambda \in \Lambda$. Therefore, the CohFT we produce is rather trivial, like in Section~\ref{Verlindeder}, and does not remember more than the dimensions of the TQFT vector spaces.

The twisted theory deals with projective representations of $G$, with cocycle determined by the class $[\alpha]$. We refer to \cite{DVVV,WV,Freed} for a full presentation. But, we will point out in Section~\ref{twis} that it gives Dehn twist eigenvalues depending in a non-trivial way on the class $[\alpha]$, and therefore leads to CohFT's which does not sit only in degree $0$.

\subsection{The untwisted theory}
\label{untwi}
\subsubsection{Frobenius algebra}
\label{untwi2f}
If $g \in G$, $C_{g}$ denotes the conjugacy class of $g$, and $Z_{g}$ the centralizer of $g$ in $G$, i.e. the set of all elements commuting with $g$. We obviously have that
$$
\# C_g = \frac{\#G}{\#Z_{g}} .
$$

The label set $\Lambda$ consists of ordered pairs $(i,\rho)$ where $C_i$ is a conjugacy class of $G$, and $\rho = \rho_{i,g_i}$ an (isomorphism class of) irreducible representation of $Z_{g_i}$ for some $g_i \in C_i$. Remark that, for any two representatives $g_i,g_i' \in C_i$, $Z_{g_i}$ and $Z_{g_i'}$ (hence their representations) are canonically isomorphic, by the following formula.
\beq
\rho_{i,kgk^{-1}}(h) = \rho_{i,g}(k^{-1}hk) .
\eeq

We denote $i^{\dagger}$ the index of the conjugacy class containing $g_i^{-1}$, and $\rho^{\dagger}$ the dual representation. Then, $(i,\rho) \mapsto (i^{\dagger},\rho^{\dagger})$ endows $\Lambda$ with an involution. The vector space $\mathcal{A} = \bigoplus_{(i,\rho)} \mathbb{C}.e_{(i,\rho)}$ is equipped with a product reflecting the operation of tensor product of representations. If $\rho_i$ and $\rho_j$ are representations of $Z_{g_i}$ and $Z_{g_j}$, one has to consider $\rho_i \otimes \rho_j$ as defining a representation of $Z_{g_k}$ for each $g_k$ that can appear as the product of two elements in $C_i$ and $C_j$. 

$\mathcal{A}$ contains two remarkable subspaces $\mathcal{A}_{{\rm conj}}$ and $\mathcal{A}_{{\rm rep}}$, which are naturally algebras, and encode respectively the product induced by the group algebra of $G$ on its center, and the representation theory of $G$. The full algebra $\mathcal{A}$ combines these two structures.

$\mathcal{A}_{{\rm rep}}$ is a subalgebra of $\mathcal{A}$, spanned by the vectors $e_{(1,\rho)}$, where $i = 1$ is reserved for the conjugacy class of the identity in $G$, and $\rho$ is an irreducible representation of $G = Z_{1}$. $\mathcal{A}_{{\rm conj}}$ is the subspace spanned by the vectors $e_{(i,1)}$, which indexes conjugacy classes $C_i$ of $G$, and the centralizer is equipped with the trivial representation, so that $\rho_{i,1}(g)$ is constant equal to $1$. It is isomorphic to the center of the group algebra of $G$, and as such it has a natural structure of commutative associative algebra, which is however not a subalgebra of $\mathcal{A}$. The structure constants in the product
\beq
\label{stru} e_{(i,1)} \mathop{\times}_{{\rm conj}} e_{(j,1)} = \sum_{k} N_{ijk^{\dagger}}\,e_{(k,1)}
\eeq
compute the number of factorizations of the identity in $G$ by elements in fixed conjugacy classes, modulo the action of $G$ by simultaneous conjugation. It can be expressed in several ways.
\bea
N_{i_1i_2i_3} & = & \#\{(g_1,g_2,g_3) \in C_{i_1} \times C_{i_2} \times C_{i_3},\quad g_1g_2g_3 = 1\big\}/G \nonumber \\
& = & \frac{\#\big\{p \in {\rm Hom}(\pi_1(\mathbf{\Sigma}_{0,3}),G)\,\,:\,\,\quad p(\mathfrak{l}_j) \in C_{i_j}\,\,\,j = 1,2,3\big\}}{\# G}\nonumber  .
\eea
Here, $\mathfrak{l}_j$ is a loop around the $j$-th puncture of $\mathbf{\Sigma}_{0,3}$. It also counts the number of (possibly disconnected) branched coverings with structure group $G$, ramified over $3$ (ordered) points on $\mathbb{S}^2$, in which the local monodromy around the $j$-th point is required to sit in the conjugacy class $C_{i_j}$.

\subsubsection{Scalars and S-matrix}

This modular functor has $\tilde{c} = 1$ and $\tilde{r}_{(i,\rho)} = 1$, but its VOA origin provides canonical log-determinations (see Section~\ref{VOADK}):
$$
c = 0,\qquad r_{\lambda} = 0 .
$$

For each $i$ labeling a conjugacy class, $g_i \in C_i$, and a representation $\rho:= \rho_{i,g_i}$ of $Z_{g_i}$, we denote by $\chi_{i,g_i}(h) = \mathrm{Tr}\,\rho_{i,g_i}(h)$ its character. We take as convention $\chi_{i,g_i}(h) = 0$ whenever $h \notin Z_{g_i}$. Let $Z_i$ be the centralizer of some (arbitrary) representative of $C_i$, and $V_{i,\rho}$ be the vector space on which $Z_i$ acts via the representation $\rho$. The dimension of this module is $\dim V_{i,\rho} = \rho_{i,g_i}(1)$. We reserve the label $i = 1$ for the conjugacy class of the identity in $G$, and then in $(1,\rho)$, $\rho$ must be a representation of $G$. Its character will be denoted $\chi_{\rho}$, and $V_{1,\rho} := V_{\rho}$ the vector space of the representation.

The S-matrix is computed in \cite{DVVV,Freed}. Keeping the notations of Section~\ref{FMF}, it reads
\beq
\label{SMFG} \underline{S}_{(i_1,\rho_1)(i_2,\rho_2)} = \frac{1}{\# G} \sum_{\substack{g_j \in C_{i_j} \\ [g_1,g_2] = 1}} \chi_{i_1,g_1}(g_2^{-1})\,\chi_{i_2,g_2}(g_1^{-1}) .
\eeq
We list for bookkeeping some special entries of the $S$-matrix.
\bea
\underline{S}_{(1,1)(1,1)} & = & \frac{1}{\# G} \nonumber \\
\underline{S}_{(i,1)(1,1)} & = & \frac{\# C_i}{\# G} \nonumber \\
\underline{S}_{(1,\rho)(1,1)} & = & \frac{\dim V_{\rho}}{\# G} \nonumber \\
\label{charc}\underline{S}_{(i,1)(1,\rho)} & = & \frac{\# C_i\,\chi_{\rho}(C_i)}{\# G}  \\
\label{charc2} \underline{S}_{(i,\rho)(1,1)} & = & \frac{\# C_{i}}{\# G} \dim V_{i,\rho} = \frac{\dim V_{i,\rho}}{\# Z_i} \\
\underline{S}_{(i,1)(j,1)} & = & \frac{\# {\rm Commutant}(C_i,C_j)}{\# G} \nonumber \\
\underline{S}_{(1,\rho)(1,\tau)} & = & \frac{\dim V_{\tau}\,\dim V_{\rho}}{\# G} \nonumber \\
\underline{S}_{(i,\rho)(1,\tau)} & = & \frac{\dim V_{i,\rho}}{\# G}\,\# C_i \cdot \chi_{\tau}(C_{i^{\dagger}}) \nonumber
\eea
$\underline{S}$ can be thought as an extension of the character table of $G$ which appears in \eqref{charc}, so that both indices $i$ and $\rho$ are treated on the same footing. This matrix is clearly symmetric, and it can be checked by direct computation that $\underline{S}^2 = C$ is the operator sending $e_{\lambda}$ to $e_{\lambda^{\dagger}}$.

\subsubsection{Local spectral curve}

Since $c = 0$ and $r_{(i,\rho)} = 0$, the CohFT just consists of several (suitably rescaled) copies of the trivial CohFT, and the local spectral curve consists of several copies of the Airy curve \eqref{AriY}. The only information left from the modular functor is the dimension of the representations, appearing as the rescaling
$$
x(z) = [\zeta_{i,\rho}(z)]^2/2,\qquad y(z) = - \frac{\dim V_{i,\rho}}{\# G}\,\zeta_{i,\rho}(z),\qquad z \in U_{i,\rho}
$$
and as we have seen in Section~\ref{Verlindeder}, this is just enough to compute the dimensions of the TQFT vector spaces.

\subsection{The twisted theory}

\label{twis}

Let us choose a $3$-cocycle $\alpha$ representing the class $[\alpha]$, and introduce for any $h \in G$, the following $2$-cocycle for $Z_{h}$
\beq
\label{chga}c_{h}(g_1,g_2) = \frac{\alpha(h,g_1,g_2)\alpha(g_1,g_2,h)}{\alpha(g_1,h,g_2)} .
\eeq
Choosing another representative $\alpha'$, we would obtain a $2$-cocycle $c_{h}'$ differing from $c_{h}$ only by a coboundary. A projective representation of $Z_{h}$ with cocycle $c_{h}$ is a map $\rho_{h}\,:\,Z_{h} \rightarrow {\rm GL}(V)$ for a finite dimensional vector space, such that $\rho_{h}(1) = 1$ is a scalar and one has that
$$
\forall (g_1,g_2) \in G^2,\qquad \rho_{h}(g_1)\rho_{h}(g_2) = c_{h}(g_1,g_2)\,\rho_{h}(g_1g_2) .
$$
In the construction of the modular functor for the twisted theory, $\rho_{i,g}$ are now the projective representations of $Z_{g_i}$ with cocycle $c_{g_i}$ \cite{Freed}. The $S$-matrix takes the form
$$
\underline{S}_{(i_1,\rho_1)(i_2,\rho_2)} = \frac{1}{\# G} \sum_{\substack{g_j \in C_{i_j} \\ [g_1,g_2] = 1}} \chi_{i_1,g_1}(g_2^{-1})\,\chi_{i_2,g_2}(g_1^{-1})\,\sigma(g_1|g_2)
$$
and the Dehn twist eigenvalues are
$$
\tilde{r}_{(i,\rho)} = \sigma(g_i|g_i)^{-1/2}
$$
for some arbitrary $g_i \in C_i$. The function $\sigma(g_1|g_2)$ depends in a non-trivial way on $[\alpha]$, but must satisfy $\sigma(g_1|g_2) = \sigma(g_2|g_1)$ and another condition of cohomological nature \cite{DVVV}. The relation between $\sigma$ and $\alpha$ is explained in full generality in \cite{Freed5}. In the simpler case where all $2$-cocycles $c_{h}$ defined in \eqref{chga} are coboundaries, let us take a $1$-cocycle $\beta_{h}$ such that
$$
c_{h}(g_1,g_2) = \frac{\beta_{h}(g_1)\beta_{h}(g_2)}{\beta_{h}(g_1g_2)} .
$$
The consistency conditions impose $\beta_{h^{-1}}(g) = \beta_{g^{-1}}(h) = [\beta_{g}(h)]^{-1}$, and the expression for $\sigma$ is then \cite{DVVV}
$$
\sigma(g_1|g_2) = \beta_{g_1}(g_2)\beta_{g_2}(g_1) .
$$

As we see, the Dehn twist eigenvalue $\tilde{r}_{(i,\rho)}$ now depends on $i$, but still not on the representation $\rho$ of $Z_{i}$. In the computation of $\omega_{0,2}$ from \eqref{finalw02}, one encounters the following terms, for a fixed $i$,
\bea
& & \sum_{\rho} \underline{S}_{(i_2,\rho_2)(i,\rho)} (\underline{S}^{-1})_{(i,\rho)(i_1,\rho_1)} \nonumber \\
& = &  \sum_{\rho} \underline{S}_{(i_1^{\dagger},\rho_1^{\dagger})(i,\rho)} \underline{S}_{(i_2,\rho_2)(i,\rho)} \nonumber \\
& = & \frac{1}{(\# G)^2} \sum_{\substack{g_1 \in C_{i_1},\,\,g_2 \in C_{i_2} \\ h_1,h_2 \in C_{i} \\ [g_j,h_j] = 0}} \sigma(g_1^{-1}|h_1)\sigma(g_2|h_2)\,\chi_{i_1,g_1}(g)\,\chi_{i_2,g_2}(h_2^{-1})\bigg(\sum_{\chi} \chi_{i,h_2}(g_2^{-1})\,\chi_{i,h_1}(g_1)\bigg) . \label{thu}
\eea
We focus on the quantity in the brackets, which is a sum over all projective representations of $Z_i$. Since there exists $k \in G$ such that $h_1 = kh_2k^{-1}$, we have $\chi_{i,h_2}(g_2^{-1}) = \chi_{i,h_1}(kg_2^{-1}k^{-1})$. The characters of projective representations of $Z_i$ form an orthonormal basis of $c_{h_1}$-class functions -- the proof, as for representations, only relies on Schur's lemma. Therefore, the sum over $\chi$ in \eqref{thu} vanishes when $g_1$ and $kg_2^{-1}k^{-1}$ do not belong to the same conjugacy class in $Z_i$. A fortiori, to have a non-zero result in \eqref{thu}, we need $g_1$ and $g_2$ to be in the same conjugacy class in $G$, i.e. $i_1 = i_2$. Therefore, $\omega_{0,2}(z_1,z_2)$ for $z_j \in U_{(i_j,\rho_j)}$ is proportional to $\delta_{i_1i_2}$, and it follows from the residue formula \eqref{toprec} that
\begin{lemma}
\label{fdkm}For $z_j \in U_{(i_j,\rho_j)}$, $\omega_{g,n}(z_1,\ldots,z_n)$ vanishes unless all $i_j$ are equal for $j \in \llbracket 1,n \rrbracket$.
\end{lemma}
We find that the CohFT does not couple the representation theory of different centralizers, or in the words of \cite{WV}, it does not couple different interaction channels. However, $\omega_{0,2}$ in a given channel will mix various $\rho$'s, and thus the CohFT is non-trivial, unlike the untwisted case.

\section{Example: WZW model for compact Lie groups}
\label{ex:WZW}

\subsection{Short presentation}
\label{Sec61}
We briefly review the definition of the Verlinde bundle that arises from the Wess-Zumino-Witten model, and the corresponding modular functor. It is based on the representation theory of a VOA, namely an algebra that incorporates the Virasoro algebra -- describing infinitesimal coordinate reparametrizations in a disk neighborhood of a puncture on a Riemann surface -- together with a central extension and extra symmetries coming from a simple complex Lie algebra $\mathfrak{g}$. Since the seminal work of Tsuchiya, Ueno and Yamada \cite{TUY}, there is a vast literature on the Verlinde bundles, that readers with different backgrounds will appreciate differently; we can suggest the nice article of Beauville \cite{Beau} giving a proof of the Verlinde formula for the non-exceptional $\mathfrak{g}$, the book of Ueno \cite{Uenobook}, and the review of Looijenga \cite{LooiWZW}. A complet proof that these theories gives modular functors was given by Andersen and Ueno in \cite{AU1,AU2}.

\subsubsection{Affine Ka\v{c}-Moody algebras}

We refer to \cite{Kacbook} for the detailed theory. Let $\mathfrak{h}$ be the Cartan subalgebra, $\Delta_+$ the set of positive roots, $\theta$ the highest root -- which is the highest weight for the adjoint representation -- and $\varrho = \frac{1}{2}\sum_{\alpha > 0} \alpha$ the Weyl vector. We normalize the Killing form $\langle\cdot,\cdot\rangle$ such that $\langle\theta,\theta\rangle = 2$. It induces an isomorphism between $\mathfrak{h}$ and $\mathfrak{h}^*$, which we use systematically to identify $\mathfrak{h}$ and $\mathfrak{h}^*$. In particular, we use the notation $\varrho$ for the Weyl vector in $\mathfrak{h}$ or the corresponding element in $\mathfrak{h}^*$. Let $(e_{i})_{i = 1}^{\dim \mathfrak{h}}$ be an orthonormal basis of $\mathfrak{h}$ for the Killing form. The quadratic Casimir is the element of the universal enveloping algebra of $\mathfrak{g}$ defined by $\mathcal{Q} = \frac{1}{2}\sum_{i = 1}^{\dim \mathfrak{h}} e_{i} \otimes e_i$. It acts as a scalar on the highest weight $\mathfrak{g}$-module $V_{\lambda}$, more precisely, one has that
$$
\mathcal{Q}|_{V_{\lambda}} = \frac{1}{2}\,\langle\lambda,\lambda + 2\varrho\rangle\,{\rm id}_{V_{\lambda}}.
$$
This scalar on the adjoint is denoted $h^{\vee}(\mathfrak{g})$, and called the dual Coxeter number of $\mathfrak{g}$, namely $h^{\vee}(\mathfrak{g}) = 1 + \langle\theta,\varrho\rangle$. Let $\mathfrak{g}_{\mathbb{R}}$ be the compact real form of $\mathfrak{g}$, and $\mathfrak{h}_{\mathbb{R}}$ its Cartan subalgebra. The weight lattice in $\mathfrak{h}^*$ is denoted $L_{W}$, it consists of all $\lambda \in \mathfrak{h}^*_{\mathbb{R}}$ such that
$$
\forall \alpha \in \Delta_+,\qquad 2\,\frac{\lambda(\alpha)}{\langle \alpha,\alpha\rangle} \in \mathbb{Z} .
$$
For a fixed $\ell \geq 1$, we introduce the set of highest weights at level $\ell$ defined by
\beq
\label{Pell}\mathcal{P}_{\ell} = \big\{\lambda \in L_{W}\mid 0 \leq \lambda(\theta) \leq \ell \big\} .
\eeq
Consider the untwisted affine Lie algebra
$$
\widehat{\mathfrak{g}} = \mathbb{C}.{\rm c} \oplus \mathfrak{g} \otimes \mathbb{C}((\xi)) .
$$
It is a central extension of $\mathfrak{g}\otimes\mathbb{C}((\xi))$, with the Lie bracket defined by
$$
\forall X_1,X_2 \in \mathfrak{g},\quad \forall f_1,f_2 \in \mathbb{C}((\xi)),\qquad [X_1 \otimes f_1,X_2 \otimes f_2] = \langle X,Y \rangle\,\big(\Res_{\xi \rightarrow 0} f_1\dd f_2\big)\cdot {\rm c} + [X_1,X_2]\otimes f_1f_2 .
$$
We can use the following decomposition
$$
\widehat{\mathfrak{g}} = \mathfrak{p}_+ \oplus \mathfrak{p}_-,\qquad \mathfrak{p}_+ = \mathfrak{g}_+ \oplus \mathfrak{g} \oplus \mathbb{C}.{\rm c},\qquad \mathfrak{g}_+ = \xi\cdot\mathfrak{g}[[\xi]].
$$
If $\lambda \in \mathcal{P}_{\ell}$, we promote the $\mathfrak{g}$-module $V_{\lambda}$ to a $\mathfrak{p}_+$ module, by declaring that $\mathfrak{g}_+$ annihilates $V_{\lambda}$, and the central element ${\rm c}$ acts as $\ell\cdot{\rm id}_{V_{\lambda}}$. Then, one introduces the Verma module as
$$
M_{\lambda} = U(\mathfrak{g}) \otimes_{\mathfrak{p}_+} V_{\lambda} .
$$
It is a left $\widehat{\mathfrak{g}}$-module, which is not irreducible. However, it contains a maximal proper submodule defined by
$$
J_{\lambda} = U(\widehat{\mathfrak{p}}_-) \cdot (\theta \otimes \xi^{-1})^{\ell - \langle \theta,\lambda \rangle + 1} .
$$
Then, $H_{\lambda} = M_{\lambda}/J_{\lambda}$ becomes irreducible, and it is actually an integrable highest weight module for $\widehat{\mathfrak{g}}$.
If $P$ is a finite set, define the multi-variable analogue of $\widehat{\mathfrak{g}}$ by
$$
\widehat{\mathfrak{g}}_{P} = \mathbb{C}.{\rm c} \oplus \bigoplus_{p \in P}  \mathfrak{g} \otimes \mathbb{C}((\xi_{p})) .
$$
Now, if $\mathfrak{X} = (\Sigma,p_1,\ldots,p_n;\xi)$ is a Riemann surface $\Sigma$ equipped with a finite marked set of points $P$ and $\xi = (\xi_p)_{p \in P}$ a set of local coordinates at the points of $P$, the Lie algebra
$$
\widehat{\mathfrak{g}}(\mathfrak{X}) = \mathfrak{g} \otimes_{\mathbb{C}} H^0(\Sigma,\mathcal{O}_{\Sigma}(*P))
$$
is naturally embedded as a subalgebra of $\widehat{\mathfrak{g}}_{P}$.

\subsubsection{From the space of vacua to the modular functor}

For any $\mathfrak{X}$ as above, and $\vec{\lambda}\,:\,P \rightarrow \mathcal{P}_{\ell}$ a set of labels, we define the space of vacua by
$$
\mathcal{V}_{\vec{\lambda}}(\mathfrak{X}) = H_{\vec{\lambda}}/\widehat{\mathfrak{g}}(\mathfrak{X})H_{\vec{\lambda}} .
$$
Then, one can show \cite{TUY,Tsuchimoto,LooiWZW} that, if $\mathfrak{X}$ and $\mathfrak{X}'$ differ by a change of coordinates $\xi \rightarrow \xi'$, there is a canonical isomorphism between $\mathcal{V}_{\vec{\lambda}}(\mathfrak{X})$ and $\mathcal{V}_{\vec{\lambda}}(\mathfrak{X}')$. This allows us to define $\mathcal{V}_{\vec{\lambda}}(\mathfrak{X})$ as a bundle over Teichm\"uller space $\tilde{\mathcal{T}}_{\Sigma,P}$. 

By the work of Tsuchiya, Ueno and Yamada \cite{TUY}, this bundle carries a projectively flat, unitary connection, and enjoys nice factorization properties over families where the surface is pinched. Exploiting this connection, Andersen and Ueno proved  \cite{AU1,AU2}, that one can make a definition independent of the complex structure: they assign unambiguously a vector space $\mathcal{V}_{\vec{\lambda}}(\mathbf{\Sigma})$ to any marked surface $\mathbf{\Sigma}$, and prove that this assignment defines a modular functor in the sense of Section~\ref{S2}.
Moreover, in \cite{AU3,AU4} Andersen and Ueno established that, for $\mathfrak{g} = \mathfrak{sl}_{N}$, this modular functor is isomorphic to the modular functor obtained from the modular tensor category of representations of the quantum group $U_{-q^{1/2}}(\mathfrak{g})$ with
\beq
\label{qvlu}q = \exp[2{\rm i}\pi/(\ell + h^{\vee}(\mathfrak{g}))] .
\eeq

\subsubsection{More notations}

Let $\mathfrak{W}$ be the Weyl group of $\mathfrak{g}$. We denote $w_0 \in \mathfrak{W}$ its longest element. Its length is $|\Delta_+|$, and it is the unique element that sends positive roots to negative roots. By definition of the Weyl vector, $w_0(\rho) = -\rho$.

We consider as label set $\Lambda := \mathcal{P}_{\ell}$, the set of representations of $\mathfrak{g}_{\mathbb{R}}$ at level $\ell$. It is equipped with the involution
$
\lambda^{\dagger} = -w_0(\lambda) .
$
Note that $\lambda^{\dagger}$ is actually the highest weight of $V_{\lambda}^*$. 

If $\beta \in \mathfrak{h}$, $q^{\beta}$ stands for the function $\mathfrak{h} \rightarrow \mathbb{C}$ defined by $x \mapsto q^{\langle\beta,x\rangle}$ on $\mathfrak{h}$. The character of $V_{\lambda}$ is denoted ${\rm ch}_{\lambda}\,:\,\mathfrak{h} \rightarrow \mathbb{C}$, and is given by the Weyl character formula
$$
{\rm ch}_{\lambda}(q^{\beta}) = \frac{\sum_{w \in \mathfrak{W}} {\rm sgn}(w)\,q^{\langle \lambda + \rho,w(\beta)\rangle}}{\sum_{w \in \mathfrak{W}} {\rm sgn}(w)\,q^{\langle \rho,w(\beta)\rangle}} .
$$
The $q$-dimension is defined as $\dim_{q} V_{\lambda} = {\rm ch}_{\lambda}(1)$, and using the Weyl denominator formula, it reads
$$
{\rm dim}_{q}\,\lambda := \prod_{\alpha > 0} \frac{[\langle\alpha,\lambda + \varrho\rangle]_{q}}{[\langle \alpha,\varrho \rangle]_{q}},\qquad \qquad [x]_{q} = q^{x/2} - q^{-x/2} .
$$
The weight lattice $L_{W}$ has a sublattice $L_{\theta}$ spanned by the elements $\langle w(\theta),\cdot \rangle$ with $w \in \mathfrak{W}$. The integer
$$
D_{\ell} := \#\,\big(L_{W}/(\ell + h^{\vee})L_{\theta}\big)
$$
appears as a normalization constant in
$$
\sum_{\lambda \in \Lambda} \big|{\rm ch}_{\lambda}(q^{\beta})\big|^2 = D_{\ell}^{-1}\cdot\bigg|\prod_{\alpha > 0} [\langle \alpha, \beta \rangle]_{q}\bigg|^2 .
$$

Consider the  matrix
\beq
\label{SLM} \mathcal{S}_{\lambda\mu} = D_{\ell}^{-1/2} \sum_{w \in \mathfrak{W}} {\rm sgn}(w)\,q^{\langle \lambda + \rho,w(\mu + \rho)\rangle} = D_{\ell}^{-1/2}\cdot \prod_{\alpha > 0} [\langle \alpha,\rho \rangle]_{q}\cdot \dim_{q} V_{\lambda}\cdot {\rm ch}_{\mu}(q^{\lambda + \rho})
\eeq
with the value $q = \exp[2{\rm i}\pi/(\ell + h^{\vee})]$. From the first equality we see that $\mathcal{S}$ is symmetric, and since the scalar product is $\mathfrak{W}$-invariant and ${\rm sgn}(w_0) = (-1)^{|\Delta_+|}$, we have that
$$
\label{S1}C\,\mathcal{S} = (-1)^{|\Delta_+|}\,\mathcal{S}^*
$$
where we recall that $C$ is the matrix of the involution $\dagger$. It is less obvious but also true using orthogonality of characters that $\mathcal{S}$ is unitary which means that
$$
\mathcal{S}\mathcal{S}^* = 1 .
$$

\subsubsection{Scalars and S-matrix}

The modularity of characters of the underlying VOA gives canonical log-determinations for the eigenvalues of the central element and the Dehn twist. These are the conformal weights and the central charge \cite{KacPeterson,TUY}
\beq
\label{logdeteqn} r_{\lambda} = \frac{\langle \lambda,\lambda + 2\rho\rangle}{2(\ell + h^{\vee})},\qquad \qquad c = \frac{\ell\,\dim \mathfrak{g}}{\ell + h^{\vee}(\mathfrak{g})}.
\eeq
The formulas manifestly satisfy $r_{1} = 0$ and $r_{\lambda} = r_{\lambda^{\dagger}}$. With Freudenthal strange formula, we can also write that
$$
\frac{c}{24} = \frac{\ell}{h^{\vee}}\,\frac{\langle \rho,\rho \rangle}{2(\ell + h^{\vee}(\mathfrak{g}))} .
$$

The S-matrix appearing in Section~\ref{FMF} is given by Ka\v{c}-Peterson formula \cite[Proposition 4.6(d)]{KacPeterson}:
\beq
\label{Fomrual}\underline{S}_{\lambda\mu} = {\rm i}^{|\Delta_+|}\,\mathcal{S}_{\lambda\mu}^*\qquad {\rm or}\,\,{\rm equivalently}\qquad \underline{S}^{-1}_{\lambda\mu} = (-{\rm i})^{|\Delta_+|}\,\mathcal{S}_{\lambda\mu} .
\eeq
Since the diversity of notations in the literature can be confusing, let us make two checks ensuring that \eqref{Fomrual} in our notations is correct. Firstly, with the properties of $\mathcal{S}$ just pointed out, we can write
$$
\underline{S}^{\top}\underline{S} = (-1)^{|\Delta_+|}\,\mathcal{S}^*\mathcal{S}^* = C\,\mathcal{S}\mathcal{S}^* = C
$$
as it should be according to \eqref{SSJ}. Secondly, the Verlinde formula \eqref{symnlm}-\eqref{Verlind2} in our notations does agree, once we insert \eqref{Fomrual}, with the Verlinde formula \cite[Corollary 9.8]{Beau}.

\subsubsection{The case $\mathfrak{g} = \mathfrak{sl}_{N}$}

\begin{figure}
\begin{center}
\includegraphics[width=0.8\textwidth]{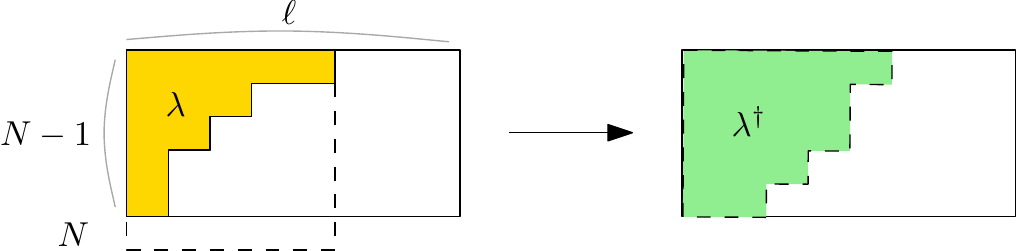}
\caption{\label{comple} The involution $\lambda^{\dagger} = -w_0(\lambda)$ in terms of Young tableaux.}
\end{center}
\end{figure}

Let $\mathbb{C}^N$ be equipped with its canonical orthonormal basis $(e_i)_{i = 1}^N$, and $e_i^*$ be the dual basis. The Cartan algebra of $\mathfrak{sl}_{N}$ can be identified with  the hyperplane in $\mathbb{C}^N$ orthogonal to $\sum_{i = 1}^N e_i$, and the Killing form is induced from the scalar product on $\mathbb{C}^N$. The positive roots are $e_i - e_j$ for $1 \leq i < j \leq N$, the highest root is $\theta = e_{1} - e_{N}$, and the Weyl vector is
$$
\rho = \frac{1}{2} \sum_{i = 1}^N (N + 1 - 2i)\,e_i .
$$
$\mathfrak{h}^*$ (resp. the weight lattice $L_{W})$ is the $\mathbb{C}$-span (resp. $\mathbb{Z}$-span) of $(e_i^*)_{i = 1}^N$ modulo the relation $\sum_{i = 1}^N e_i^* = 0$. If $\lambda \in \mathfrak{h}^*$, let us denote $|\lambda| = \sum_{i = 1}^N \lambda_i$. The Killing form induced on $\mathfrak{h}^*$ is \cite{Fultonrep}
$$
\Big\langle \sum_{i = 1}^N \lambda_i\,e_i^*\,,\, \sum_{j = 1}^N \mu_j\,e_j^*\Big\rangle = \sum_{i = 1}^N \lambda_i\mu_i - \frac{|\lambda||\mu|}{N} .
$$
The element representing the Weyl vector in $\mathfrak{h}^*$ is
$$
\rho = \frac{1}{2} \sum_{i = 1}^N (N + 1 - 2i)\,e_i^* = \sum_{i = 1}^{N - 1} (N - i)e_i^*\,\,{\rm mod}\,\,\Big(\sum_{i = 1}^N e_i^*\Big) . 
$$
The fundamental weights are
$$
1 \leq i \leq N - 1,\qquad w_i = \sum_{j = 1}^i e_j^*
$$
and the corresponding highest weight $\mathfrak{sl}_{N}$-module is $\bigwedge^i\mathbb{C}^N$. Irreducible highest weight representations of $\mathfrak{sl}_{N}$ are encoded in an $(N - 1)$-tuple $\lambda = (\lambda_1,\ldots,\lambda_{N - 1})$ such that $\lambda_1 \geq \cdots \geq \lambda_{N - 1} \geq 0$, which corresponds to the highest weight vector
$$
\lambda = \sum_{j = 1}^N \lambda_j\,e_j^* = \sum_{j = 1}^{N - 1} (\lambda_{j} - \lambda_{j + 1})\,w_j
$$
with the convention $\lambda_{N} = 0$. The representations at level $\ell$ are those with $\lambda_1 \leq \ell$. The Weyl group permutes the $(e_i)_{i = 1}^N$, and its longest element is the permutation $e_i \mapsto e_{N + 1 - i}$ for all $i \in \llbracket 1,N \rrbracket$. Since $e_{N}^* = -\sum_{i = 1}^{N - 1} e_i^*$ on the hyperplane $\mathfrak{h}$, we see that the involution takes $\lambda$ to its "complement" as in Figure~\ref{comple}, i.e. $\lambda^{\dagger}_i = \lambda_1 - \lambda_{N - i + 1}$. The characters ${\rm ch}_{\lambda}$ are the Schur polynomials, and the central charge and Dehn twist eigenvalues are
$$
h^{\vee}(\mathfrak{sl}_{N}) = N,\qquad c = \frac{\ell(N^2 - 1)}{\ell + N},\qquad r_{\lambda} = \frac{1}{2(\ell + N)}\bigg( \sum_{i = 1}^{N - 1} \lambda_i(\lambda_i - 2i + 1)  + N|\lambda|- \frac{|\lambda|^2}{N}\bigg) .
$$
Since the Weyl group acts transitively on the set of roots, $L_{\theta}$ is just the root lattice
$$
L_{\theta} = \big\{\sum_{i = 1}^N a_i e_i^*,\quad \sum_{j = 1}^N a_j = 0\big\}\big/\big\{\sum_{i = 1}^N e_i^* = 0\big\}
$$
and it has index $N$ in the weight lattice. Therefore, the normalization integer reads
$$
D_{\ell} = N\,(\ell + N)^{N - 1} .
$$

In the case $N = 2$ for level $\ell \geq 0$, $\Lambda$ is the set of integers from $0$ to $\ell$, the involution is identity, and $D_{\ell} = 2(\ell + 2)$. The scalars and $S$-matrix read
$$
\forall \lambda,\mu \in \llbracket 0,\ell \rrbracket \qquad c = \frac{3\ell}{\ell + 2},\qquad r_{\lambda} = \frac{\lambda(\lambda + 2)}{4(\ell + 2)},\qquad \underline{S}^{-1}_{\lambda\mu} = \sqrt{\frac{2}{\ell + 2}}\,\sin\Big[\frac{\pi(\lambda + 1)(\mu + 1)}{\ell + 2}\Big] .
$$

\subsection{CohFT and double Hurwitz numbers}

The Chern character of the Verlinde bundle has already been studied in \cite{Tsuchimoto,MarianOP}, and \cite{Zvonkin} showed that it defines a CohFT -- this was in fact the example motivating our work. In this particular case, our work only completes  \cite{Zvonkin} by remarking that the CohFT correlation functions are computed by the topological recursion with the local spectral curve of Theorem~\ref{TH42}. It would be interesting to know the meaning of this CohFT from an enumerative geometry perspective. For $\mathfrak{g} = \mathfrak{sl}_{N}$, we show that $\omega_{0,2}$ given in \eqref{finalw02} is related in some way to double Hurwitz numbers. 

\subsubsection{Hurwitz numbers}

Let us first review the definition of Hurwitz numbers. For any finite group $G$, we consider the number of $G$-principal bundles over a surface of genus $g$ and $n$ punctures, whose monodromy around the $j$-th punctures belongs to the conjugacy class $C_{i_j}$ of $G$. It is computed by the Frobenius formula \cite{Zagierapp}
\beq
\label{fchar} \frac{\#\big\{R \in {\rm Hom}(\pi_1(\mathbf{\Sigma}_{g,n}),G)\mid R(\mathfrak{l}_j) \in C_{j}\big\}}{\# G} = \sum_{\nu} \Big[\frac{\chi_{\nu}(1)}{\# G}\Big]^{2 - 2g}\,\prod_{j = 1}^n f_{\nu}(C_{j})
\eeq
where the sum ranges over irreducible representations of $G$, and
$$
f_{\nu}(C) := \frac{\# C\,\chi_{\nu}(C)}{\chi_{\nu}(1)}.
$$

When $G = \mathfrak{S}_{d}$, this is the number of (possibly disconnected) branched coverings of degree $d$ over a surface of genus $g$. Conjugacy classes of $\mathfrak{S}_{d}$ are labeled by partitions $\mu$ of $d$: the parts of $\mu$ are the lengths of the cycles of a representative of $C_{\mu}$. For a collection of partitions $\mu_1,\ldots,\mu_n \vdash d$, the numbers
$$
H_{g}^{d}(\vec{\mu}) = \frac{\#\big\{R \in {\rm Hom}(\pi_1(\mathbf{\Sigma}_{g,n}),\mathfrak{S}_{d})\mid R(\mathfrak{l}_j) \in C_{\mu_j}\big\}}{d!}
$$
are called "Hurwitz numbers" of genus $g$. Though it is a good starting point, the formula \eqref{fchar} involving the character table of $\mathfrak{S}_{d}$ is not the end of the story.

Let $C_{(2)}$  be the conjugacy class of a transposition. One often would like to count branched coverings with arbitrary number $b$ of simple ramification points.The corresponding Hurwitz numbers are
$$
H_{g}(\vec{\mu}\,|\,T) := \sum_{b \geq 0} \frac{T^b}{b!}\,H_{g}^{d}(\vec{\mu},\underbrace{C_{(2)},\ldots,C_{(2)}}_{b\,\,{\rm times}}) .
$$
If we keep $n$ points with arbitrary ramifications $\vec{\mu}$, this is the generating series of $n$-tuples Hurwitz numbers in genus\footnote{For connected coverings, $g$ should not be confused with the genus $\tilde{g}$ of the total space, which is given by the Riemann-Hurwitz formula. 
$$
d(2 - 2g) = 2 - 2\tilde{g} - b - \sum_{i = 1}^n \big\{|\mu_i| - \ell(\mu_i)\big\}.
$$} $g$. For instance, the generating series of double Hurwitz numbers in genus $0$ and degree $d$ is, according to \eqref{fchar},
\beq
\label{H2d2} H_{0}^{d}(\mu_1,\mu_2\,|\,T) = \frac{1}{d!^2} \sum_{\nu \vdash d} \# C_{\mu_1}\# C_{\mu_2}\,\chi_{\nu}(C_{\mu_1})\chi_{\nu}(C_{\mu_2})\,e^{T\,f_{\nu}(C_{(2)})} .
\eeq
The generating series of genus $0$ simple \cite{GJ,GJV,BMconj,BEMS,EMS} and double \cite{Okoun,MarkHur,Johnsondouble,HarnadOrlov,HarnadGP} Hurwitz numbers in genus $0$ have been intensively studied from the point of view of combinatorics, mirror symmetry, and integrable systems. The generating series of Hurwitz numbers in genus $1$ is somewhat simpler because the coupling in \eqref{fchar} disappears, and there is a nice theory relating them to quasimodular forms \cite{DijkE,KanZ}. The realm of genus $g \geq 2$ seems uncharted.

\subsubsection{Rewriting of the $2$-point function}

We recall two elementary facts on symmetric functions \cite{MacDo}. By Schur-Weyl duality, for a partition with $|\lambda| = d$ boxes, the Schur polynomials decompose on the power sums as
$$
{\rm ch}_{\lambda}(x) = \frac{1}{d!} \sum_{\mu \vdash d} \# C_{\mu}\,\chi_{\lambda}(C_{\mu})\,p_{\mu}(x) .
$$
Further, $f_{\lambda}(C_{(2)}) = \frac{1}{2}\sum_{i = 1}^N \lambda_i(\lambda_i - 2i + 1)$ is related to the quadratic Casimir, hence to the Dehn twist eigenvalues by the formula
$$
r_{\lambda} = \frac{1}{\ell + N}\Big[f_{\lambda}(C_{(2)}) + \frac{N|\lambda|}{2} - \frac{|\lambda|^2}{2N}\Big] .
$$
Putting together the expression \eqref{SLM}-\eqref{Fomrual} of the S-matrix and these two observations, we can compute with \eqref{bu2bu} the $2$-point function 
$$
B(u_1,u_2) = \sum_{\lambda_1,\lambda_2 \in \Lambda} \frac{\delta_{\lambda_1\lambda_2} - B_{\lambda_1 \lambda_2}(u_1 + u_2)}{u_1 + u_2}\,\underline{\epsilon}_{\lambda_1}\otimes\underline{\epsilon}_{\lambda_2} .
$$
We recall that $B_{\lambda_1\lambda_2}(u_1 + u_2)$ is essentially the Laplace transform of $\omega_{0,2}$ in $U_{\lambda_1}\times U_{\lambda_2}$.

\begin{lemma}
With $T = ut/(\ell + N)$, we have that
\bea
B_{\lambda_1\lambda_2}(u) & = & D_{\ell}^{-1}\Big|\prod_{\alpha > 0} [\langle \alpha,\rho \rangle]_{q}\Big|^2 \dim_{q} V_{\lambda_1}\,(\dim_{q} V_{\lambda_2})^* \nonumber \\
& & \cdot\bigg\{\sum_{d \geq 0} e^{(Nd/2 - d^2/2N + c/24)T} \sum_{\mu_1,\mu_2 \vdash d} p_{\mu_1}(q^{\lambda_1 + \rho})p_{\mu_2}(q^{-(\lambda_2 + \rho)})\,H^{d,[\ell,N]}_{0}(\mu_1,\mu_2\,|\,T)\bigg\} \nonumber
\eea
with
$$
H_{0}^{d,[\ell,N]}(\mu_1,\mu_2\,|\,T) = \frac{1}{d!^2} \sum_{\substack{\nu \, \vdash d \\ \ell(\nu) \leq N - 1 \\ \nu_{1} \leq \ell}} \# C_{\mu_1}\# C_{\mu_2}\,\chi_{\nu}(C_{\mu_1})\chi_{\nu}(C_{\mu_2})\,e^{T\,f_{\nu}(C_{(2)})}.
$$
\end{lemma}
The generating series $H_{0}^{d,[\ell,N]}(\mu_1,\mu_2\,|\,T)$ only differs from the generating series of double Hurwitz numbers in genus $0$ \eqref{H2d2} by the fact that, instead of using Frobenius formula \eqref{fchar}, we only sum over partitions $\nu \vdash d$ which belong to $\Lambda$, i.e. included in the rectangle of size $\ell \times (N - 1)$. For $d \leq \min(\ell,N - 1)$, this is not a restriction, namely $H_{0}^{d,[\ell,N]} = H_{0}^{d}$ exactly encodes double Hurwitz numbers. 

It would be interesting to devise a combinatorial meaning for $H_{0}^{d,[\ell,N]}$, maybe by counting coverings with extra geometric constraints, or counting paths between $C_{\mu_1}$ and $C_{\mu_2}$ in $\mathfrak{S}_{d}$ with certain properties \cite{GJ,HarnadGP}, or to relate it to tau functions \cite{HarnadGP}. It is natural to ask if such a combinatorial interpretation could be extended to higher genus $g$ of the base. Since we already know that we have a CohFT -- or equivalently since we have the topological recursion formula -- all correlation functions $\omega_{g,n}$ are determined by these $[\ell,N]$-restricted double Hurwitz numbers. As there is already a good knowledge of the Chern class of the Verlinde bundle -- including its expression on the boundary for $n = 0$ \cite{MarianOP} -- this would provide an ELSV-like formula \cite{ELSV} for a combinatorial problem yet to be found.

\subsubsection{Remark} In \cite{Karev}, Karev considered a generating series for the numbers \eqref{fchar}, and by combinatorial means, he expressed it in terms of an Airy-like integral over the center of the group algebra of $G$, which can be reexpressed as a product of $1d$ Airy integrals after a suitable change of basis. We shall explain how his results square from the CohFT perspective.

The product on $\mathbb{C}[G]$ and the trace $\sum_{k \in G} c_{k}\,k \mapsto c_{1}$ induces the structure of a semi-simple Frobenius algebra on the center $Z(\mathbb{C}[G])$ -- which is the $\mathcal{A}_{{\rm conj}}$ of Section~\ref{untwi2f}. The conjugacy classes define a basis
$$
e_{\lambda} = \frac{1}{\sqrt{\# C_{\lambda}}} \sum_{k \in C_{\lambda}} k
$$
while the characters define another basis 
$$
\underline{\epsilon}_{\lambda} = \sum_{k \in G} \frac{\chi_{\lambda}(k)}{\sqrt{\# G}}\,k
$$
and by these notations we mean that the properties of Section~\ref{defFb} are satisfied, with $\Delta_{\lambda}^{-1} = (\dim \lambda)^2$. The change of basis matrix is
$$
e_{\lambda} = \sum_{\mu} \underline{S}_{\lambda\mu}^{-1}\,\underline{\epsilon}_{\mu},\qquad \underline{S}_{\lambda\mu}^{-1} = \sqrt{\frac{\# C_{\lambda}}{\# G}}\,\chi_{\mu}(C_{\lambda}) 
$$
and up to rescaling, this is a (unitary) submatrix of the S-matrix of the modular functor described in Section~\ref{defFb}. Then, we consider the trivial  CohFT on $Z(\mathbb{C}[G])$, and apply a rescaling to define
$$
\Omega_{g,n}^*(\underline{\epsilon}_{\lambda_1},\ldots,\underline{\epsilon}_{\lambda_n}) = \Big(\frac{\# G}{\dim \lambda}\Big)^{2g - 2 + n}\,[\overline{\mathcal{M}}_{g,n}] .
$$
If we change basis, we obtain, according to Frobenius formula \eqref{fchar},
$$
\Omega_{g,n}^*\Big(\sum_{k_1 \in C_{\lambda_1}} \frac{k_1}{\# G} \otimes \cdots \otimes \sum_{k_n \in C_{\lambda_n}} \frac{k_n}{\# G}\Big) = \frac{\#\big\{R \in {\rm Hom}(\pi_{1}(\mathbf{\Sigma}_{g,n}),G)\mid R(\mathfrak{l}_i) \in C_{\lambda_i}\big\}}{\# G}\,\cdot [\overline{\mathcal{M}}_{g,n}] .
$$
These numbers are encoded in the CohFT partition function
\beq
\label{CohfZ} Z^{{\rm CohFT}} := \exp\bigg(\sum_{n \geq 1} \sum_{g \geq 0} \frac{\hbar^{g - 1}}{n!} \sum_{\substack{d_1,\ldots,d_n \geq 0 \\ \lambda_1,\ldots,\lambda_n \geq 0}} \bigg\{\int_{\overline{\mathcal{M}}_{g,n}} \Omega_{g,n}^*\big(\bigotimes_{j = 1}^n \underline{\epsilon}_{\lambda_j}\big) \prod_{j = 1}^n \psi_{j}^{d_j} \bigg\} \prod_{j = 1}^n t_{d_j,\lambda_j}\bigg) .
\eeq
Since we have several copies of the rescaled trivial CohFT, $Z^{{\rm CohFT}}$ is a product of suitably rescaled matrix Airy function \cite{Kontsevich}. Karev's partition function is \eqref{CohfZ} after specialization of the times to $t_{d,\lambda} = \xi^{d}p_{\lambda}$ for a set of formal variables $p_{\lambda}$.

\subsection{Bundle comparison and families index formulae}

The Wess-Zumino-Witten ${\rm SU}(N)_{\ell}$ theory can also be approached from the perspective of geometric quantization of moduli spaces of flat ${\rm SU}(N)$ connections. This allows us to give an interpretation of  the CohFT correlations functions via families index formulae -- \eqref{indexform} below -- modulo some identifications (Hypothesis \textbf{I} below) which have not been proved yet to the best of our knowledge. As the statement of current vs. unknown results is subtle, we thought it useful to explain it in detail.

For $g \geq 2$, let $\mathcal{U}_{g,0}^{(N)}$ be the moduli space of pairs $(X,E)$ where $X$ is a Riemann surface of genus $g$ and $E$ a semistable holomorphic vector bundle on $X$ of rank $N$ with trivial determinant. By Narasimhan-Seshadri results \cite{NSth}, $\mathcal{U}_{g,0}^{(N)}$ is homeomorphic to $\mathcal{M}_{g,0}\times M_{g,0}^{{\rm SU}(N)}$, where the last factor is the moduli space of flat ${\rm SU}(N)$ connections on a smooth surface of genus $g$. The fibration $\pi\,:\,\mathcal{U}_{g,0}^{(N)} \rightarrow \mathcal{M}_{g,0}$ is proper, in particular for each choice of complex structure $\sigma$ on the smooth surface of genus $g$, $\pi^{-1}(\sigma)$ is the moduli space of semi-stable bundles of rank $N$ and trivial determinant, which compactifies the smooth part of the fiber consisting of the moduli space of stable rank $N$ bundles with trivial determinant. Caporaso for $N = 1$ \cite{Caporaso}, and Pandharipande \cite{Pandhss} for general $N$, constructed a compactification $\overline{\mathcal{U}}_{g,0}^{(N)}$ of this moduli space, as a projective variety together with a proper fibration $\pi\,:\,\overline{\mathcal{U}}_{g,0}^{(N)} \rightarrow \overline{\mathcal{M}}_{g,0}$.

This picture was later generalized to genus $g$ surfaces with $n \geq 1$ punctures. Here, one fixes a sequence of conjugacy classes $\mathbf{c} = (c_1,\ldots,c_n)$ with finite order in ${\rm SU}(N)$, and consider, on the one hand the moduli space $M_{g,n}^{{\rm SU}(N)}(\mathbf{c})$ of flat ${\rm SU}(N)$ flat connections on a smooth surface such that the holonomies of the connection around the $i$-th puncture belongs to $c_i$ and on the other hand, the moduli space $\mathcal{U}_{g,n}^{(N)}(\mathbf{c})$ of pairs $(X,E)$ where $X$ is a Riemann surface, $E$ is a semistable parabolic holomorphic vector bundle of rank $N$ with trivial determinant with a parabolic structure at the $i$-th puncture prescribed by $c_i$. By Mehta-Seshadri results \cite{MehtaSeshadri}, $\mathcal{U}_{g,n}^{(N)}(\mathbf{c})$ is homeomorphic to $\mathcal{M}_{g,n}\times M_{g,n}^{{\rm SU}(N)}(\mathbf{c})$. In a recent thesis, Schl\"uter \cite{DSchluter} constructs a compactification $\overline{\mathcal{U}}_{g,n}^{(N)}(\mathbf{c})$, giving a fibration of projective varieties $\pi\,:\,\overline{\mathcal{U}}_{g,n}^{(N)}(\mathbf{c}) \rightarrow \overline{\mathcal{M}}_{g,n}$.

For the remaining, we fix the level $\ell \geq 1$. For any $\vec{\lambda} \in \Lambda^n$, we consider the sequence of conjugacy classes $\mathbf{c}_{\vec{\lambda}} := ([\exp(\lambda_i/\ell)])_{i = 1}^{n}$, recalling that $\lambda_i$ is a highest weight at level $\ell$. We shall denote
$$
M_{g,n,\vec{\lambda}}^{(N)} := M_{g,n}^{{\rm SU}(N)}(\mathbf{c}_{\vec{\lambda}}),\qquad \mathcal{U}_{g,n,\vec{\lambda}}^{(N)} := \mathcal{U}_{g,n}^{(N)}(\mathbf{c}_{\vec{\lambda}})
$$

\subsubsection{Geometric quantization via Quillen line bundle}

The moduli space $\overline{\mathcal{U}}_{g,n,\vec{\lambda}}^{(N)}$ carries the Quillen determinant line bundle $\mathscr{L}_{Q}$, and one can consider the $\ell$-th power
\begin{figure}[h!]
\begin{center}
\begin{tikzpicture}[node distance = 2.5cm,auto]
\node (L) {$\mathscr{L}_{Q}^{\otimes \ell}$} ;
\node (Mtop) [right = 1cm of L] {$\overline{\mathcal{U}}_{g,n,\vec{\lambda}}^{(N)}$} ;
\node (Mdown) [below = 1cm of Mtop] {$\overline{\mathcal{M}}_{g,n}$} ;
\draw[->] (L) to node {} (Mtop) ;
\draw[->] (Mtop) to node {$\pi$} (Mdown) ;
\end{tikzpicture}
\end{center}
\end{figure}

Then, by push-forward we obtain the coherent sheaf
$$
[\mathcal{Z}_{\vec{\lambda}}^{(Q)}]_{g,n} = \pi_{!}(\mathscr{L}_{Q}^{\otimes \ell}).
$$
As the restriction of $\mathscr{L}_{Q}^{\otimes \ell}$ to fibers of $\pi$ over $\mathcal{M}_{g,n}$ have no higher cohomology, the restriction of this sheaf to $\mathcal{M}_{g,n}$ defines a bundle. This bundle by its very definition extends to a sheaf over $\overline{\mathcal{M}}_{g,n}$ (it is expected to remain locally free, but we do not use this). By the version of Grothendieck-Riemann-Roch of \cite{BaumFultonMacPherson}, we have that
\beq
\label{GRRindex}
{\rm Ch}\big([\mathcal{Z}_{\vec{\lambda}}^{(Q)}]_{g,n}) =  \pi_*\big( e^{\ell c_{1}(\mathscr{L}_Q)} \cap {\rm Td}({\rm Ker}\,\dd\pi)\big).
\eeq
If we restrict to the degree $0$ part, we retrieve the index formula for the rank of the bundles
\beq
\label{dimZQ} {\rm dim}\,[\mathcal{Z}_{\vec{\lambda}}^{(Q)}]_{g,n} = e^{\ell\varpi} \cap {\rm Td}\big(M_{g,n,\vec{\lambda}}^{(N)}\big) 
\eeq
where $\varpi$ is the first Chern class of the restriction of $\mathscr{L}_{Q}$ to a fiber of $\pi$. We return to $\varpi$ in the next paragraph, but we can already state that it does not depend on the complex structure, compatible with that the dimension is independent of the complex structure, since the higher cohomology groups of the restriction of ${\mathscr L}_Q$ to the fibers of $\pi$ vanishes. 

\subsubsection{Geometric quantization via Chern-Simons and comparison}
\label{GQSEC}

The moduli space of flat ${\rm SU}(N)$ connections over a surface $\Sigma$ of genus $g$ minus the $n$ punctures, with prescribed holonomies $\mathbf{c}$ around the punctures, is equipped with the Goldman symplectic form $\varpi$, and for each $\sigma \in \mathcal{M}_{g,n}$, with a complex structure coming from the Riemann surface $\Sigma_{\sigma}$, making it a K\"ahler variety $(M_{g,n,\vec{\lambda}}^{(N)})_{\sigma}$ with K\"ahler form $\varpi$. It further supports the Chern-Simons line bundle $\mathscr{L}_{{\rm CS}}$, which supports the Chern-Simons connection whose curvature is  $-i\varpi$. The geometric quantization procedure associates to it the bundle $[\mathcal{Z}^{({\rm CS})}_{\vec{\lambda}}]_{g,n} \longrightarrow \mathcal{M}_{g,n}$ whose fiber above $\sigma$ is $H^0\big((M_{g,n,\vec{\lambda}}^{(N)})_{\sigma}, \mathscr{L}^k_{\rm CS}\big)$. 
For each $\sigma \in \mathcal{M}_{g,n}$, using the isomorphism of Narasimhan-Seshadri, one finds that the Chern-Simons line bundle $\mathscr{L}_{{\rm CS}}$ is isomorphic to $\mathscr{L}_{Q}$, as topological bundles on $M_{g,n,\vec{\lambda}}^{(N)}$ and as smooth line bundles over the smooth part of the moduli space. Therefore, $c_1(\mathscr{L}_{Q})$ restricted to the smooth locus of  $M_{g,n,\vec{\lambda}}^{(N)}$ is represented by $ \varpi$. Over $M_{g,n,\vec{\lambda}}^{(N)}$, one has the comparison isomorphism \cite{Fujita} for $n=0$ of holomorphic vector bundles
\beq
\label{CSQcompar}\iota^*\mathscr{L}_{Q} \otimes \pi^*\mathcal{L}_{D}^{\otimes N} \simeq \mathscr{L}_{{\rm CS}}.
\eeq
From this we conclude that over $\mathcal{M}_{g,0}$ we have the isomorphism
\beq
\label{isoQCS} [\mathcal{Z}^{(Q)}]_{g,0} \otimes {\mathcal L}_{D}^{\otimes N\ell} \simeq [\mathcal{Z}^{({\rm CS})}]_{g,0}.
\eeq

It is expected, though currently unproved, that these results also hold for $n > 0$, maybe upon twisting $[\mathcal{Z}_{\vec{\lambda}}^{(Q)}]_{g,n}$ by $\bigotimes_{i = 1}^n \mathcal{L}_{i}^{\ell n_{\lambda_i}}$ for some integers $n_i \in \mathbb{Z}$.

\subsubsection{Comparison with our bundle}

It is clear from the construction of the WZW modular functor in \cite{AU2} that $[\mathcal{Z}_{\vec{\lambda}}^{({\rm CFT})}]_{g,n}$ is isomorphic to the bundle that we obtain in Theorem~\ref{thaa3} by specializing to the WZW modular functor and for the specific choice \eqref{logdeteqn} of log-determinations of twists and central charge. Besides, for $n = 0$ combining the isomorphism \eqref{isoQCS} with the isomorphism from \cite[Section 5.3]{MarianOP}, we have the following isomorphism of holomorphic bundles over $\mathcal{M}_{g,0}$
\beq
\label{isoCSCFT} [\mathcal{Z}^{({\rm CS})}]_{g,0} \simeq [\mathcal{Z}^{(Q)}]_{g,0} \otimes {\mathcal L}_{D}^{\otimes N\ell} \simeq [\mathcal{Z}^{({\rm CFT})}]_{g,0}.
\eeq
The fact that the first isomorphism hold projectively over $\mathcal{M}_{g,0}$ follows directly from the fact that the Picard group of the moduli space semi-stable bundles of rank $n$ bundles with trivial determinant is a copy of the integers. That the second isomorphisms holds projectively over $\mathcal{M}_{g,0}$ is an older result of Laszlo \cite{Laszlo}, which actually also identifies projective the Hitchin connection \cite{Hi} in the second bundle with the TUY-connection in the last bundle \cite{TUY}. See also \cite{A1} for the projectively identification of the Axelrod, Della-Pietra and Witten \cite{ADW} connection in the first bundle and the Hitchin connection in the second bundle. 

Restricting over $\mathcal{M}_{g,0}$, we see that the bundle $[\mathcal{Z}_{\vec{\lambda}}^{(Q)}]_{g,0}$ coincide with our bundle of Theorem~\ref{thaa3} for a non-standard choice (compare to \eqref{logdeteqn}) of log-determination $c'$ of the central charge
\beq
\label{cnons} c' = \frac{\ell(N^2 - 1)}{\ell + N} + 2N\ell
\eeq

It is expected, but to our knowledge currently unproved, that
\begin{itemize}
\item[$\bullet$] the sheaf $[\mathcal{Z}^{(Q)}]_{g,0}$ over $\overline{\mathcal{M}}_{g,0}$ remains locally free over the Deligne-Mumford compactification, i.e. exists as a bundle over $\overline{\mathcal{M}}_{g,0}$.
\item[$\bullet$] the second isomorphism in \eqref{isoCSCFT} extends to $\overline{\mathcal{M}}_{g,0}$.
\item[$\bullet$] these results extend for $n > 0$, so that there is an isomorphism of bundles over $\overline{\mathcal{M}}_{g,n}$:
$$
[\mathcal{Z}^{(Q)}_{\vec{\lambda}}]_{g,n} \otimes {\mathcal L}_{D}^{\otimes N\ell} \otimes \bigotimes_{i = 1}^n \mathcal{L}_{i}^{\ell n_{\lambda_i}} \simeq [\mathcal{Z}^{({\rm CFT})}_{\vec{\lambda}}]_{g,n}.
$$
for some integers $(n_{\lambda}')_{\lambda \in \Lambda}$.
\end{itemize}
For convenience, we call this expectation Hypothesis \textbf{I}. If it were true, it would mean that $[\mathcal{Z}^{(Q)}_{\vec{\lambda}}]_{g,n}$ coincide with the bundle we construct in Theorems~\ref{th2}-\ref{thaa3} for the choice \eqref{cnons} of log-determination of the central charge, and for the non-standard choice (compare to \eqref{logdeteqn}) of log-determination of the Dehn twist eigenvalues
$$
r'_{\lambda} = \frac{\langle \lambda,\lambda + 2\rho \rangle}{2(\ell + N)} - n_{\lambda}'.
$$

\subsubsection{Families Index interpretation of the CohFT correlation functions}

If Hypothesis \textbf{I} holds, we deduce that the $\omega_{g,n}$ of the CohFT we produce for the specific choice of log-determinations \eqref{logdeteqn}, have an index interpretation. By homogeneity, it is enough to state it for $t = 1$ and $t = 0$. For $t = 1$, following \eqref{GRRindex} we have for $n \geq 1$
\beq
\label{indexform} \sum_{\vec{\lambda} \in \Lambda^n} \prod_{i = 1}^n \underline{S}^{-1}_{\mu_i\lambda_i} \omega_{g,n}^{[\vec{\lambda}]}(\zeta_1,\ldots,\zeta_n) =\tilde\pi_*\left( \prod_{i = 1}^n \Xi(\psi_{i},h_{\mu_i}^{1/2}\zeta_i) \cup \pi_*\big(e^{\ell c_{1}(\mathscr{L}_Q)} \cap {\rm Td}({\rm Ker}\,\dd\pi)\big)\right)
\eeq
where $\tilde \pi$ is the morphism from $\overline{\mathcal{M}}_{g,n}$ to a point. On the left-hand side, we use the notations of \S~\ref{sec:basischange}, and on the right-hand side, we recall that
$$
h_{\mu} = r_{\mu} + \frac{c}{24},\qquad \Xi(u,\zeta) = \sum_{d \geq 0} \frac{\Gamma[2d + 2;\zeta]\,\dd\zeta}{2^{d}d!\,\zeta^{2d + 2}}\,u^{d}
$$
And, for $n = 0$ and $g \geq 2$
$$
F_{g} = \tilde\pi_*\big( e^{\ell c_{1}(\mathscr{L}_Q)} \cap {\rm Td}({\rm Ker}\,\dd\pi) \big).
$$

For $t = 0$, we do not need Hypothesis \textbf{I}. As explained in \S~\ref{Sec432} the left-hand side computed by topological recursion factors into the Verlinde dimensions and the intersection numbers over $\overline{\mathcal{M}}_{g,n}$
$$
\sum_{\vec{\lambda} \in \Lambda^n} \prod_{i = 1}^n \underline{S}^{-1}_{\mu_i\lambda_i} \omega_{g,n}^{[\vec{\lambda}]}(\zeta_1,\ldots,\zeta_n) = D_{\vec{\mu}}(\mathbf{\Sigma}_{g,n}) \omega_{g,n}^{{\rm KdV}}(\zeta_1,\ldots,\zeta_n).
$$
On the other hand, it is known, see e.g. \cite{A}, Theorem 8.1, that $D_{\vec{\mu}}(\mathbf{\Sigma}_{g,n})$ are given by the index formula \eqref{dimZQ}. Recalling that $\omega_{g,n}^{{\rm KdV}}$ is a generating series of intersection of $\psi$-classes over $\overline{\mathcal{M}}_{g,n}$, this can be rewritten as follows
\beq
\label{t0form} \sum_{\vec{\lambda} \in \Lambda^n} \prod_{i = 1}^n \underline{S}^{-1}_{\mu_i\lambda_i} \omega_{g,n}^{[\vec{\lambda}]}(\zeta_1,\ldots,\zeta_n) = \bigg(e^{\ell\varpi} \cap {\rm Td}\big(M_{g,n,\vec{\lambda}}^{(N)}\big) \bigg)\bigg( \int_{\overline{\mathcal{M}}_{g,n}} \prod_{i = 1}^n \Xi(\psi_i,\zeta_i)\bigg)
\eeq
with
$$
\Xi(u,\zeta) = \sum_{d \geq 0} \frac{(2d + 1)!!\dd\zeta}{\zeta^{2d + 2}}\,u^{d}.
$$
The factorization between the moduli space of curves, and the moduli space of flat connections over a smooth surface, is nicely displayed in \eqref{t0form}, while for general $t$ the $\omega_{g,n}$ is interpreted by (\ref{indexform}) as a families index of the determinant line bundle over the universal moduli space of bundles, seen as a family over $\overline{\mathcal M}_{g,n}$.

\section{Discussion about global spectral curves}
\label{sec:glob}
\label{Globals}

A  global spectral curve is by definition a Riemann surface $\Sigma$ equipped with a branched cover $x\,:\,\Sigma \rightarrow \mathbb{P}^1$. One retrieves a local spectral curve by considering the disconnected neighborhoods of the ramification points of $x$. The only difference in the axiomatics is that, in the global case, we require that\footnote{Although not strictly necessary, one often adds the assumption that $\omega_{0,1} \in H^0(\Sigma',K_{\Sigma'})$ for a dense open subset $\Sigma' \subseteq \Sigma$.}
$$
\omega_{0,2} \in H^0(\Sigma^2,K_{\Sigma}^{\boxtimes 2}(-2 \Delta))^{\mathfrak{S}_{2}}
$$
and therefore \eqref{Xidef1} defines meromorphic $1$-forms on $\Sigma$, \textit{i.e.} $\Xi_{d,i} \in H^0(\Sigma,K_{\Sigma}((2d + 2)\{o_i\}))$.

Although the topological recursion is well-defined for local spectral curves as presented in Section~\ref{Locals}, it was originally defined on global spectral curves and this allows further properties and manipulations \cite{EOFg,BEthink}. It is thus desirable, whenever we have a local spectral curve, to realize it embed in a global spectral curve. This is not always possible, but given the geometric origin of modular functors and general ideas from mirror symmetry, we may still hope it admits a global description.

\subsection{Landau-Ginzburg models and Frobenius manifolds}

Among the different realizations of a Frobenius manifold, let us focus on a particular one which gives rise in a natural way to global spectral curves -- or more generally, varieties.

In singularity theory, one is interested in the study of Landau-Ginzburg (LG) models. They are defined by a family of germ of functions
$$
x_a\,:\, \mathbb{C}^d \longrightarrow \mathbb{C}, 
$$
called "potential" and parameterized by a point $a$ living in a $k$-dimensional manifold $A$ obtained as a miniversal deformations of a function $x_0$ with an isolated singularity at $z = 0$ with Milnor number $k$. The tangent spaces $\mathcal{A}_{a} := T_{a} A$ are naturally identified with the Jacobi ring 
$$
{\rm Jac}(x_a) := \frac{\mathbb{C}[z_1,\ldots,z_{d}]}{\langle \partial_{z_1} x_{a} ,\ldots, \partial_{z_d} x_{a} \rangle} .
$$
The multiplication in this polynomial ring is denoted $\times$, and the unit is $1$. For any holomorphic volume form $\Omega$, one can define the residue pairing by
\beq
\label{metri} \forall (\phi,\psi) \in {\rm Jac}(x_a),\qquad  b(\phi,\psi) := \frac{1}{(2{\rm i}\pi)^{d}}\,\oint_{|\partial x_a/\partial z_1| = \epsilon_1} \!\!\!\!\!\!\!\!\!\!\!\!\! \cdots \,\,\,\,\,\, \oint_{|\partial x_a/\partial z_d|  = \epsilon_d} \frac{\phi(z)\,\psi(z)\,\Omega(z)}{\prod_{i=1}^d \partial x_a/\partial z_i}
\eeq
for small enough $\epsilon_i$'s. According to Saito \cite{Saitoprimitive}, there exists a volume form -- called primitive -- such that \eqref{metri} is a flat metric. It thus provides ${\rm Jac}(x_a)$ with a Frobenius structure, promoting $(\mathcal{A}_{a})_{a}$ to a Frobenius manifold $A$. If for a generic $a \in A$, $x_a$ has only isolated Morse singularities, the associated CohFT is semi-simple. We denote by $(t_i)_{i = 1}^k$ flat coordinates on an open set  $A' \subseteq A$, $(\varphi_{i})_{i = 1}^k$ the corresponding frame in the tangent bundle, $\nabla$ the Levi-Civita connection, and
$$
\varphi_i \times \varphi_j = \sum_{\ell = 1}^k N_{ij}^{\ell}(t)\,\varphi_{\ell}
$$
the multiplication in the Jacobi ring. We also denote $(a_i)_{i = 1}^k$ canonical coordinates (that we can also assume to be defined on $A'$), i.e. such that $(\partial_{a_i})_{i = 1}^k$ form a canonical basis of the tangent bundle. To connect with the notations of Section~\ref{defFb}, $\partial_{a_i} = \tilde{\epsilon}_i$ and $\Delta_i = 1/b(\partial_{a_i},\partial_{a_i})$, and let $\Psi \in {\rm End}(T^*A')$ the change of basis from the flat to the canonical coordinates
$$
\Psi(\dd t_i) = \Delta_i^{-1/2}\,\dd a_i .
$$

In this example of Frobenius manifold, the $R$-matrix can be written in terms of oscillating integrals. Indeed, given a Frobenius manifold $A$, the axioms of a CohFT ensure that
$
\nabla_i^{(u)}:= \nabla_i + u^{-1}\varphi_i \times
$
forms a flat pencil of connections on $A$ parametrized by $u \in \mathbb{C}^*$. This ensures the compatibility of the following PDEs for a section $J(u) = \sum_{i = 1}^k J_i(u)\dd t_i$ of $T^*A'$.
$$
\forall i,j \in \llbracket 1 , k\rrbracket,\qquad  u \partial_{t_i} J_j(u;t)  = \sum_{\ell =1}^k N_{ij}^{l}(t)\,J_{l}(u;t) .
$$
If $A'$ is semi-simple and conformal\footnote{Conformal here means there exists a vector field $E$ on $A'$ such that the Lie derivative $\mathfrak{L}_{E}$ acts by multiplication by a scalar on the metric tensor, on the product tensor, and on the unit vector field.}, there is a unique basis of solutions \cite{Dubrovin,Giventals} such that, in matrix form one has that
\beq
\label{JR} J(u;t) =  \Psi R(u;t) \exp(\mathbf{a}(t)/u)
\eeq
where $\mathbf{a} \in {\rm End}(T^*A')$ is defined such that $\textbf{a}(\dd a_i) = a_i\,\dd a_i$ and $R(u;t) \in {\rm End}(T^*A')[[u]]$ is an operator satisfying a unitarity and homogeneity condition. This $R$-matrix is the one appearing in Section~\ref{Rmatriact}.

For the Frobenius manifold attached to a Landau-Ginzburg model with potential $x_t$, one can find a simple integral representation of this solution of the form (in flat coordinates)
\beq
J_{ij}(u;t) = \frac{1}{(2{\rm i}\pi u)^{d/2}}\,\int_{\Gamma_j} \exp[x_t(z)/u]\,\phi_i(z)\,\Omega(z)
\eeq 
where $(\Gamma_j)_{j = 1}^k$ are cycles in $\mathbb{C}^d$ which can be constructed through Morse theory of the function $\mathrm{Re}[x_t/u]$. By the Fubini theorem, this integral can be rewritten as an integral along the path $\mathcal{X}_j = x_{t}(\Gamma_j) \subset \mathbb{C}$, whose integrand itself is an integral over some vanishing cycle $\gamma_j(X) \subseteq x_{t}^{-1}\{X\}$ over a point $X \in \mathcal{X}_j$
$$
J_{ij}(u;t) = \frac{1}{(2{\rm i}\pi u)^{d/2}}\,\int_{\mathcal{X}_j} e^{X/u}\,\int_{\gamma_j(X)} \phi_i\,\Omega .
$$

Comparing with \eqref{JR}, this provides us with an integral representation of the $R$-matrix of the form
$$
R_{ij}(u;t) = \frac{\Delta_i^{1/2}}{(2{\rm i}\pi u)^{d/2}}\,\int_{\mathcal{X}_j} e^{(X - a_j(t))/u}\int_{\gamma_j(X)} \phi_i\,\Omega
$$
and the condition that $R(u = 0) = {\rm id}$ implies that the paths $\mathcal{X}_j$ start from $a_j(t)$, and extends to $-\infty$.

\subsection{$1d$ Landau-Ginzburg models}

The realization of a given Frobenius manifold via a LG model, when it exists, has no reason to be unique, and several realizations may have different dimensions. It is particularly interesting, for tractability, if a $1$-dimensional LG model can be found. This means having a family of Riemann surfaces $\Sigma$ together with a family of potentials $x_t\,:\,\Sigma \rightarrow \mathbb{C}$ such that the local algebra of $x_t$ at the ramification points is isomorphic to the tangent space $T_tA$ of our Frobenius manifold $A$. In this case, the vanishing cycle $\gamma_j(X)$ is just an ordered pair of points $\{z(X),\sigma_j(z(X))\} \subseteq \Sigma$ above $X \in \mathbb{C}$, related to each other by the (analytic continuation of the) local involution $\sigma_j$ permuting the two sheets that meet at the ramification point above $a_j$. In flat coordinates, the $R$-matrix then reads
\beq
\label{RRRR}R_{ij}(u;t) = \frac{\Delta_i^{1/2}}{(2{\rm i}\pi u)^{1/2}}\,\oint_{\mathcal{X}_j} e^{(X - a_j(t))/u}\,[\varphi_i\,\omega]^{{\rm odd}}_{\sigma_j}
\eeq
for some $1$-form $\om$ on $\Sigma$, and thimbles $(\mathcal{X}_j)_{j = 1}^k$. Here we have used the notation $[f]^{{\rm odd}}_{\sigma} := f - \sigma^* f$.
 
\subsection{Fusion potentials for modular functors}

In this paragraph, we give an easy argument to obtain a $1d$ LG model for the Frobenius algebra $\mathcal{A}$ of any modular functor (see Section~\ref{FMF}), i.e. a function $x_0$ -- called the "fusion potential" -- whose Jacobi ring is isomorphic to $\mathcal{A}$. Then the Frobenius manifold will be (at least locally) defined by considering miniversal deformations $(x_a)_a$ of the potential. For the Wess-Zumino-Witten models (Section~\ref{ex:WZW}), this has been posed as a conjecture by Gepner \cite{Gepner}, which was later answered by Di Francesco and Zuber \cite{DFZ} and further studied by Aharony \cite{Aha}. We borrow the idea of \cite{DFZ} to propose a family of fusion potentials for the Frobenius algebra of any modular functor, which hinges on the commutativity of the curve operators $(\mathcal{C}[\beta;\lambda])_{\lambda \in \Lambda}$.

For any vector $f \in  \mathbb{C}[\Lambda]$, let us define
$$
\mathcal{C}_{f}[\beta] := \sum_{\lambda \in \Lambda} f_{\lambda}\,\mathcal{C}[\beta;\lambda] .
$$
Its characteristic polynomial reads
$$
P_{f}(\eta) := \det(\eta - \mathcal{C}_{f}[\beta]) = \prod_{\mu \in \Lambda} (\eta - \mathfrak{c}_{\mu,f}),\qquad \mathfrak{c}_{\mu,f} = \sum_{\lambda \in \Lambda} f_{\lambda}\,\mathfrak{c}_{\mu}[\lambda]
$$
with the eigenvalues $\mathfrak{c}_{\mu}[\lambda]$ of the curve operators given in \eqref{eignde}. Let $Q_{f}(\eta)$ be any polynomial such that $Q_{f}'(\eta) = P_{f}(\eta)$.

\begin{lemma}
For generic $f \in \mathbb{C}[\Lambda]$, the Jacobi ring of $x_0\,:\,\mathbb{C} \rightarrow \mathbb{C}$ defined by $x_0 = Q_{f}(\eta)$ is isomorphic to the Frobenius algebra $\mathcal{A}$ of the modular functor.
\end{lemma}

\noindent \textbf{Proof.} The Jacobi ring is ${\rm Jac}(x_0) = \mathbb{C}[\eta]/\langle P_{f}(\eta) \rangle$. For generic $f$, the roots $(c_{\mu,f}[\mu])_{\mu \in \Lambda}$ are pairwise distinct. Therefore, we can construct by Lagrange interpolation a unique set of polynomials $(\varphi_{\lambda,f})_{\lambda \in \Lambda}$ such that
$$
\forall \mu \in \Lambda,\qquad \varphi_{\lambda,f}(\mathfrak{c}_{\mu,f}) = \mathfrak{c}_{\mu}[\lambda].
$$
This is actually a basis of the Jacobi ring. Indeed, if there is a relation $\sum_{\lambda \in \Lambda} k_{\lambda}\,P_{\lambda}(\eta) = 0$, by evaluation at $\mathfrak{c}_{\mu,f}$ and using the expression \eqref{eignde} we would have that
$$
\forall \mu \in \Lambda,\qquad \sum_{\lambda \in \Lambda} k_{\lambda}\,(\underline{S}^{-1})_{\lambda\mu} = 0
$$
which implies $k_{\lambda} = 0$ for any $\lambda \in \Lambda$ since $(\underline{S}^{-1})^{\top}$ is invertible. Then, by definition and evaluation at the eigenvalues, their multiplication modulo the annihilating polynomial $P_{f}$ reads
$$
\varphi_{\lambda}(\eta) \varphi_{\mu}(\eta) = \sum_{\nu \in \Lambda} N_{\lambda\mu\nu^{\dagger}}\,\varphi_{\nu}(\eta)\,\,\, {\rm mod}\,\,P_{f}(\eta) .
$$
Thus, we have build a basis of the Jacobi ring in which we identify the Frobenius algebra of Section~\ref{FMF}. \hfill $\Box$

\vspace{0.2cm}

Notice that, knowing $(\mathfrak{c}_{\mu,f})_{\mu}$, the position of the ramification points appearing in \eqref{RRRR} is
$$
a_{\mu} := x(o_{\mu}) = Q_{f}(\mathfrak{c}_{\mu,f}) .
$$
Alternatively, one can decide to impose the values $(a_{\mu})_{\mu \in \Lambda}$, and ask for the determination of roots $(\eta_{\mu})_{\mu \in \Lambda}$ and a polynomial $Q$ of degree $|\Lambda| + 1$ such that
\beq
\label{rerp}\forall \mu \in \Lambda,\qquad Q(\eta_{\mu}) = a_{\mu}\qquad {\rm and} \qquad Q'(\eta_{\mu}) = 0 .
\eeq
Remark that  $Q(\eta),(\eta_{\mu})_{\mu}$ is a solution iff $Q(\eta) \leftarrow Q(\gamma\eta + \gamma')$ and $\eta_{\mu} \leftarrow \gamma^{-1}(\eta_{\mu} - \gamma')$ is a solution. Hence, \eqref{rerp} represents $2|\Lambda|$ constraints for the same number of independent unknowns, and can be solved for generic $a$'s, and we can find the corresponding $f$'s by the formula
$$
f_{\lambda} = \sum_{\mu \in \Lambda} \underline{S}_{\mu\lambda}\,\eta_{\mu}\,(\underline{S}^{-1})_{1\mu} .
$$

\subsection{Open question: global spectral curve for modular functors}

This defines, at least locally around an (arbitrary, generic) origin $a^0$, a $1d$ LG model $(x_{a})_{a}$ that we can try to use to describe as a "global curve" $\Sigma_{a}$ for the modular functor. However, the explicit construction of a primitive form and a good set of integration cycles is a serious issue in general which we have not been able address so far. To restate (a part of) the problem in an elementary way, it is not easy to find a $1$-form $\omega_{0,1}$ on $\Sigma_{a}$ whose expansion at the ramification point is equal to \eqref{finalw02} modulo the even part -- at least when $a := a(t)$ is coupled in some way to the single parameter $t$ on which our local curve depend.

In a handful of examples, one can rely on an indirect derivation of a (higher-dimensional) Landau-Ginzburg model through mirror symmetry. For instance, for the $\mathfrak{sl}_{N}$, level $\ell$ WZW models recapped in Section~\ref{ex:WZW}, Witten \cite{Wittengrass} proved that the Frobenius algebra $\mathcal{A}$ is isomorphic to the quantum cohomology of a Grassmannian. On the other hand, the Landau-Ginzburg model mirror dual to Grassmannians has been built in \cite{MR}, but, in most cases, the issue of finding a primitive form and a good basis of cycles has not been solved yet, except in some very simple examples such as $\mathbb{P}^2$ \cite{Gross}. The complete construction of a global spectral curve thus remains an open problem, even in simple examples. It must be addressed if, for instance, one wishes to take advantage of the topological recursion to study the level $\ell \rightarrow \infty$ limit in WZW models.

\appendix

\section{Extra properties of the $S$-matrix}
\label{Sextra}

In this section we derive the symmetries of the $S$-matrix in the case the modular functor in question has duality or is unitary. 

\subsubsection*{MF-U}

Let us first assume that the modular functor $\mathcal{V}$ is unitary. We recall from Section~\ref{Referencespace} that $\Sigma_{1}$ is a closed oriented surface of genus one and $\alpha,\beta$ two oriented simple closed curves on $\Sigma_{1}$, such that they intersect in exactly one point with intersection number one. Then we have two marked surfaces $\Sib^{(\alpha)}_{1} = (\Sigma_{1}, \{\alpha\})$ and $\Sib_1^{(\beta)} = (\Sigma_{1}, \{\beta\})$. Both $\mathcal{V}(\Sib^{(\alpha)}_{1})$ and $\mathcal{V}(\Sib^{(\beta)}_{1})$ are equipped with hermitian inner products, which are compatible with the factorization isomorphisms (Section~\ref{FMFV})
$$\mathcal{V}({\Sib}^{(\alpha)}_{1}) \cong \bigoplus_{\m \in \L}
\mathcal{V}({\Sib}_{0,2},\m,\m^{\dagger})
$$
and 
$$\mathcal{V}({\Sib}^{(\beta)}_{1}) \cong \bigoplus_{\m \in \L}
\mathcal{V}({\Sib}_{0,2},\m,\m^{\dagger}).
$$
The basis $\zeta[\mu] \in \mathcal{V}({\Sib}_{0,2},\m,\m^{\dagger})$ induces via these isomorphisms the basis $e_{\mu}$ of $\mathcal{V}({\Sib}^{(\alpha)}_{1})$ and $\varepsilon_{\mu}$ of $\mathcal{V}({\Sib}^{(\beta)}_{1})$. The unique orientation preserving diffeomorphism $S$ that sends $(\alpha,\beta)$ to $(\beta,-\alpha)$ sends the $e$-basis to the $\varepsilon$-basis:
$$
\varepsilon_{\mu} = \sum_{\lambda \in \Lambda} S_{\mu\lambda}\,e_{\lambda}\,.
$$
Now we simply compute
\begin{eqnarray*}
\delta_{\lambda\mu} & = & \big(\zeta[\lambda],\zeta[\mu]\big)\\
& = & \big(e_{\lambda}, e_{\mu}\big)\\
& = & \big(\mathcal{V}(S)(e_{\lambda}), \mathcal{V}(S)(e_{\mu})\big)\\
& = & \sum_{\rho,\tilde{\rho} \in \Lambda}S_{\lambda\rho}^* S_{\mu\tilde{\rho}}\,\big(\varepsilon_\rho, \varepsilon_{\tilde\rho}\big)\\
& = & \sum_{\rho \in \Lambda} S_{\lambda\rho}^*\overline{S}_{\mu\rho}\,.
\end{eqnarray*}
where ${}^*$ here denotes complex conjugation. Hence we get that the $S$-matrix is unitary
$$ (S^*)^{\top}S = 1$$
when the modular functor is unitary.

\subsubsection*{MF-D}

Let us now instead assume that the modular functor $\mathcal{V}$ satisfies the orientation reversal axiom. Let now $C_\alpha$ be the orientation reversing self-diffeomorphism of $\mathbf{\Sigma}_{1}$ which maps $(\alpha,\beta)$ to $(\alpha,-\beta)$. Let $C_\beta$ the same as $C_\alpha$, except that the roles of $\alpha$ and $\beta$ are exchanged. Then, we have the following commutative diagram, where ${}^{\star}$ denotes the dual operation

$$\begin{CD} \mathcal{V}({\Sib_1^{(\alpha)}}) @>{\mathcal V}(S)>>
 \mathcal{V}({\Sib_{1}^{(\beta)}})\\ 
@V\mathcal{V}(C_\alpha)VV @VV\mathcal{V}(C_\beta)V\\
 \mathcal{V}({-\Sib_1^{(\alpha)}}) @>{\mathcal V}(S)>>  \mathcal{V}({-{\Sib}_{1}^{(\beta)}})\\
@V{\cong}VV @VV{\cong}V\\ 
\mathcal{V}({\Sib_1^{(\alpha)}})^{\star} @>{\mathcal V}(S^{-1})^{\star} >>
 \mathcal{V}({\Sib_{1}^{(\beta)}})^{\star}.
\end{CD}$$

\vspace{0.2cm}

Let us denote the dual basis of $(e_\lambda)_{\lambda}$ by $(e_\lambda^{\star})_{\lambda}$. We see, by compatibility between the glueing isomorphism and the orientation reversal isomorphism, that $e_\lambda$, under the composition of the two maps in the first column in the above diagram, is taken to $(e_{\lambda^\dagger}^{\star})_{\lambda}$ and likewise for $\beta$ in the last column. But since we have the following easy computation
\begin{eqnarray*}
{\mathcal V}(S^{-1})^{\star}(e_{\lambda^\dagger}^{\star})(\varepsilon_{\mu}) & = & e_{\lambda^\dagger}^{\star}({\mathcal V}(S^{-1})(\varepsilon_{\mu}))\\
& = & e_{\lambda^\dagger}^{\star} \left(\sum_{\rho \in \Lambda} S^{-1}_{\mu\rho}e_{\rho}\right) \\
& = & S^{-1}_{\mu\lambda^\dagger},
\end{eqnarray*}
we see that the orientation reversal axiom implies that
$$ S_{\lambda\mu} = S^{-1}_{\mu\lambda^\dagger}.$$

\newpage

\providecommand{\bysame}{\leavevmode\hbox to3em{\hrulefill}\thinspace}
\providecommand{\MR}{\relax\ifhmode\unskip\space\fi MR }
\providecommand{\MRhref}[2]{%
  \href{http://www.ams.org/mathscinet-getitem?mr=#1}{#2}
}
\providecommand{\href}[2]{#2}

\end{document}